\documentclass[
    aps,
    superscriptaddress,
    10pt,
    prd,
    notitlepage,
    twocolumn,s
    tightenlines,
    nofootinbib,
    floatfix]{revtex4-2}

\usepackage[linktocpage,breaklinks]{hyperref}
\usepackage[usenames,dvipsnames]{xcolor}

\usepackage{aas_macros}
\usepackage{amsmath}
\usepackage{amssymb}
\usepackage{amsfonts}
\usepackage{txfonts}
\usepackage{bm}
\usepackage{stmaryrd}
\usepackage{tensor}
\usepackage{mathrsfs}
\usepackage[utf8]{inputenc}
\usepackage{url}
\usepackage{makecell}

\usepackage{graphicx}
\usepackage{epsfig}
\usepackage{epstopdf}
\usepackage[normalem]{ulem}

\usepackage{natbib}
\usepackage{hyperref}
\usepackage{cleveref}
\hypersetup{colorlinks=true,
            citecolor=Black,
            linkcolor=Black,
            urlcolor=Black}

\newcommand{\paramvec}{\boldsymbol{\theta}}
\newcommand{\software}[1]{\texttt{#1}}

\newcommand{\GR}{\mbox{\tiny GR}}

\makeatletter
\newcommand{\pseudoeqref}[1]{\textup{\tagform@{#1}}}
\makeatother

\graphicspath{{figures/}}

\newcommand{\icasu}{Illinois Center for Advanced Study of the Universe \&  Department of Physics, University of Illinois at Urbana-Champaign, Champaign, Illinois 61820 USA}
\newcommand{\caps}{Center for AstroPhysical Surveys, National Center for Supercomputing Applications, Urbana, Illinois 61801 USA}

\begin{document}
\title{Are Parametrized Tests of General Relativity with Gravitational Waves Robust \\ to Unknown Higher Post-Newtonian Order Effects?}
\begin{abstract}
Gravitational wave observations have great potential to reveal new information about the fundamental nature of gravity, but extracting that information can be difficult.
One popular technique is the parametrized inspiral test of general relativity (a realization of the parametrized post-Einsteinian framework), where the gravitational waveform, as calculated in Einstein's theory as a series expansion in the orbital velocity, is parametrically deformed at a given set of orders in velocity.
%
%
%
However, most current approaches usually only analyze the data while considering a single, specific modification at a time. 
Are then constraints placed with a single modification robust to our ignorance of higher post-Newtonian order corrections? 
We show here that for a wide class of theories, specifically those that admit a post-Newtonian expansion, single-parameter tests are indeed robust. 
In particular, through a series of full Bayesian parameter estimation studies on several different sets of synthetic data, we show that single-parameter constraints are not degraded but rather are improved by the inclusion of multiple parameters, provided one includes information about the mathematical structure of the series.
We then exemplify this with a specific theory of gravity, shift-symmetric scalar Gauss-Bonnet theory, where the waveform has been calculated to higher post-Newtonian orders than leading. 
We show that the inclusion of these higher order terms strengthens single-parameter constraints, instead of weakening them, and that the strengthening is very mild.
This analysis therefore provides strong evidence that single-parameter post-Einsteinian tests of general relativity are robust to ignorance of high post-Newtonian order terms in the general relativistic deformations. 

\end{abstract}
\author{Scott Perkins}
\email{scottep3@illinois.edu}
\affiliation{\icasu}
\affiliation{\caps}
\author{Nicol\'as Yunes}
\email{nyunes@illinois.edu}
\affiliation{\icasu}
\date{\today}
\maketitle

\section{Introduction}\label{sec:intro}

As the field of gravitational wave (GW) astrophysics matures, the growing amount of data is being used in a variety of ways and in different contexts to solve problems in topics like fundamental physics, astrophysics, and cosmology.
One critical subject that has drastically benefited by the observation of GWs is the study of the gravitational interaction.
General relativity (GR), our best theory of gravity to date, has some shortcomings, if we may call them that~\cite{Will:2014kxa,Berti:2015itd}.
These include the incompatibility of GR with quantum mechanics (c.f. Ref.~\cite{Shomer:2007vq}) and the singularity problem~\cite{Senovilla:2014gza,Penrose:1964wq}, as examples.
Aside from these shortcomings, the theory also struggles to adequately explain certain phenomena observed in our universe without the inclusion of additional structure, like the late time acceleration of the universe (unless one includes an ``unnaturally'' small cosmological constant) ~\cite{SupernovaSearchTeam:1998fmf,SupernovaCosmologyProject:1998vns}, the rotation curves of galaxies (unless one includes dark matter particles that have so far been undetected via direct experiment)~\cite{Sofue:2000jx,Bertone:2016nfn}, and the matter-antimatter asymmetry of the universe (unless one includes new parity-violating interactions that satisfy the Sakharov conditions)~\cite{Canetti:2012zc}.
Modifications to GR in the ultraviolet or in the infrared may potentially resolve some of these theoretical and observational issues, which is why they have attracted attention lately in light of new GW data.

If we are searching for or attempting to constrain ``alternatives'' to GR, GW observations are an excellent place to look~\cite{Carson:2020rea,Yunes:2016jcc}.
The dynamics of the system that generates them are highly relativistic (reaching characteristic speeds up to a substantial fraction of the speed of light), and involve large gravitational potentials and fields.
However, deducing information about fundamental physics from GW data is a nuanced business, where the information encoded inside the waveform can, in some cases, be very faint and buried in noise.
One possible philosophy is to take a maximally ``ignorant'' approach, where one seeks to involve as little outside information as possible.
This approach has its appeal in terms of its perceived robustness (uninfluenced by any external, \emph{a priori} information), but the exact method with which one encodes this ``ignorance'' is not a trivial or well defined matter.
Furthermore, a maximally uninformed approach can minimize and sometimes even erase the strength of any inferences about fundamental physics one might hope to extract~\cite{LIGOScientific:2016lio}. 
While it would be appealing to conduct these analyses independently of our priors, this seems to be a quixotic approach. 

For example, one maximally uninformed approach would be to reconstruct the signal with an orthogonal basis of wavelets, as is normally done with \software{BayesWave}~\cite{Cornish:2014kda,Cornish:2020dwh}. 
The fact that this method is totally agnostic to what physics produced the signal allows one to reconstruct the signal to almost arbitrary precision, which is useful in many contexts. However, such an approach usually provides little information about the theory of gravity that led to the signal that is being reconstructed, or more specifically about the fundamental physics involved in the generation and propagation of GW in our universe (although there have been recent attempts to tie in to certain topics in fundamental physics~\cite{Chatziioannou:2021mij}). 
Given this, it's useful to remember that general tests like ``residual tests,'' in which one studies the signal-to-noise ratio contained in the difference between the data and the best re-constructed signal~\cite{Cornish:2011ys,Vallisneri:2012qq}, are inherently a consistency check of the reconstruction procedure, and should not be interpreted as direct tests of the underlying physics.
The value of such tests is only as great as the accuracy of the underlying reconstruction models, regardless of what physics those models are trying to represent.

Meanwhile, a less uninformative approach, though still agnostic, might attempt to model the signal as a GW produced in GR, but with some \textit{small deformation} that is represented parametrically.
This type of approach defines \textit{the parametrized post-Einsteinian (ppE) formalism}~\cite{Yunes:2009ke,Cornish:2011ys,Sampson:2013lpa,Chatziioannou:2012rf}, which is the general framework that the ``parametrized inspiral tests of GR,'' used by the LIGO/Virgo collaboration (LVC), is built from~\cite{Mishra:2010tp,Arun:2006yw,LIGOScientific:2019fpa,LIGOScientific:2020tif,Abbott:2020jks,LIGOScientific:2021sio,Li:2011cg,Agathos:2013upa,Meidam:2017dgf,LIGOScientific:2018dkp,LIGOScientific:2016lio}. 
In this approach, deformations from GR are incorporated directly at the level of the model for the signal (i.e.~the GW template or GW model) that one will use to compare against the data. Moreover, these deformations are represented parametrically through a particular basis (i.e.~in the inspiral, a polynomial in orbital velocity), with the polynomial exponent of the leading post-Newtonian (PN) order term encoding the type of modification one is considering, and the polynomial coefficient the strength of the deformation. 

The resulting functional form of the deformation of the signal is not arbitrary. In fact, one can easily show that such a representation derives from parametrically deforming the binding energy of the binary, the rate of change of this binding energy, or the dispersion relation of the propagating GW through a polynomial in velocity~\cite{Yunes:2009ke}. Such an analysis also reveals that there is a \textit{mapping} between deformations in these physical quantities and those that are introduced in the waveform, thus allowing for constraints on the waveform deformations to then be \textit{mapped back} to constraints on theoretical physics~\cite{Yunes:2009ke,Yunes:2016jcc,Tahura:2018zuq,Perkins:2020tra}. Moreover, a polynomial in velocity is a natural basis for expansions in the inspiral phase, where the PN expansion is expected to hold to approximate the solution to the field equations. Obviously, such a representation need not apply in the merger, and would have to be changed in the ringdown, as discussed extensively in~\cite{Yunes:2009ke}, but we are here focusing on inspiral tests only.

The initial and simplest ppE proposal was to include a single deformation at a time~\cite{Yunes:2009ke}. As explained above, the deformation is represented by adding a term of the form $\beta \, v^b$ in the GW Fourier phase (and a similar term in the Fourier amplitude), where $\beta$ is a real parameter one is attempting to estimate, $v$ is the orbital velocity of the binary (which is connected to the GW frequency via Kepler's third law), and $b$ is a fixed constant. One attempts to measure or constrain $\beta$ (sometimes called the ppE amplitude coefficient) because it determines the strength of the GR deformation, and it is connected to the coupling constant of modified GR theories. The quantity $b$ (sometimes called the ppE exponent coefficient) is not a parameter, but rather it is a real number that determines the type of modification one is considering; therefore, one chooses a value of $b$ \textit{a priori}, before carrying out parameter estimation or model selection. By picking a set of such $b$ constants, one can then derive posterior probability distributions for each $\beta$ (corresponding to each value of $b$ that one chose), to then construct the so-called ``violin plots'' the LVC generates in their testing GR papers~\cite{LIGOScientific:2016lio,LIGOScientific:2018dkp,LIGOScientific:2019fpa,LIGOScientific:2020tif}.

However, as explained when the ppE framework was first developed~\cite{Yunes:2009ke}, deformations to GR in the inspiral should not consist of a single PN modification. Rather, one expects modifications to enter at some leading PN order (a so-called ``controlling factor'' of an asymptotic expansion), multiplied by an entire PN series in velocity, i.e.~$\beta_0 \, v^b (1 + \sum_{i=n} \beta_n v^n)$. One way to incorporate this idea into an inspiral test in GR would be to truncate the series at some PN order, allow every $\beta_n$ coefficient to be independent, and attempt to estimate all of them simultaneously~\cite{Arun:2006yw,Arun:2006hn}. As argued in~\cite{Sampson:2013lpa}, and then verified in~\cite{LIGOScientific:2016lio}, this approach is doomed to fail because of large covariances between the $\beta_n$ parameters that prevents one from estimating any one of them with any precision (i.e.~the marginalized posteriors of $\beta_n$ that one recovers are very similar to the priors one chooses for them). If allowing for multiple, independent deformation terms at once erases all useful information one may glean from the GW signal, are constraints one obtains with a single deformation at a time robust?

At first sight, one may think the answer to this question is no. After all, we know that a modified gravity signal will not contain just a single modification in the phase. Therefore, forcing a template to be of this form ought to introduce uncontrolled systematic error. Moreover, the velocities at merger are close to $0.3$--$0.5$ the speed of light, so PN terms at higher-than-leading order need not be negligible. Therefore, by neglecting these higher order terms one could be ignoring ``degeneracies'' that could deteriorate constraints if they were properly modeled. If this were the case, single-parameter ppE tests or single-parameter parametrized inspiral tests could over-confidently predict and place constraints that are more stringent than what we have any right in claiming.

We are of course not the first to study the concerns mentioned above. To our knowledge, one of the first studies to do this for aLIGO observations appeared in an Appendix~\cite{Yunes:2016jcc}, where the authors investigated how constraints on Brans-Dicke theory are improved as higher PN order terms are added to the inspiral phase in the extreme mass-ratio limit. The authors found that constraints on the Brans-Dicke coupling parameter are very stable to the inclusion of these higher PN order terms, with relative fractional changes of at most a few tens of percent. A more recent study of this same topic was by~\cite{Perkins:2021mhb}, where constraints on a specific theory of gravity, shift-symmetric scalar Gauss-Bonnet (ssGB), were studied with the O2 and O3 GW catalog. In this work, the authors modeled the higher PN order terms as unknown, and then marginalized over them; they found that the inclusion of these higher order terms only changed the constraints again by only tens of percent. 

In spite of these studies, there has not yet been a dedicated study of the concerns mentioned above, which is what motivated this paper. As we will show in detail here, and in agreement with the previous work described above, the pessimistic viewpoint expressed above regarding single- versus multiple-parameter tests of GR is completely unwarranted. For a multitude of reasons, which we will elaborate on in this work, constraints obtained by using single-parameter models are both robust and reliable. In particular, we will show that although systematic error is incurred when one uses a single-parameter ppE model, the error is small and tends to predict a \textit{less} stringent constraint than what one would obtain if one included higher PN order corrections subject to a reasonable PN prior; in this sense, single-parameter ppE and parametrized inspiral tests are \textit{conservative}. Moreover, we will show that the improvement one gains by including higher PN order terms in the GR deformation makes constraints only slightly more stringent (of at most one order of magnitude in the most extreme cases). 

We arrive at these conclusions through a detailed and full Bayesian analysis of a multitude of synthetic signals. We first carry out a fully Bayesian Markov-chain Monte-Carlo (MCMC) parameter estimation study, in which we inject a synthetic GR signal in stationary and Gaussian noise, and then we extract it with single-parameter ppE models, as well as multi-parameter models. For the latter, we develop a new prior based on the theoretical foundations of PN theory~\cite{Blanchet:2013haa,Damour2016}, which ensures that the modifications one introduces lead to a non-GR PN series that has similar or better convergence properties than that of GR. If the modified theory of gravity one is considering accepts a PN expansion, then the prior we develop is guaranteed to be valid. It is the development and use of this prior that restricts the impact of the covariances found in previous studies~\cite{LIGOScientific:2016lio,Arun:2006yw,Arun:2006hn}.

We then examine our conclusions under the light of a specific example: constraints on ssGB theory. Higher-than-leading PN order waveforms have been recently calculated in this theory~\cite{Shiralilou:2021mfl}, thus allowing us to verify our prior. With this at hand, we carry out again a fully Bayesian test of GR, injecting synthetic GR signals in stationary and Gaussian noise, and extracting them with the new, high-PN-order ssGB model, the leading-PN-order ssGB model, a single-parameter ppE model, and a multiple-parameter ppE model with the PN-inspired prior developed above. In all cases, we map constraints on the parameters we search over to constraints on the coupling constant of ssGB gravity. We find that the the leading PN order ssGB model, the single-parameter ppE model, and the multiple-parameter ppE model with the PN-inspired prior all lead to roughly the same constraints, in agreement with previous work~\cite{Perkins:2021mhb}. The new, high-PN-order ssGB model leads to constraints that are stronger (not weaker) than the other constraint,  but only by a factor of roughly 3 (at 90\% confidence), because this model adds new physical information (at higher PN orders). Therefore, our analysis proves that, at least in this theory (and very likely in all theories of this type), higher-PN order corrections to the ssGB waveform improve constraints instead of deteriorating them, and the improvement is only very mild. 

The conclusions we arrive at in this paper are admittedly strong, so they come with a couple of caveats. One of them is that the signal one analyzes is dominated by the inspiral (and not the merger) phase of the coalescence. If one can only hear the last few cycles of coalescence and the ringdown, then one is outside the regime of validity of the PN approximation, rendering some of the arguments presented above invalid. In practice, this implies that our conclusions apply only to binaries of sufficiently small total mass, with the maximum mass allowed dependent on the characteristics of the instrument (and in particular, the low frequency seismic wall of the noise spectral density of the detector); for the Advanced LIGO detectors at design sensitivity, the validity of our conclusions require that at least the total mass of the binary be less than roughly $40 M_\odot$, so that at least the frequency of the innermost stable circular orbit (of the effective one-body problem) be above in the sensitivity bucket of aLIGO at 100 Hz. Another caveat is that we consider modifications to GR that lead to ``persistent'' effects in the inspiral, as opposed to modifications that turn on suddenly during the inspiral. The latter can occur in theories with additional length scales that lead to screening (like massive gravity theories~\cite{deRham:2014zqa}), or theories with additional fields that activate during the inspiral due to non-linear effects (like dynamical scalarization~\cite{Damour:1993hw,Damour:1996ke,Doneva:2017bvd,Silva:2017uqg} or vectorization~\cite{Ramazanoglu:2017xbl,Silva:2021jya}). As shown in~\cite{Sampson:2013jpa}, however, even for such theories a single-parameter ppE test is sufficient to detect such sudden deviations (at the cost of deteriorating the effectiveness of the test).   

The rest of this paper will present the details of the calculations that have led us to the conclusions we described above, and it is structured as follows. We begin by discussing some of basics of Bayesian inference with GWs in Sec.~\ref{sec:bayesian_analysis}. This is followed by a discussion of the current methodology for testing GR in the inspiral phase of GW binaries in Sec.~\ref{sec:current_methods}.
We continue with a discussion of a proposed improvement on those methods in Sec.~\ref{sec:improved_methods}, in which we outline a framework informed by the PN formalism.
After this, we expand on the experimental design we use to test this new framework in Sec.~\ref{sec:exper_design}.
With the experiment summarized, we move on to discuss the results and implications of those experiments in Sec.~\ref{sec:exper_results}.
In Sec.~\ref{sec:alt_params}, we investigate alternative parametrizations to determine the robustness of our conclusions.
From here, we focus on a specific theory of gravity, ssGB, to determine how realistic our conclusions are in the context of actual modifications to GR in Sec.~\ref{sec:ssGB}.
Finally, we summarize our findings and propose future research in Sec.~\ref{sec:conclusions}.
Throughout this work, we will work in geometric units, where $G=1=c$.

\section{Testing GR with GWs}\label{sec:tests_of_gr}
Inferences about fundamental physics from GW data begins with matched filtering and Bayesian signal analysis, which we review in Sec.~\ref{sec:bayesian_analysis}.
We then move on to one of the currently implemented methodologies used to test GR with GW data (the LVC implementation of the ppE framework, which they dub a parametrized inspiral test) in Sec.~\ref{sec:current_methods}.

\subsection{Bayesian Analysis of GW Data}\label{sec:bayesian_analysis}
The foundation of any Bayesian analysis is, of course, Bayes' theorem. 
In the context of GW analysis, this can be written as 
\begin{equation}\label{eq:bayes_theorem}
p(\paramvec| D) = \frac{p(\paramvec) p(D| \paramvec) }{p(D)}\,,
\end{equation}
where $p(\paramvec| D)$ is the posterior probability of the source parameters $\paramvec = \{\theta_1,\theta_2,\ldots,\theta_N\}$ given some dataset $D$, $p(\paramvec)$ is the prior probability of $\paramvec$, $p(D| \paramvec)$ is the likelihood that the data $D$ is described by the model and parameter vector $\paramvec$, and $p(D)$ is the evidence for the model.
The posterior distribution is the quantity of interest in this study, as we are  primarily interested in deducing the values additional, beyond-GR parameters can take while still being consistent with data observed in our Universe. 
The evidence can also be of interest when calculating Bayes' factors between models, but we will not be focused on model selection in this study.
The information gained from the current experiment at hand is encoded in the likelihood distribution, while the beliefs about the different parameters held \textit{before the experiment began} is embodied by the prior distribution.

In the case of inference with GW data, the likelihood function is calculated by matched-filtering. 
With this method, a template response function defined by a model and a set of parameters is compared against the data. 
This function can be succinctly written in the Fourier domain (assuming Gaussian, stationary noise) as
\begin{equation}
    \log p(D| \paramvec) = -\frac{1}{2} \sum_i^N \left(D_i - h_i | D_i - h_i \right) \,,
\end{equation}
for $N$ detectors, where $h_i$ is the response template of the $i$-detector given a GW metric perturbation (the perturbation template contracted on to the $i$-th detector $h_i = h_+ F_{+,i} + h_\times F_{\times,i}$) and the inner product is defined as
\begin{equation}
    \left(A| B \right) = 4 \operatorname{Re} \int_0^{\infty} \frac{A B^{\ast}}{S(f)}df\,,
\end{equation}
where $({}^{\ast})$ denotes complex conjugation and $S(f)$ is the one-sided power spectral density of the noise in the detector.

As a general rule in Bayesian inference, one should be careful in how the prior information is specified.
Either in the case of encoding information from past experiments or through mathematical intuition, or in how one attempts to encode one's ``ignorance'' about the parameter in question, an overly restrictive prior can lead to biased posterior distributions.
Standard choices used in GW science include a uniform prior in volume, a uniform prior on the component masses and the spin magnitudes, with a prior for the spin directions, sky location, and binary orientation uniform on the unit two-sphere.
These choices of prior encode a certain amount of basic information through certain assumptions we have made about these parameters (e.g.~that the universe is uniform and isotropic on large scales, or that the masses of compact objects cannot be negative), but are considered to be ``uninformative.''

Ultimately, we seek to determine the form of the posterior defined in Eq.~\eqref{eq:bayes_theorem}, because the degree to which the final marginalized posterior differs from the prior determines how much new information we have gained by performing the experiment. 
Once the posterior has been calculated, we can deduce many useful quantities, such as the confidence regions and maximum likelihood values for various parameters.
However, without making any simplifying assumptions, this function is not analytic.
Therefore, we are typically forced to explore these distributions through sampling techniques like Markov-Chain Monte-Carlo (MCMC) algorithms.
Analyses such as these produce samples from the distribution of interest, which can then be binned into a histogram representation of the distribution.

Throughout this work, we will sample the likelihood, and therefore construct the posterior, through an MCMC exploration of the parameter space using software written for and implemented in previous works~\cite{Perkins:2021mhb}. 
The Bayesian exploration of parameter space is based on the Metropolis-Hasting algorithm, and implements a variety of techniques to optimize the sampling efficiency, which we will briefly outline here. 
A critical component of the sampler is the use of parallel-tempering~\cite{PhysRevLett.57.2607,Earl2005ParallelTT}, in which multiple MCMC chains are run in parallel. 
The first chain in the series samples from an unmodified posterior distribution, but all of the other chains in the series sample from a ``tempered'' distribution, in which the likelihood is raised to some power, $1/T_i$, where $T_i\geq1$.
The modified posterior is given by
\begin{equation}
    p_i(\paramvec | D) \propto p(\paramvec) p(D| \paramvec)^{1/T_i}\,,
\end{equation}
for some ``temperature'' $T_i$.
Periodically, the chains are allowed to exchange information by proposing a swap of the parameter coordinates of two of the chains.
Each temperature ladder has a certain set of $\{T_i\}$ comprised of 15 (30) separate chains for the 2g (3g) injections (where the 2g and 3g injection parameters are discussed in Sec.~\ref{sec:exper_design}). The temperatures run from $T_0 = 1$ (untempered) to $T_{14 ~(29)} \approx \infty$ (effectively sampling the prior).
To increase efficiency, we run 14 (5) copies of the above set of chains in parallel, so that we have multiple chains sampling at each temperature, $T_i$.
All chains in the ensemble are allowed to swap with all other chains to improve convergence.
Samples are then harvested from the $T = 1$ chains, which are sampling from the unmodified, target posterior.
While this does introduce some correlations between the samples from the cold chains, we harvest more than enough samples to mitigate the impact of this, as discussed below.
Furthermore, correlations of this type are analogous to those that would be present with other ensemble samplers, which base proposal distributions on the position of other chains in the ensemble.

In terms of the proposals we have implemented, we used two types: steps drawn from a random normal distribution centered at the current location and steps along the eigenvectors of the Fisher information matrix. 
We ``burn-in'' approximately 80,000 (28,000) samples per chain, during which we allow the sampler to optimize itself to the target distribution by tuning the step widths, continuously updating the Fisher matrix, and optimizing the temperature spacing of the various chains using the algorithm by Vousden et.al.~\cite{2016MNRAS.455.1919V}. 
After this initial burn-in phase, we then fix all the parameters of the sampler and continue to run for an additional 100 independent samples per untempered chain, as determined by the auto-correlation length, to make sure that indeed the chains have exited the burn-in phase. 
Then, we proceed to harvest samples that we will use to build the distribution, while thinning out by the autocorrelation length.
We sample until we have harvested approximately 4,000 independent samples per untempered chain.
This produces on the order of 56,000 (20,000) total, independent samples per 2g (3g) injections.
With the samples drawn, we calculate the 1-$\sigma$ confidence intervals by calculating the $16\%$ and $84\%$ quantiles from the data using \software{Numpy} routines~\cite{harris2020array}, and taking the difference then dividing by 2.

\subsection{Current Parametric Tests of GR}\label{sec:current_methods}

With the general framework outlined above, we can now discuss some of the current methodologies used in searches for physics beyond GR in GW data.
In the course of this paper, we will always work with small deformations away from GR, consistent with our experience that GR describes gravity extremely well up to (and possibly beyond) the most extreme energy scales we have observed so far. 
In the spirit of this philosophy, we use a base GR waveform model (\software{IMRPhenomD}~\cite{Husa:2015iqa,Khan:2015jqa}) and append modifications on top through a phenomenological framework. We could have used a more advanced base GR model, like \software{IMRPhenomPv3}~\cite{Chatziioannou:2017tdw}, but doing so will probably not impact our conclusions qualitatively.

The LVC parametrized inspiral tests are a realization of the ppE framework~\cite{Yunes:2016jcc}, restricted to single-parameter deformations and deformations that enter at half-integer PN orders, and thus, there are two ``standard'' ways to implement them. In the first implementation, the modifications are added as fractional deviations at each PN order directly to the GW phase. Explicitly, let us write the GW Fourier phase of the dominant mode as as~\cite{Husa:2015iqa,Khan:2015jqa,Blanchet:2013haa}
\begin{align}\nonumber
    \Psi_{\rm GW}(f) &= \frac{3}{128\eta} v^{-5}  \left(1 + \sum_{i=2}^{7} p_i(\boldsymbol{\Xi}) v^{i}\right)\\
    &+ \frac{3}{128\eta} v^{-5}\left(\sum_{i=5}^6 p_{i,l}(\boldsymbol{\Xi}) \log\!\left(v^3\right) \, v^i\right) \,,
    \label{eq:ppE1}
\end{align}
where $v\equiv (\pi m f)^{1/3}$, $m\equiv m_1+m_2$ is the total mass for a binary with component masses $m_1$ and $m_2$, and $f$ is the GW frequency of the dominant mode, while $p_i \equiv p_{i,\GR} ( 1+ \delta_i^{\bar{i}} \, \delta p_i)$ and $p_{i,l} \equiv p_{i,l,\GR} ( 1+ \delta_i^{\bar{j}}  \, \delta p_{i,l})$. The parameters $p_{i,\GR}$ are the coefficients of the PN expansion in GR at the $(i/2)$-th PN order, and $p_{i,l,\GR}$ are the coefficients proportional to logarithmic terms; these coefficients are purely functions of the source parameters, $\boldsymbol{\Xi}$, and not of frequency. The parameters $\delta p_i$ and $\delta p_{i,l}$ are the non-GR deformation parameters that one attempts to estimate or constrain~\cite{Li:2011cg,Arun:2006yw,Mishra:2010tp,Meidam:2017dgf,Agathos:2013upa}. The constants $\bar{i}$ and $\bar{j}$ determine the PN order of the single deformation that is turned on, with the Kronecker delta guaranteeing that all other deformations are turned off.
This is the parametrization that has been largely adopted by the LVC in their parametrized inspiral tests of GR (c.f. ~\cite{LIGOScientific:2019fpa,LIGOScientific:2020tif,Abbott:2020jks,LIGOScientific:2021sio,LIGOScientific:2018dkp,LIGOScientific:2016lio}). 

A second implementation is to write the GW Fourier phase of the dominant mode as  
\begin{align}\nonumber
    \Psi_{\rm GW}(f) &= \frac{3}{128 \eta} v^{-5}  \left(1 + \sum_{i=2}^{7} p_{i,\GR}(\boldsymbol{\Xi}) \, v^{i}\right)\\ \nonumber
    &+ \frac{3}{128\eta} v^{-5}\left(\sum_{i=5}^6 p_{i,l,\GR}(\boldsymbol{\Xi}) \log\!\left(v^3\right) \, v^i\right) \\
    &+ \beta \, \mathcal{U}^{(b-5)}\,,
    \label{eq:ppE2}
\end{align}
where $\mathcal{U} = (\pi \mathcal{M} f)^{1/3}$, $\mathcal{M}= \eta^{3/5} m$ is the chirp mass of the binary, with $\eta = m_1 m_2/m^2$ the symmetric mass ratio, $\beta$ is a deformation parameter and $b$ is a real number one fixes before carrying out parameter estimation. This is closer to the original ppE framework~\cite{Yunes:2009ke} and it is employed by researchers because it allows for a slightly more direct mapping to constraints on specific modified theories~\cite{Cornish:2011ys,Sampson:2013lpa,Chatziioannou:2012rf,Tahura:2018zuq}. As it should be painstakingly clear by comparing Eqs.~\eqref{eq:ppE1} and~\eqref{eq:ppE2}, these two implementations are equivalent to each other with $\bar{i} = b - 5$, $\bar{j} = 0$, a reparametrization between $\delta p_i$ and $\beta$, and the transformation of their priors, as shown in detail in~\cite{Yunes:2016jcc}.

Regardless of the preferred implementation, both of the ones presented above focus on a single modification at a time (e.g. a single $\beta$ or $\delta p_i$ or $\delta p_{i,l}$ at a time).
This is more a choice made out of practicality than one made because of a physically-motivated reason.
In reality, solving for the phase of the gravitational waveform, even in GR, leads to a series with an infinite number of terms when using the PN formalism, as shown in Eqs.~\eqref{eq:ppE1} and~\eqref{eq:ppE2}.
As these generic parametrizations represent unknown corrections to these terms in the PN series (whether in the GW phase directly or in the energy flux, binding energy or dispersion relations), there should also be an infinite number of deformations when the calculation is done for any specific modified theory of gravity. 
One would therefore expect a Fourier waveform phase of the form of Eq.~\eqref{eq:ppE1} but without the Kronecker deltas in the definitions of $p_i$ and $\delta p_{i,l}$ for the first implementation.
For the second implementation shown in Eq.~\eqref{eq:ppE2}, the natural generalization would be
\begin{align}\nonumber
    \Psi_{\rm GW}(f) &= \frac{3}{128\eta} v^{-5}  \left(1 + \sum_{i=2}^{7} p_i(\boldsymbol{\Xi}) v^{i}\right)\\ \nonumber
    &+ \frac{3}{128\eta} v^{-5}\left(\sum_{i=5}^6 p_{i,l}(\boldsymbol{\Xi}) \log\left( \pi m f \right) v^i\right) \\
    &+ \beta \, \mathcal{U}^{(b-5)}\left( 1 + \sum_{i=1}^7 \beta_i \mathcal{U}^i \right)\,,
    \label{eq:ppE2_multi}
\end{align}
where we now include $\beta$ along with higher terms in the PN expansion, $\beta_i$

To be maximally robust, one would then think that one should simultaneously fit for \textit{all} the $\delta p_i$ and $\delta p_{i,l}$ or all the $\beta_{i}$ and $\beta$ parameters at all known PN orders (for waveforms used in current parameter estimation analysis, this would go up to 3.5PN order).
However, when this was attempted in the past, the constraints on any one $\delta p_i$ or $\beta_i$ parameter quickly degraded to the point that the posterior devolved into the prior~\cite{LIGOScientific:2016lio}, and no new information could be gleaned about the underlying physics. This was predicted back in~\cite{Sampson:2013lpa} and then verified recently in~\cite{LIGOScientific:2016lio}. 
As is commonly understood, this is due to overfitting~\cite{Arun:2006yw,Arun:2006hn,Gupta:2020lxa}.
By increasing the dimensions of the parameter space and allowing a large enough prior range that significant degeneracies surface, any insight into the individual modifications is washed out by covariances.
But if information is lost when including multiple deformations, are the constraints found while using a single deformation robust, as these covariances are completely neglected in this restricted case?
If all one wishes to establish is whether a deviation \textit{exists}, then previous work has shown that indeed a single-term deformation is sufficient~\cite{Sampson:2013lpa}.
Here, however, we are not concerned with detecting a deviation, but rather determining the accuracy to which deformation coefficients can be estimated, and thus, constraints can be placed. 

\section{Improved Parametric Tests of GR}\label{sec:improved_methods}
From the standpoint of testing for new physics, is the situation truly so bleak?
There are some critical details that are completely ignored by the overly agnostic approach of varying over all deformation parameters simultaneously, and which one could may be use to perform robust parametric tests. 
We discuss the critical, additional information we will incorporate into our analysis in Sec.~\ref{sec:math_structure} and the details of the exact implementation in Sec.~\ref{sec:improved_methods_implementation}.

\subsection{Restrictions Based on Mathematical Structure of Modified Theories}\label{sec:math_structure}

There exists at least two pieces of basic information that one can infer and use about the exact form of the deformations introduced into the GW Fourier phase by real modified theories of gravity.

The first piece of information concerns the PN structure of the deformations. Let us consider theories in which the early inspiral waveform can be calculated with the PN approximation, as we do in GR, where the phase is written as a series in the orbital velocity.
If this is the case, there is a hierarchy in the magnitude of the modifications at each PN order, so that the PN series is convergent in certain regimes of parameter space. For example, using the parametrization of Eq.~\eqref{eq:ppE1} 
\begin{equation}
    \delta p_i p_{i,{\rm GR}} v^i \gg \delta p_{i+1} p_{i+1,{\rm GR}}v^{i+1}\,,
\end{equation}
while using the parametrization of Eq.~\eqref{eq:ppE2_multi}
\begin{equation}
    \beta_i v^i \gg \beta_{i+1} v^{i+1}\,.
\end{equation}
These conditions then lead to the asymptotic inequalities $\delta p_{i+1} \, v \ll \delta p$ and  $\beta_{i+1} \, v \ll \beta_i$.
The implication here is that allowing the $\delta p_i$ or the $\beta_i$ parameters at each PN order to vary over completely independent ranges is not representative of any PN-compatible theory of gravity beyond GR. For example, if the 2 PN deformation coefficient $\delta p_4$ or $\beta_4$ had a prior range $(-10^7,+10^7)$, then there would be many choices for which the 2PN terms would completely dominate over the 1PN and the Newtonian term, rendering the PN approximation inaccurate in the regime of velocities of interest to ground-based detectors.  

Whether the PN series is convergent, asymptotic, or divergent has not been formally established due to the lack of an exact solution one can compare against. For comparable masses, Ref.~\cite{Blanchet:2002xy,Blanchet:2013haa} computed the radius of convergence of the PN series using the PN binding energy, and up to 2PN order the author found it to be $v/c \approx 1/2$. In the extreme mass-ratio limit, Ref.~\cite{Yunes:2008tw} have shown the PN series is asymptotic, and determined the optimal radius of convergence as a function of PN order to be $v/c \approx (1/5,1/3)$ depending on the PN order studied. Either way, the growth of the coefficients of the PN approximation cannot grow with PN order too rapidly if the solution is to remain accurate in the late inspiral (at velocities of roughly $1/3 c$). This is true not just in GR, but also in modified theories of gravity, since it is a mathematical statement about the validity of a series solution. 

The second piece of information one can use relates to the fact that the mathematical structure of all physical theories guarantee that each modified GR term will be a function of the coupling constants of the theory. Let us restrict attention to modified theories that have a continuous GR limit, i.e.~they reduce to GR continuously as their coupling constants tend to a specific value. Then, for small deformation from GR, each modified GR term can be Taylor expanded in the coupling constants of the theory about their GR values. To make this concrete, let us focus on theories with a single coupling constant, and let this constant be called $\ell$. Then, Taylor expanding the deformation coefficients at any given PN order, we have
\begin{equation}
    \delta p_i(\ell,\boldsymbol{\Xi}) = \delta \bar{p}_i(\boldsymbol{\Xi}) \, \ell^p\,,
\end{equation}
or
\begin{equation}
    \beta_i(\ell,\boldsymbol{\Xi}) = \bar{\beta}_i(\boldsymbol{\Xi}) \, \ell^p\,,
\end{equation}
where $\delta \bar{p}_i(\boldsymbol{\Xi})$ and $\bar{\beta}_i(\boldsymbol{\Xi})$ are functions of the source parameter vector $\boldsymbol{\Xi}$ only (like the masses, spins, etc), and $p$ is a positive real number. Note that although the functional dependence on the source parameters may change with PN order, the functional dependence on the coupling parameter will not in the small deformation limit (as otherwise the Taylor expansion of the full series would not be well-defined). One can now introduce a new parameter $\gamma = \ell^p$ to write all deformation coefficients as linear functions in $\gamma$ to leading-order in small deformations. This conclusion is not only valid for theories with a single coupling constant, but rather, it can be generalized to theories with multiple couplings if there is one that dominates due to others being already constrained by other experiments or observations.

In summary, the PN expansion of most theories of gravity should give a GW Fourier phase $\Psi_{\rm GW}$ that takes the form
\begin{align}\nonumber \label{eq:PN_form}
\Psi_{\rm GW}(f, \gamma, \boldsymbol{\Xi}) &= \Psi_{\rm GR}(f,\boldsymbol{\Xi}) + \gamma \, \delta \psi_0  ( \boldsymbol{\Xi}) \, v^{b}  \\
&\times \left[ 1 + \delta \psi_{2}(\boldsymbol{\Xi}) \, v^2 + \delta \psi_{3 }(\boldsymbol{\Xi}) \, v^3 +\ldots   \right]\,,
\end{align}
where we use the new parameter $\delta \psi_i$ to denote the deformation amplitudes because we now choose to use the PN expansion parameter $v=(\pi m f)^{1/3}$; this is a slight deviation from the notation of the standard ppE PN expansion, which uses the parameter $\mathcal{U}=(\pi \mathcal{M} f)^{1/3}$. Here, $\Psi_{\rm GR}$ is the GW phase in GR, and $\gamma$ generically represents the coupling constant in a given theory of gravity. The subscripts $i$ in $\delta \psi_i$ represent the order in $v$ or the $i/2$ PN order at which the deformation enters the GW Fourier phase relative to the leading PN order term in the non-GR sector. The 0PN modification could enter the phase at any PN order relative to GR, as controlled by the $b$ parameter. As required by the first piece of information presented above, the range of the $\delta \psi_{i}$ functions above must satisfy $\delta \psi_{i+1} \, v \ll \delta \psi_{i} $. As required by the second piece of information, the coupling constant of the modified theory shows up as a pre-factor that multiplies all deformations, so that in the GR limit, the non-GR terms vanish. When explicitly attempting to search for physics beyond GR, the real information comes from the first term in the series, where the coupling constant $\gamma$ appears.
Therefore, estimates on our ability to discern new information about physics should be quantified purely in terms of constraints on this constant, regardless of the number of higher order terms we include.

One of the first things one notices from Eq.~\eqref{eq:PN_form} is that the $\delta \psi_{i}$ are not constants but rather functions of the source parameters $\boldsymbol{\Xi}$. Given a specific modified theory, one knows what these functions are, and therefore, the $\delta \psi_{i}$ are not new parameters. In fact, in this case, the only new parameter to the model is simply the coupling constant $\gamma$. If one is attempting to carry out an agnostic test of GR, however, one is not privy to the functional form of $\delta \psi_{i}$. In this case, one can replace Eq.~\eqref{eq:PN_form} via 
\begin{align}
\nonumber 
\label{eq:PN_form_2}
\Psi_{\rm GW}(f, \gamma, \boldsymbol{\Xi}) &= \Psi_{\GR}(f,\boldsymbol{\Xi}) + \gamma \, \delta \bar{\psi}_0 \, v^{b}  \\
&\times \left[ 1 + \delta \bar{\psi}_{2} \, v^2 + \delta \bar{\psi}_{3} \, v^3 +\ldots   \right]\,,
\end{align}
\begin{equation}
    \delta \bar{\psi}_{i} = \frac{\int \delta {\psi}_{i} \; d^N {{\Xi}}}{\int  d^N {{\Xi}}}\,,
\end{equation}
where $N$ is the dimensionality of the source parameter space and $d^N {\Xi}$ is the parameter space volume factor. Clearly then, the coefficients $\delta \bar{\psi}_{i}$ are now constants that become new independent parameters to search over. Moreover, since $\gamma$ enters multiplied by $\delta \bar{\psi}_0$, these two parameters are 100\% degenerate, so one must employ the reparametrization $\bar{\gamma} = \gamma \, \delta \bar{\psi}_0$. The waveform parameter space is then enlarged to $\boldsymbol{\Xi} \cup \boldsymbol{\Theta}$, where $\boldsymbol{\Theta} = \{\bar{\gamma}, \delta \bar{\psi}_{2},\delta \bar{\psi}_{3},\ldots \}$. As before, the GR deviations are all still controlled by a single parameter $\bar{\gamma}$, while the range of the deformation coefficients is restricted by $\delta \bar{\psi}_{i+1}\, v \ll \delta \bar{\psi}_{i} $.

\subsection{Implementation of Improvements \\ in Parametric Inspiral Tests}\label{sec:improved_methods_implementation}

Using the representation of Eq.~\eqref{eq:PN_form_2} and enforcing convergence of the series, we can now attempt to include multiple deformations at once.  
However, care must be taken in the way one enforces convergence, as the criteria $\delta \bar{\psi}_{i+1} \, v \ll \delta \bar{\psi}_{i} $ depends on the orbital velocity $v$.

To do this, we will implement a specialized prior that ensures the series of $\delta \bar{\psi}_i$ are convergent at a ``reasonable'' value of $v$.
That is, we choose the prior to be 
\begin{equation}\label{eq:prior_boundary}
|\delta \bar{\psi}_{i} | > |\delta \bar{\psi}_{(i+1)}\left(v_{\text{eval}}\right)  |\,,
\end{equation}
for $i \in [2,3,4,6,7]$, where $v_{\text{eval}} = (\pi m f_{\text{eval}})^{1/3}$; the exact value of $f_{\text{eval}}$ and thus $v_{\text{eval}}$ that we choose are discussed below.
For the 1PN term, the prior must be handled separately, and we choose it to be
\begin{equation}\label{eq:prior_boundary_NLO}
|\delta \bar{ \psi}_{2} \left(v_{\text{eval}} \right)^{ 2  }| < 1 \,,
\end{equation}
as this is the next-to leading order (NLO) term due to our omission of the $0.5$PN term.

We must now choose the value of the velocity, or equivalently the frequency, at which to evaluate this prior. We do so by choosing the GW frequency and orbital velocity that would correspond to the waves emitted by a binary at an orbital separation of $r_{12, \rm eval} = 100m$ at leading-order in PN theory. Note that $r_{12,\rm eval}$ is \emph{not} the formal radius of convergence of the series, assuming the series were convergent. Rather, this quantity represents the orbital separation at which we are confident that adding higher order PN terms improves the PN approximation to the (unknown) exact answer. More aggressive (smaller) choices of the orbital separation could be viable options as well, but we will remain conservative in this work, as we will explain below. Using the Newtonian version of Kepler's third law for quasi-circular orbits, we then have
\begin{align}
    v_{\rm eval} &= \left(\frac{m}{ r_{12, \rm eval}}\right)^{1/2} = 10^{-1}\,,
    \\
f_{\text eval} &= \frac{(r_{12, \rm eval}/m)^{-3/2} }{ (\pi m )} \approx \frac{3 \times 10^{-4}}{m}\,.
\end{align}
This choice in frequency is inspired by the convergence properties of the GW phase in GR, as studied in~\cite{Yunes:2008tw} and revisited below.
GR is the obvious model for us to base our new prior on, as it is much better understood than any modified theory and it is our null hypothesis.

\begin{figure}
    \centering
    \includegraphics[width=\linewidth]{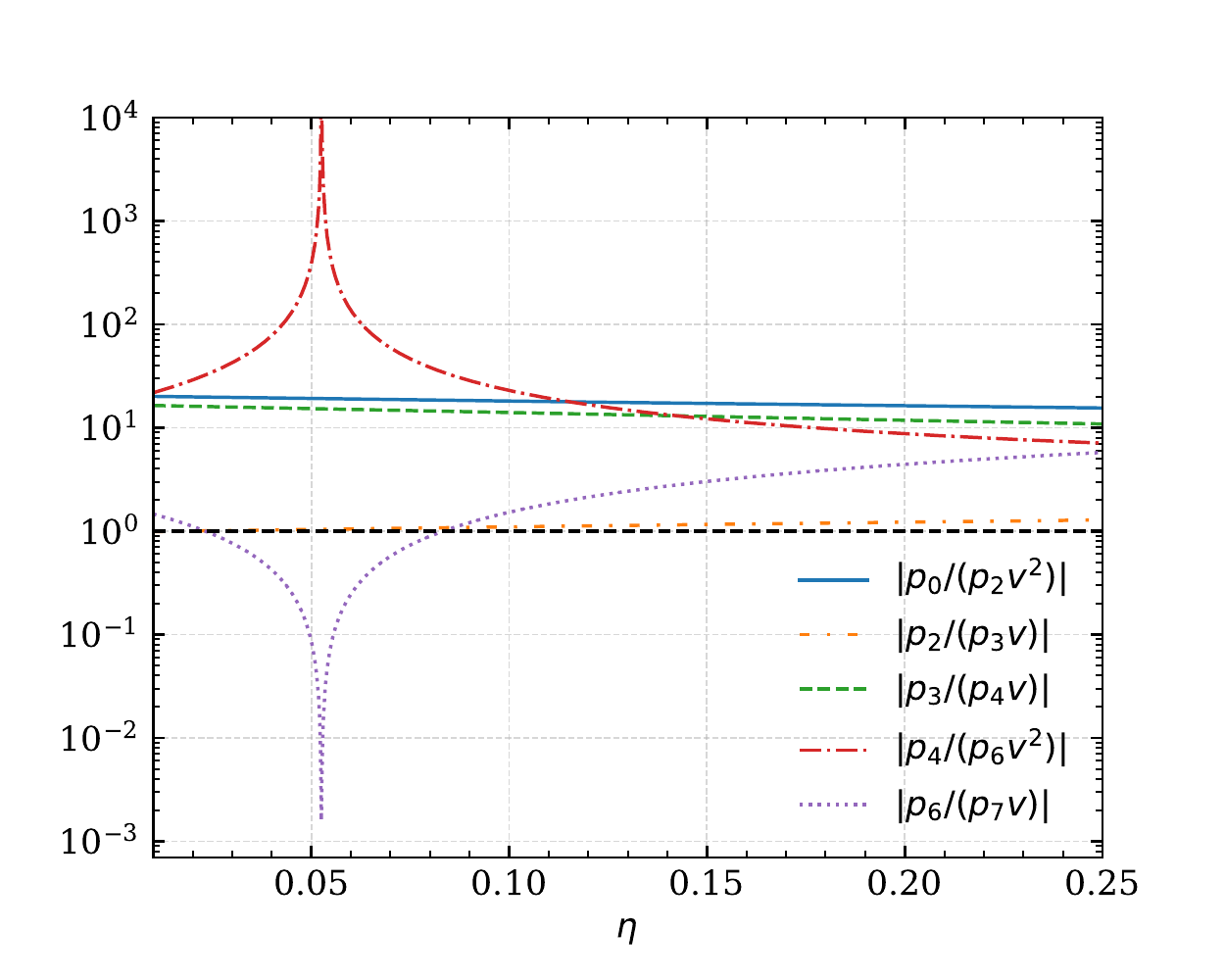}
    \caption{
        Ratio of the PN coefficients of the Fourier GW phase (Eq.~\eqref{eq:ppE1}) for a non-spinning binary as functions of the symmetric mass ratio $\eta$.
        The ratios are evaluated at an orbital separation of $r_{\rm eval} = 100m$, corresponding to $v_{\rm eval}=0.1$.
        Each line corresponds to a different pair of coefficients, and the dotted black line simply identifies the threshold we use for convergence ($|p_i / (p_{i+1} u_{\rm eval}|>1$). 
         The spikes shown here correspond to when the coefficient $p_6 = 0$ at $\eta \sim 0.053$.
        Observe that all ratios are above unity for symmetric mass ratios $\eta > 0.09$, indicating that our choice of $v_{\rm eval}$ has led to a convergent structure of PN coefficients.
        }
    \label{fig:gr_convergence}
\end{figure}

As we now show, with this choice of separation (or equivalently frequency or velocity) the coefficients of the PN expansion in GR present a convergent structure for most mass ratios. 
The phase in GR (via the Taylor F2 waveform~\cite{Damour:2000zb,Damour:2002kr,Arun:2004hn}, accurate to 3.5PN) was already presented in Eq.~\eqref{eq:ppE1}, and the exact forms of the $p_i$ and $p_{i,l}$ parameters can be found in various places in the literature (c.f.~\cite{Blanchet:2013haa,Husa:2015iqa,Khan:2015jqa}).
For a non-spinning binary in GR, the ratios $p_i/(p_{i+1} v_{\rm eval})$ only depend on the symmetric mass ratio $\eta$ alone. If these ratios are larger than unity, then the PN coefficient exhibit a convergent structure\footnote{This is in the sense that $v_{\rm eval}$ would then be within the radius of convergence of the series if the series were convergent. If the series is asymptotic, then this would be an indication that one is evaluating the series in the regime where the asymptotic series is a good approximation to the exact answer.}. Figure~\ref{fig:gr_convergence} presents this ratio as a function of $\eta$, where we observe that our choice of $v_{\rm eval}$ leads to a convergent structure of the series for all $\eta > 0.09$. The spikes at around $\eta \sim  0.053$ are because $p_6$ vanishes, but this occurs outside the comparable-mass regime, which is what we focus on here.

As the parametrization in Eq.~\eqref{eq:PN_form_2} has not been used extensively in previous literature, we will also study a related parametrization, closely associated with the standard ppE formalism. 
Let us then denote absolute deviations as $\bar{\Delta}_{i}$, and relative deviations as $\delta \bar{\psi}_{i}$ (as introduced in Eq.~\eqref{eq:PN_form_2}) for the $i/2$-th PN order.
These new $\bar{\Delta}_{i}$ parameters then enter the GW phase at various PN orders via  
\begin{align}\nonumber \label{eq:PN_form_alternate}
\Psi_{\rm GW}(f, \gamma, \boldsymbol{Xi}) &= \Psi_{{\rm GW},GR}(f,\boldsymbol{\Xi}) + \bar{\gamma} \, v^{b} +\\
& \bar{\Delta}_{2} \, v^{(2+b)} + \bar{\Delta}_{3} \, v^{(3+b)} +\ldots   \,,
\end{align}
where each modification ($\bar{\gamma},\bar{\Delta}_i$) now explicitly depends on the coupling constant of the theory.
The mapping between Eq.~\eqref{eq:PN_form_alternate} and Eq.~\eqref{eq:PN_form_2} is simply given by
\begin{equation}
\bar{\Delta}_i = \delta \bar{\psi}_{i} \times \bar{\gamma}\,,
\end{equation}
and $\bar{\gamma}$ is unchanged.

The priors we have defined above, although simple to state mathematically, are non-trivial.
To illustrate this, we sampled from the prior directly using a variation of the sampling methods outlined in Sec.~\ref{sec:bayesian_analysis}. 
In effect, we drew each of the deformation parameters from a uniform distribution $U(-B,+B)$ with some boundaries $-B$ and $+B$. 
For the leading order and NLO deformation, $|B| = 100$, while for the other, higher order parameters, $|B|$ is set to a value large enough to ensure the final distributions are not affected by $|B|$. 
With these random draws, we excluded any sample that violated Eq.~\eqref{eq:prior_boundary}.
Although each parameter began as a uniform distribution, our prior in Eq.~\eqref{eq:prior_boundary} imposed very non-trivial structure to the histogramed distributions, as the boundary of the prior for each deformation depends on the value of other deformations in the series.

\begin{figure*}[!h]
    \includegraphics[width=.45\linewidth]{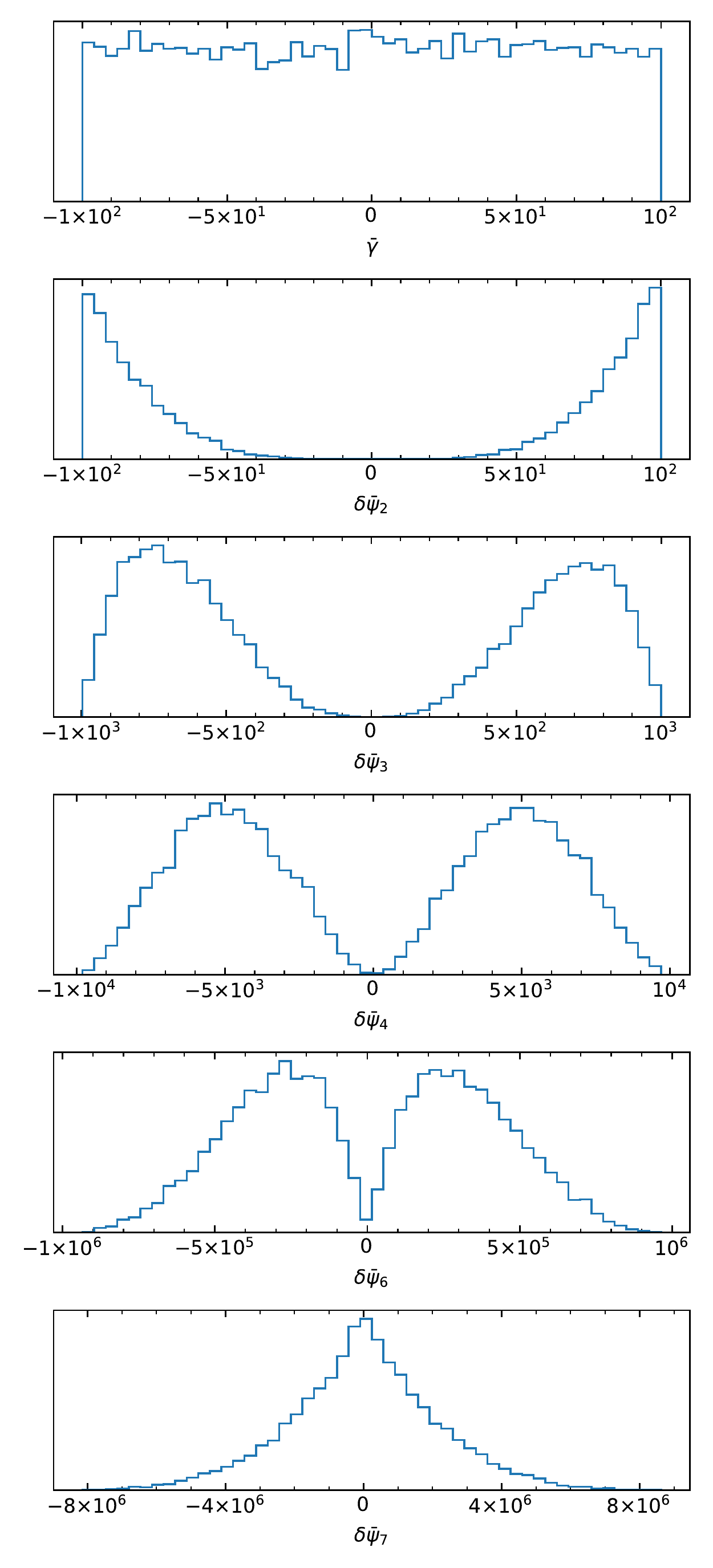}
     \includegraphics[width=.45\linewidth]{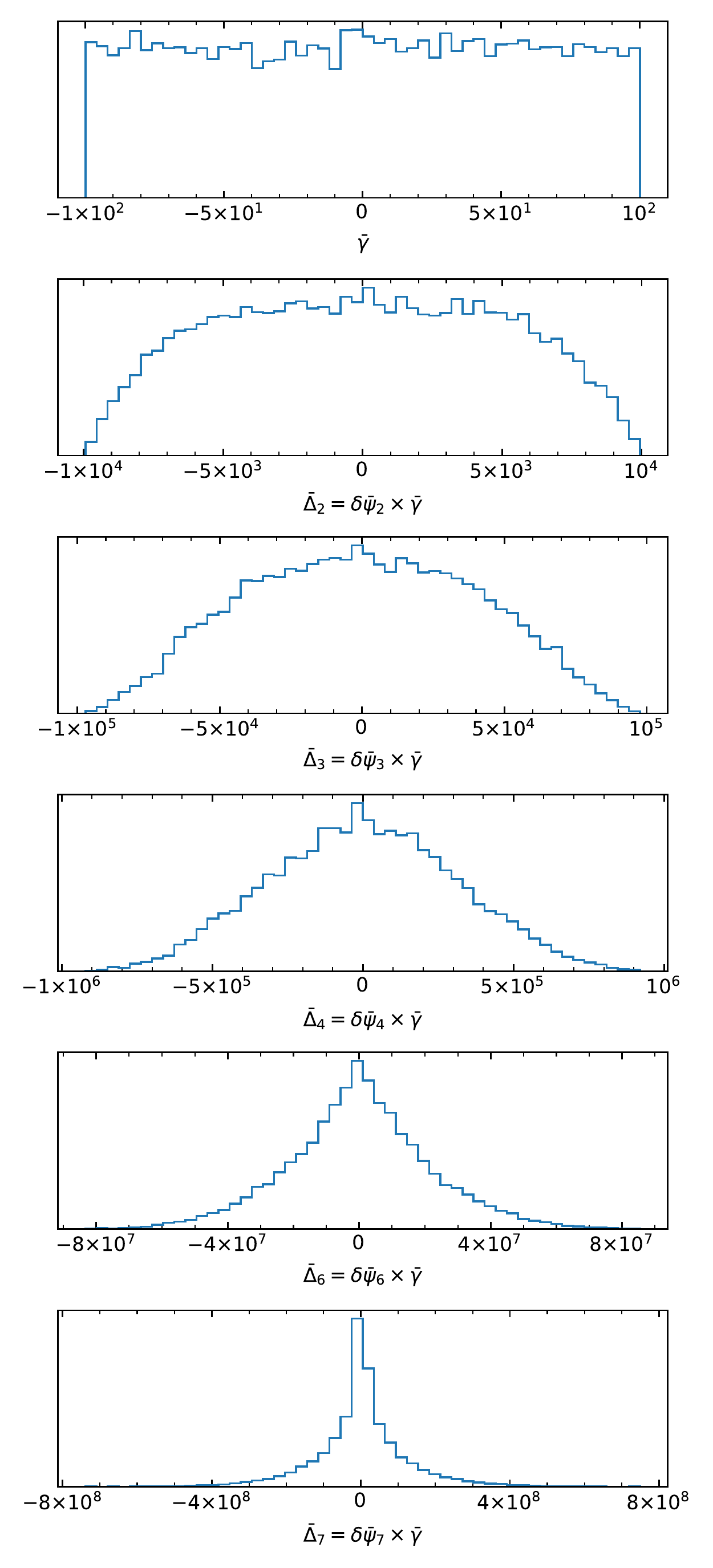}
    \caption{
    The one dimensional, marginalized prior on a set of deformations.
    The leading order deformation enters at Newtonian order relative to GR, and each of the subsequent deformations are labeled as the $i/2$PN order relative to this leading order deformation.
    The priors for this parametrization are non-trivial, and can be misleading. 
    They appear to disallow the GR limit $\delta \bar{\psi}_i =0$ for several of the different PN orders.
    This is resolved when looking at the transformed priors of the right panel, rewritten in terms of the absolute modifications, $\bar{\Delta}_i$.
    }\label{fig:prior_example}
\end{figure*}

The priors on $\delta \bar{\psi}_i$ with 6 deformations and beginning at 0PN order relative to GR (proportional to $v^{-5}$) are shown in the left panels of Fig.~\ref{fig:prior_example}, while those transformed to the absolute deformations $\bar{\Delta}_i$  are in the right panel. Observe that the prior on the leading PN order term in the series (which encodes most of the information we are interested in) is indeed flat, because Eq.~\eqref{eq:prior_boundary} does not affect it. Observe also that the larger PN order the relative deviation, the larger the range of the prior. This range, however, is indeed \textit{finite}, since if it were any larger, then the relative deviations would begin to affect the convergent structure of the PN series. Finally, observe that although the priors on $\delta \bar{\psi}_i$ seem to be pushing modifications away from GR ($\delta \bar{\psi}_i = 0$), this is just an artifact of looking at relative deviations; the priors on the absolute modifications $\bar{\Delta}_i$ are fully consistent with GR ($\bar{\Delta}_i = 0$).

\section{Experimental Design}
\label{sec:exper_design}

To determine the impact of these higher order deformations on our inferences about fundamental physics, we now conduct a series of parameter estimation analyses on synthetic signals (injections) with a variety of different recovery models.
Each of the recovery models has the same base GR waveform (\software{IMRPhenomD}), but they differ in the different number of phase deformations overlaid on top of that base model.
For each model, the modifications are added in ascending order beyond leading, beginning with the 1PN term (relative to the leading order deformation), and increasing by 0.5PN order up to 3.5PN order relative to GR, or relative to the leading order term, which ever criteria is met first.
We skip terms of 2.5PN order relative to GR in all models, as they are 100\% degenerate with the coalescence phase (which is an arbitrary constant).
When skipping terms in the series due to these total-degeneracies, we also skip them in the enforcement of the prior. 
For example, if we were to skip the $\delta\bar\psi_i$ term in the series because it was degenerate with the coalescence phase, we would update our prior for $\delta\bar\psi_{i-1}$ and $\delta\bar\psi_{i+1}$ to satisfy $\delta\bar\psi_{i-1}\geq \delta \bar\psi_{i+1} v^2$.
Therefore, every model is uniquely defined by the PN order of the leading deformation (relative to GR) and the number of subsequent terms.

We will specifically target leading order GR deformations that enter at $[-1,0,1,1.5,2,3,3.5]$PN orders, relative to Newtonian order in GR, and we include up to 5 additional deformations (again, never exceeding 3.5PN order relative to Newtonian order in GR).
For example, one model might have a leading order deformation at 1PN order relative to Newtonian order in GR, with 3 higher PN order deformations at $[2,3,3.5]$PN orders (again relative to Newtonian order in GR).
As another example, another model might have a leading order deformation that enters at 1.5PN order relative to the Newtonian term in GR, with only a single higher PN order deformation. 
This would leave us with deformations at $[1.5,3]$PN order relative to Newtonian order in GR.

As expected, all inferences will depend on the astrophysical source that produced the GW we are assuming has been detected, as well as the detector network that observes the signals.
To attempt to explore these aspects as much as possible (while still keeping the scope of this work computational tractable), we focus on two sources: a ``heavy'' source and a ``light'' source, whose properties are listed in Table~\ref{tab:inj_param}.
The spin configurations also differ slightly, with one binary having both spins aligned with the orbital angular momentum, while the other binary has one spin anti-aligned. In both cases, we focus on binary black hole inspirals, and do not consider neutron star inspirals or mixed binaries. We expect the qualitative conclusions we will find to also hold for those systems.

For the detector networks, we also focus on two configurations.
For the first configuration, and as a proxy for a 2g network, we use a network comprised of LIGO Hanford~\cite{TheLIGOScientific:2014jea}, LIGO Livingston~\cite{TheLIGOScientific:2014jea}, and Virgo~\cite{VIRGO:2014yos}.
For their sensitivities, we use analytic approximations to the aLIGO design sensitivity~\cite{ligo_SN_forecast} and the first phase of Advanced Virgo's sensitivity estimate~\cite{ligo_SN_forecast}.
For the second configuration, and as a proxy for a 3g network, we use a network comprised of Cosmic Explorer (CE)~\cite{2015PhRvD..91h2001D} and the Einstein Telescope (ET)~\cite{Punturo:2010zza}.
For their sensitivities, we use the first phase of the CE noise curve~\cite{CE_psd} and the ET-D configuration of the ET noise curve~\cite{Hild:2010id}.
To reduce the impact of the uncertainties concerning the noise curves we use, the luminosity distances of the injected sources are all scaled such that the SNR is exactly 20 (when observed by the entire 2g detector network). 
The distances are then left fixed at these values when we transitioned to the 3g network, so as to isolate the impact of a pure boost in SNR.

With all these considerations in mind, we then have 21 separate analyses (one for each combination of multi-deformation-parameter ppE model) for each detector network and source combination , for which there are four combinations. This comes out to a total of 84 separate Bayesian parameter estimation studies, whose results we summarize next.

\begin{table}
\begin{tabular}{c | c c c c c c c}
\hline \hline
\makecell{Source \\ Identifier} & $m_1$ ($M_{\odot}$) & $m_2$ ($M_{\odot}$)& $\chi_1$ & $\chi_2$ & $D_L$ (Mpc) & SNR${}_{\rm 2g}$ & SNR${}_{\rm 3g}$ \\ 
\hline \hline
Heavy & 25 & 25 & 0.3  & -0.1 & 1613 & 20 & 517.798 \\ \hline
Light & 5 & 5 & 0.3 &0.1 & 466 & 20  & 496.355\\ \hline
\hline
\end{tabular}
\caption{
    Choices of GW sources for injection campaigns.
    The source parameters are the following: $m_1$ and $m_2$ are the masses of the larger and smaller black holes, respectively, $D_L$ is the luminosity distance from Earth to the source, $\chi_i$ is the aligned, dimensionless spin of the $i$-th black hole, SNR${}_{\rm 2g} $ and SNR${}_{\rm 3g}$ are the SNRs of the source as measured by a 2g and 3g detector network respectively.
    The ``heavy'' source and the light ``source'' both have an SNR of 20 (for the 2g detector network), but have different total masses and different spin configurations.
    All source parameters were kept the same between the analyses involving the 2g and the 3g detector networks.
}\label{tab:inj_param}
\end{table}

\section{Bayesian Results}\label{sec:exper_results}

Let us now present the results of our experiments.
Because the leading PN order deformation contains all the relevant constants controlling the magnitude of GR deviations, we focus on constraints on this leading-order term, presenting results both visually and in tabular form.  

\begin{figure*}
\includegraphics[width=0.45\linewidth]{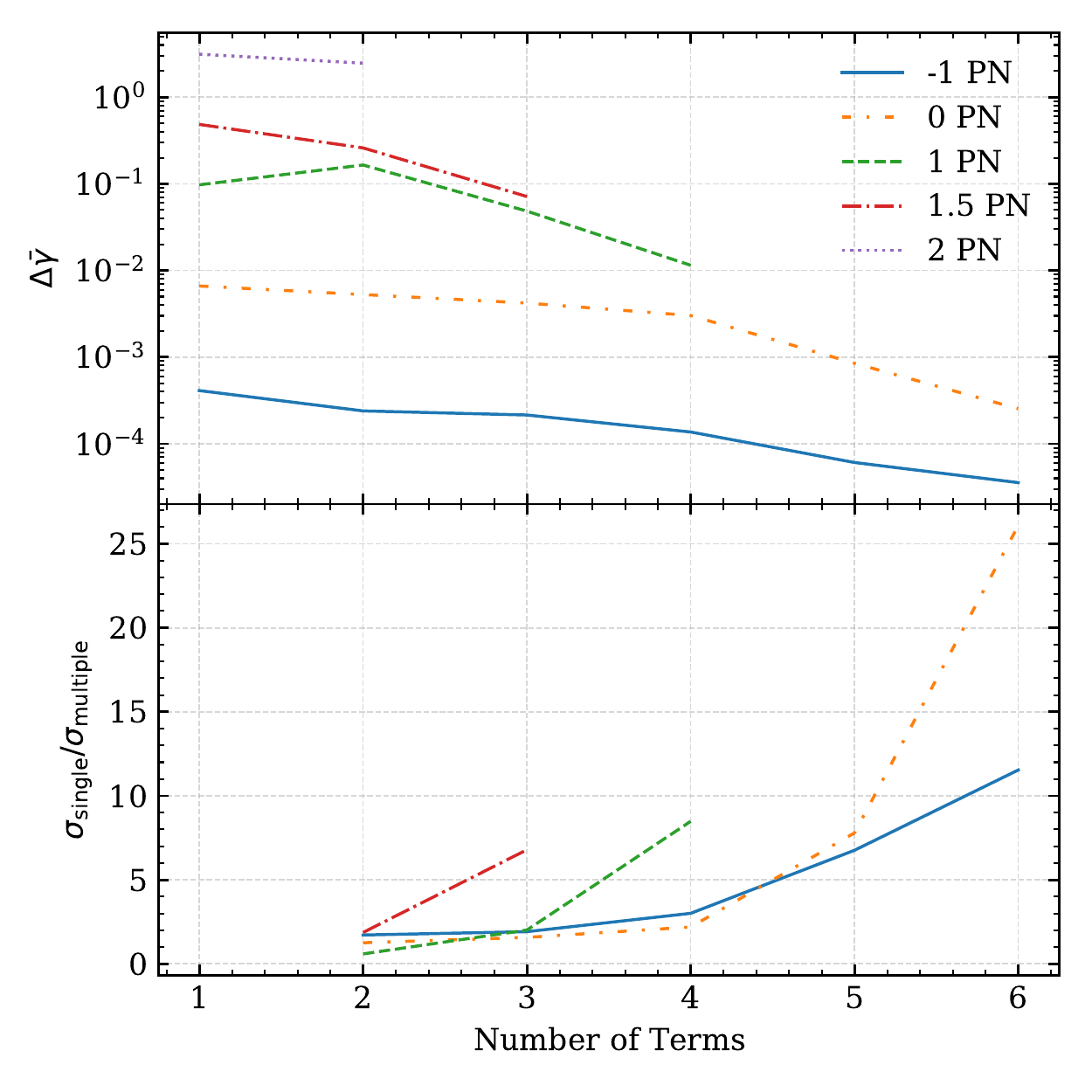}
\qquad
\includegraphics[width=0.45\linewidth]{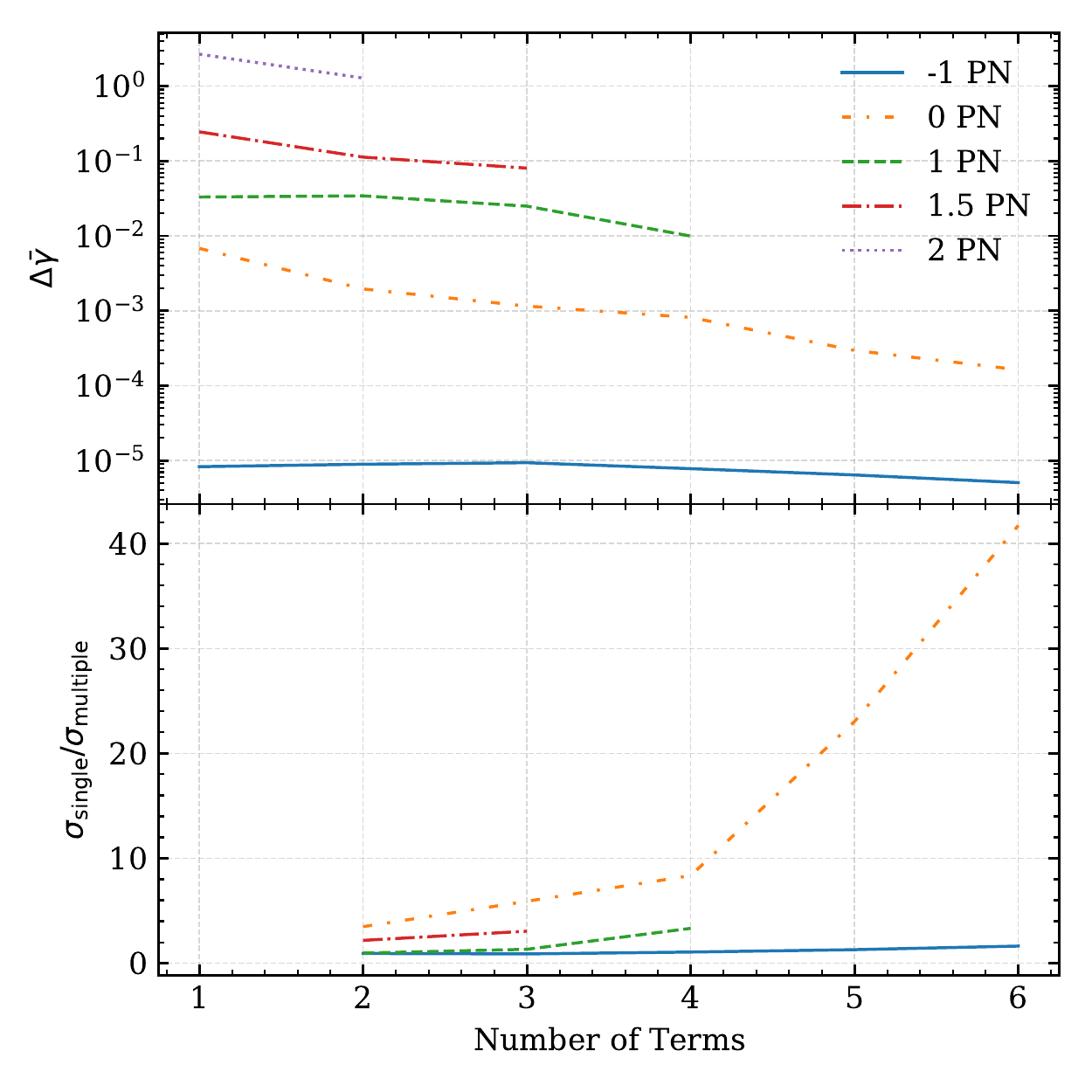}
\caption{Marginalized $1\sigma$ constraints on the leading PN order deformations for the ``heavy'' source (left) and the ``light'' source (right) using the 2g network (see Sec.~\ref{sec:exper_design} for details) as a function of the number of sub-leading PN order terms added in the modified sector (in ascending order). Each line corresponds to modifications that start at different leading PN order. The functional form of GW phase is given in Eq.~\eqref{eq:PN_form_2}, and the priors used are those presented in Sec.~\ref{sec:improved_methods}. The top panels show the $1\sigma$ constraint on the leading PN order deformation, while the lower panel shows the strengthening factor, as defined in Eq.~\eqref{eq:fractional_change}. Observe how the lines in the upper panels trend downward, which means that constraints with just a single parameter deviation are \textit{conservative} (i.e.~adding higher PN order corrections strengthens the constraint obtained with a single parametric deviation). Moreover, observe how the slope of the lines are small, which means that the strengthening of the constraint is mild, with improvements of at most roughly one order of magnitude.
}\label{fig:heavy_constraints_2g}
\end{figure*}

\begin{table*}
\begin{tabular}{ c | c c c c c}
Terms/LO & -1 & 0 & 1 & 1.5 & 2  \\ \hline \hline
1 & $4.1\times 10^{-4}$& $6.6\times 10^{-3}$& $9.7\times 10^{-2}$ & $4.8\times 10^{-1}$ & $3.1$\\ \hline
2 & $2.4\times 10^{-4}$& $5.3\times 10^{-3}$& $1.6\times 10^{-1}$ & $2.6\times 10^{-1}$ & $2.5$\\ \hline
3 & $2.2\times 10^{-4}$& $4.2\times 10^{-3}$& $4.8\times 10^{-2}$ & $7.1\times 10^{-2}$ & \\ \hline
4 & $1.4\times 10^{-4}$& $3.0\times 10^{-3}$& $1.1\times 10^{-2}$ &  & \\ \hline
5 & $6.1\times 10^{-5}$& $8.5\times 10^{-4}$&  &  & \\ \hline
6 & $3.6\times 10^{-5}$& $2.5\times 10^{-4}$&  &  & \\ \hline
\end{tabular}
\qquad  \qquad 
\begin{tabular}{ c | c c c c c }
Terms/LO & -1 & 0 & 1 & 1.5 & 2  \\ \hline \hline
1 & $8.3\times 10^{-6}$ & $6.8\times 10^{-3}$& $3.3\times 10^{-2}$& $2.5\times 10^{-1}$& $2.7$\\ \hline
2 & $8.9\times 10^{-6}$ & $2.0\times 10^{-3}$& $3.4\times 10^{-2}$& $1.1\times 10^{-1}$& $1.3$\\ \hline
3 & $9.4\times 10^{-6}$ & $1.2\times 10^{-3}$& $2.5\times 10^{-2}$& $8.0\times 10^{-2}$& \\ \hline
4 & $7.8\times 10^{-6}$ & $8.2\times 10^{-4}$& $9.9\times 10^{-3}$&  & \\ \hline
5 & $6.4\times 10^{-6}$ & $3.0\times 10^{-4}$&  &  & \\ \hline
6 & $5.1\times 10^{-6}$ & $1.6\times 10^{-4}$&  &  & \\ \hline
\end{tabular}
\caption{
    Marginalized $1\sigma$ constraints on the leading PN order deformations for the ``heavy'' source (left) and the ``light'' source (right) using the 2g network (see Sec.~\ref{sec:exper_design} for details). The columns represent the PN order (relative to the Newtonian term in GR) at which the GR deformation is first introduced. The rows corresponds to the number of PN corrections that are added on top of the leading PN order one in ascending PN order. The functional form of the phase is given in Eq.~\eqref{eq:PN_form_2}, where the priors are presented in Sec.~\ref{sec:improved_methods}. Observe the (order of magnitude) consistency of the constraints as one includes more and more sub-leading PN order deviations (i.e.~as one moves down the column for a fixed leading-order term). This indicates that higher PN order corrections to the current modified gravity ppE waveforms will not invalidate current bounds placed on modified theories with only leading PN order deformations.
}\label{tab:heavy_2g}
\end{table*}

The results for the 2g network injections are shown in Fig.~\ref{fig:heavy_constraints_2g} and Table~\ref{tab:heavy_2g} for the ``heavy'' and ``light'' sources.
The top panels show the marginalized $1\sigma$ constraint on the leading PN order deformation as a function of the number of terms included in the series, beginning with 1 (only the leading PN order deformation) and up to 6 total terms.
The lower panel shows the \textit{strengthening factor}, which we define here as 
\begin{equation}\label{eq:fractional_change}
    (\text{strengthening factor}) = \frac{\sigma_{\rm Single}}{\sigma_{\rm Multiple}}\,.
\end{equation}
If this number is larger (smaller) than unity, then adding higher PN order deformations strengthens (weakens) the constraint one obtains by using a single parameter deformation. The bottom panels of Fig.~\ref{fig:heavy_constraints_2g} show the strengthening factor as a function of PN order. 

Two main conclusions can be drawn from these figures and tables. First, observe how all curves in the top panels trend downwards (have a negative slope). This implies that adding higher PN order corrections to the modified sector \textit{strengthens} the constraints one would have gotten if one included only a single parametric deformation. This is confirmed in the bottom panels, which shows the strengthening factor is always greater than one. In this sense, single-parameter constraints are therefore \textit{conservative}. Second, observe how the slopes of the curves are small. This implies that the strengthening one obtains is somewhat mild, with improvements of only up to a factor of 40 in the most extreme case, as shown in the bottom panels of the figure.

These results and conclusions may seem counter-intuitive and, in fact, opposite to what one would expect. Adding additional, independent parameters to a model has been shown to increase degeneracies, and therefore deteriorate our ability to estimate any given parameter~\cite{Arun:2006yw,Arun:2006hn,Sampson:2013lpa,LIGOScientific:2016lio}. The reason this does not happen here is because of the series structure of the deformations in the model.
Equation~\eqref{eq:PN_form_2} is different from how other work has modeled deviations, because the strength of all deviations is here controlled by the leading PN order term (as expected from most modified theories of gravity). If an MCMC chain keeps the leading PN order deformation small, then the overall modification at each PN order ($\bar{\Delta} =  \delta \bar{\psi}_i \times \bar{\gamma}$) can remain small even if the chain visits large values of $\delta \bar{\psi}_i $.
This opens up more of the prior volume for chain exploration, which can thus be preferred  more by the posterior distribution.
In other words, there are more ``states'' for the model to take with small leading PN order deformation than with large leading PN order deformations, putting ``pressure'' on the leading coefficient to remain small.

But if the higher PN order terms were allowed to explore infinitely large values, then this would put infinite ``pressure'' on the leading PN order coefficient, forcing its posterior to be artificially tight around zero. Whether this happens or not depends on the prior one chooses on the higher PN order parameters. Naively, one may think that the most conservative prior is one that is infinitely wide and flat (this would correspond to our PN-based priors but evaluated at an infinite orbital separation.)
Choosing such a flat, uniform prior, however, would allow the magnitude of the higher PN order deformation parameters to increase indefinitely. In turn, this would put infinite ``pressure'' on the leading PN order coefficient and push its posterior to zero. The result would be an overly confident or overly aggressive constraint that is generated not by the information contained in the data, but by the prior choice. Our PN-based prior prevents this from happening. By ensuring the higher order deformations are not infinitely large (as otherwise the PN series would break down), the prior volume remains a reasonable size, and the ``pressure'' on the leading PN order deformation, sourced by this prior volume, is therefore kept to a reasonable amount. This then means that the constraint on the leading PN order parameter is governed by the information contained in the data (and the mathematical requirement that the PN approximation be valid in the inspiral for the modified theory), and not by the imposition of an overly restrictive prior.

\begin{figure}
    \centering
    \includegraphics[width=\linewidth]{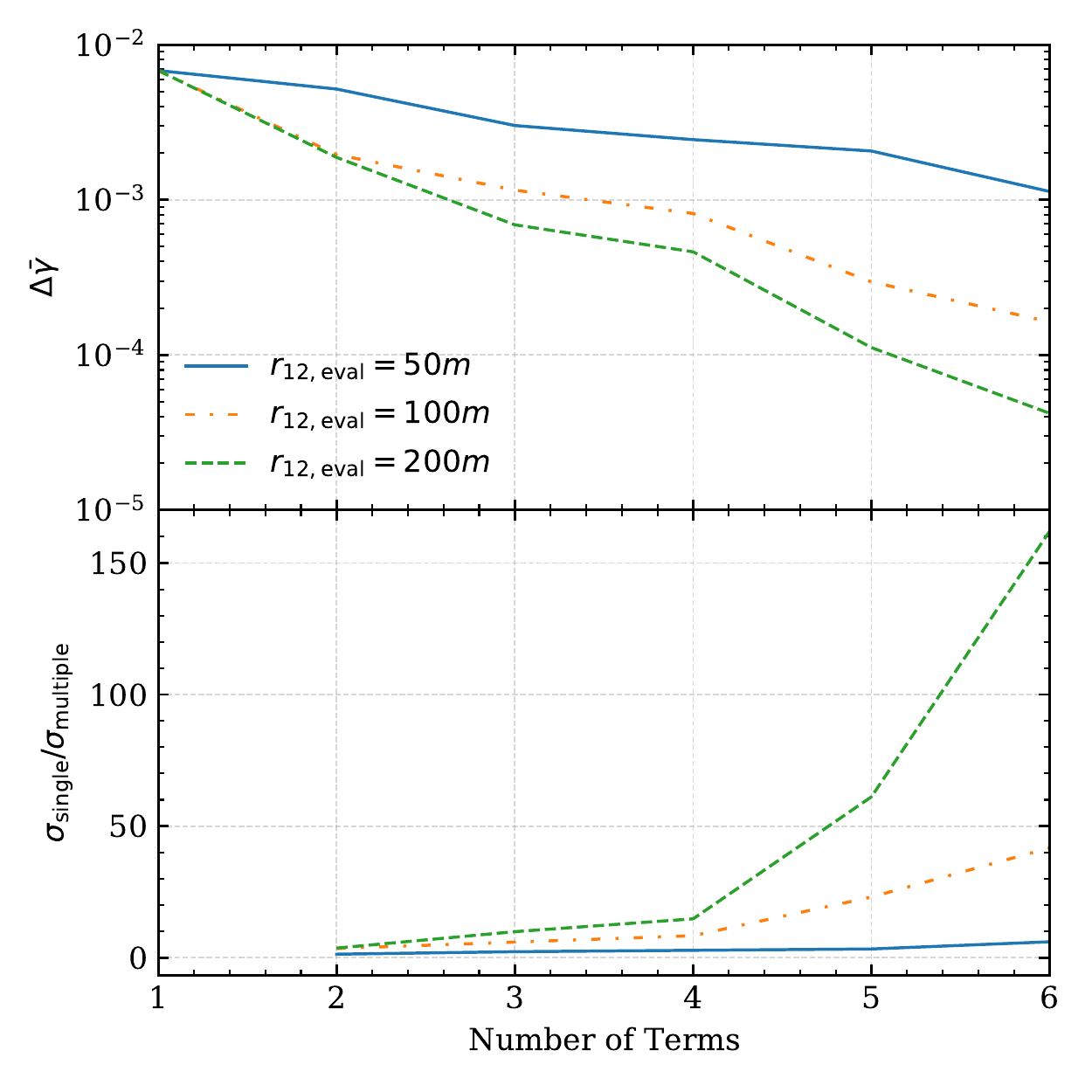}
    \caption{
    Marginalized $1\sigma$ constraint on $\bar{\gamma}$ for a modification that first enters at 0PN order as a function of number of PN terms kept in the modified sector, using 3 different choices of $r_{12,\rm eval}$ for the PN prior. In all cases, we here focus on the light source and the 2g detector network. Observe that in all cases the constraint on $\bar{\gamma}$ becomes stronger the more PN order terms are added. Observe also that the larger we choose $r_{12,\rm eval}$ to be, the stronger the ``pressure'' on $\bar{\gamma}$ and thus the stronger the constraint.  
    }
    \label{fig:prior_comparison}
\end{figure}

Given how important the PN prior is to prevent overly aggressively constraints on the leading PN order term, one may wonder whether these are robust to variations of the details associated with this prior. The main quantity we can vary to change the PN prior is the choice of velocity $v_{\rm eval}$ (or orbital separation $r_{12,eval}$) at which the prior is evaluated. Figure~\ref{fig:prior_comparison} shows the constraints on $\bar{\gamma}$ and the strengthening factor as a function of the number of higher PN order terms kept in the modified sector, but this time obtained with PN-based priors evaluated at 3 different orbital radii. Observe that in all cases the constraints improve as you add higher PN order terms in the modified sector. Therefore, the first main conclusion of our paper (i.e.~that single-parameter constraints are conservative) is robust to modifications in the PN-based prior. 

Figure~\ref{fig:prior_comparison} also shows that the degree of importance of the higher PN order terms does depend on the PN-based prior. Indeed, the strengthening factor is larger, the larger the value of $r_{12,\rm eval}$ that one chooses for the PN-based prior. This is consistent with our explanation above that as $r_{12,\rm eval} \to \infty$, then the prior on the higher PN order terms becomes flat and infinite, and, thus, $\Delta \bar{\gamma} \to 0$ because of the infinite ``pressure'' created by the higher PN order terms. The choice of $r_{12,\rm eval}$ should then be the smallest value of the orbital separation (or the orbital velocity) at which one expects the PN approximation to still be valid in the inspiral of compact binaries in the modified theory. Without specifying a particular modified theory, one can therefore not choose $r_{12, \rm eval}$ precisely. This is why we used the properties of the PN series in GR to set $r_{12,\rm eval} = 100 m$ in this paper, since we are sure that for such large values of $r_{12,\rm eval}$, the PN series in GR is still a good approximation. Had we chosen a smaller value, our conclusions about the importance of the higher PN order terms would have been even stronger (i.e.~we would have concluded the higher PN order terms are even less important than stated so far).

\begin{figure*}
\includegraphics[width=0.45\linewidth]{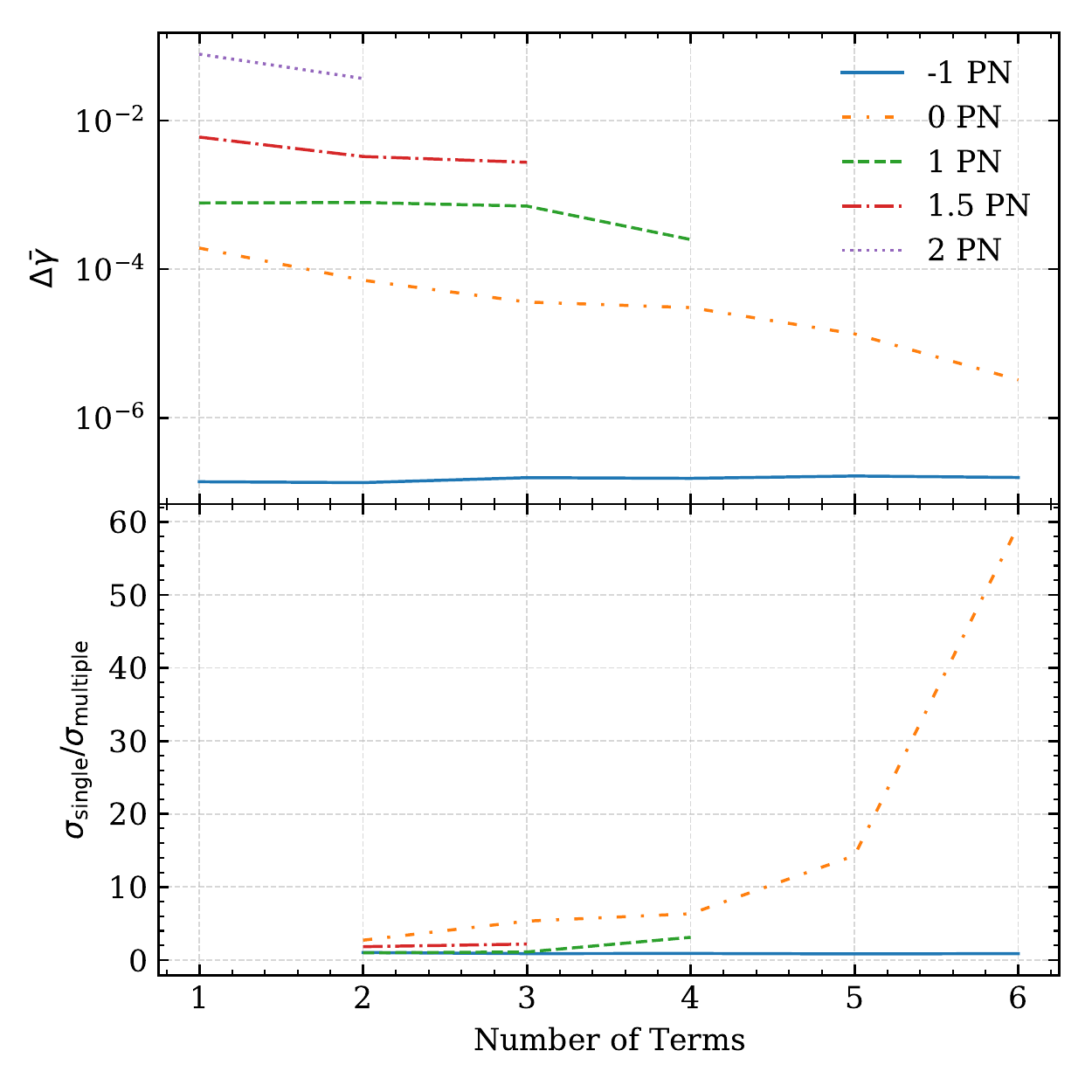}
\qquad 
\includegraphics[width=0.45\linewidth]{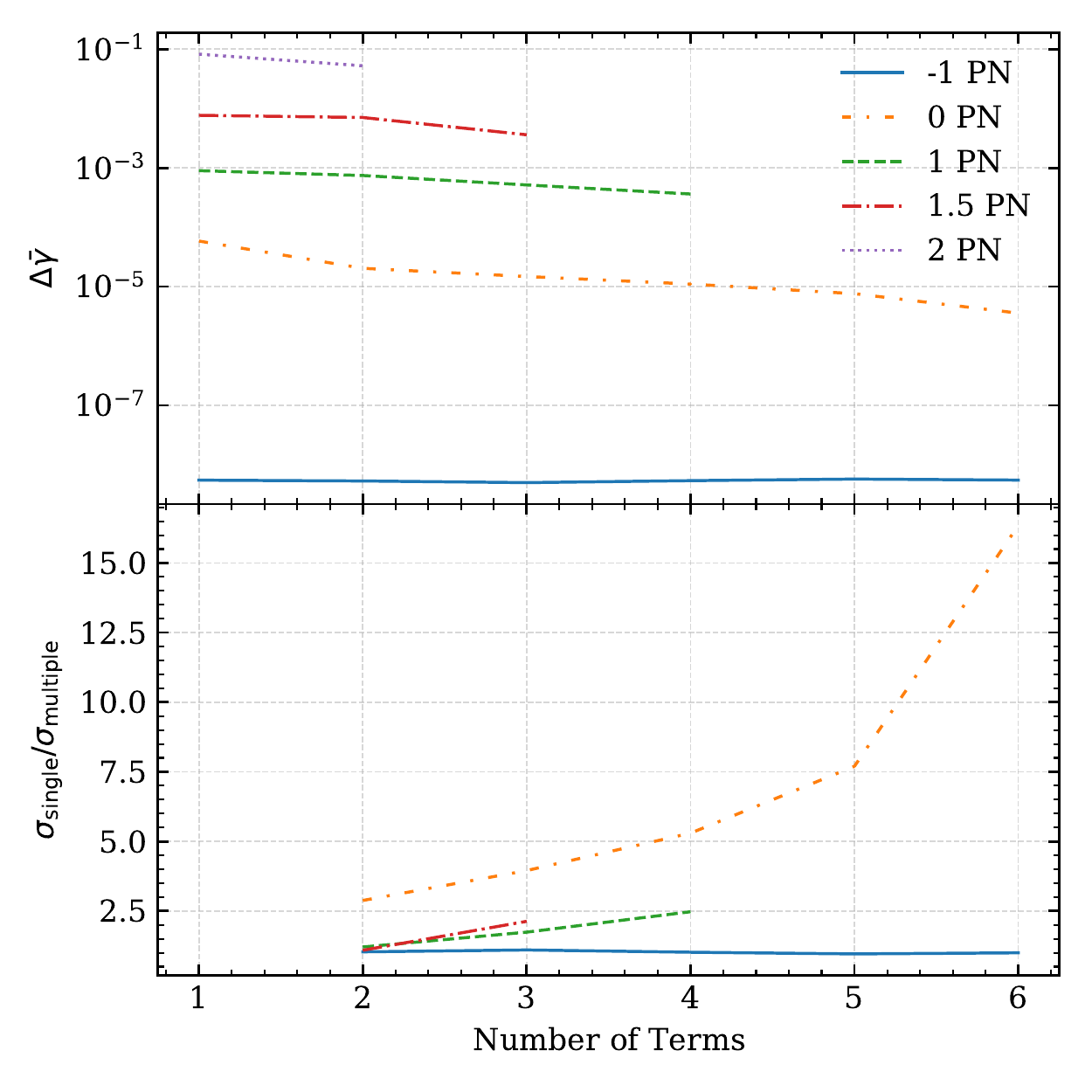}
\caption{
    Same as Fig.~\ref{fig:heavy_constraints_2g}, but for a 3g network, using a ``heavy'' source (left) and a ``light'' source (right).
    Observe that the trends found with the 2g detector networks continue when considering 3g detector networks.
}\label{fig:heavy_constraints_3g}
\end{figure*}

Our conclusions thus far are the following. First, single-parameter tests of GR are conservative, and would become stronger if higher PN order terms are included in the modified sector. Second, the improvement of these higher PN order terms is relatively mild, with enhancements of at most 1--2 orders of magnitude, depending on the inspiral signal observed. Given these conclusions, we then infer that current constraints on modified theories of gravity are robust to uncertainties in waveform modeling related to unknown, higher PN order corrections.
But are these conclusions robust also to an improvement in the detector's sensitivity, or alternatively, to an increase in the signal-to-noise ratio of the signals detected? Motivated by this, we repeated the analysis described above for the proxy of a 3g network described in Sec.~\ref{sec:exper_design}, and we arrived at very similar conclusions, as shown in Fig.~\ref{fig:heavy_constraints_3g} and Table~\ref{tab:heavy_3g}. Observe that the increased SNR of the injections does not change the trends we have described above for the less sensitive 2g networks. 
While the bounds become stronger because of the more sensitive detectors, the inferences made with single-parameter models are still robust to uncertainties in the higher PN order terms.

\begin{table*}
\begin{tabular}{ c | c c c c c}
Terms/LO & -1 & 0 & 1 & 1.5 & 2  \\ \hline \hline
1 & $1.4\times 10^{-7}$ & $1.9\times 10^{-4}$ & $7.8\times 10^{-4}$ & $6.0\times 10^{-3}$ & $7.8\times 10^{-2}$\\ \hline
2 & $1.3\times 10^{-7}$ & $7.1\times 10^{-5}$ & $7.9\times 10^{-4}$ & $3.3\times 10^{-3}$ & $3.7\times 10^{-2}$\\ \hline
3 & $1.6\times 10^{-7}$ & $3.6\times 10^{-5}$ & $7.1\times 10^{-4}$ & $2.7\times 10^{-3}$ & \\ \hline
4 & $1.5\times 10^{-7}$ & $3.0\times 10^{-5}$ & $2.5\times 10^{-4}$ &  & \\ \hline
5 & $1.6\times 10^{-7}$ & $1.3\times 10^{-5}$ &  &  & \\ \hline
6 & $1.6\times 10^{-7}$ & $3.2\times 10^{-6}$ &  &  & \\ \hline
\end{tabular}
\qquad
\begin{tabular}{ c | c c c c c}
Terms/LO & -1 & 0 & 1 & 1.5 & 2  \\ \hline \hline
1 & $5.5\times 10^{-9}$ & $5.8\times 10^{-5}$ & $9.0\times 10^{-4}$ & $7.7\times 10^{-3}$ & $8.2\times 10^{-2}$\\ \hline
2 & $5.3\times 10^{-9}$ & $2.0\times 10^{-5}$ & $7.4\times 10^{-4}$ & $7.0\times 10^{-3}$ & $5.2\times 10^{-2}$\\ \hline
3 & $5.0\times 10^{-9}$ & $1.5\times 10^{-5}$ & $5.2\times 10^{-4}$ & $3.6\times 10^{-3}$ & \\ \hline
4 & $5.4\times 10^{-9}$ & $1.1\times 10^{-5}$ & $3.6\times 10^{-4}$ &  & \\ \hline
5 & $5.7\times 10^{-9}$ & $7.6\times 10^{-6}$ &  &  & \\ \hline
6 & $5.5\times 10^{-9}$ & $3.6\times 10^{-6}$ &  &  & \\ \hline
\end{tabular}
\caption{
    Same as Table~\ref{tab:heavy_2g}, but for a 3g network, using the ``heavy'' source (top) and the ``light'' source (bottom).
    Once more, the trends found with the 2g detector networks continue when considering a 3g detector network.
}\label{tab:heavy_3g}
\end{table*}

\section{Alternative Parametrizations}\label{sec:alt_params}

One may wonder how robust the conclusions presented in Sec.~\ref{sec:improved_methods} are to the exact form of the parametrization we implemented in that section. 
Therefore, before continuing on to a specific theory of gravity, let us briefly examine two other reasonable parameterizations to determine how our conclusions are affected. 
In what follows, we will re-analyze the light system with the 2g network, exactly as defined earlier, using 6 deformations and starting with the Newtonian-order term relative to GR, i.e. terms at [0,1,1.5,2,3,3.5]PN orders relative to GR. The difference, however, will be in the exact form of the parametrization of the deformations at each of those orders and how the prior for those deformations is imposed.

\begin{figure*}[!h]
    \includegraphics[width=.45\linewidth]{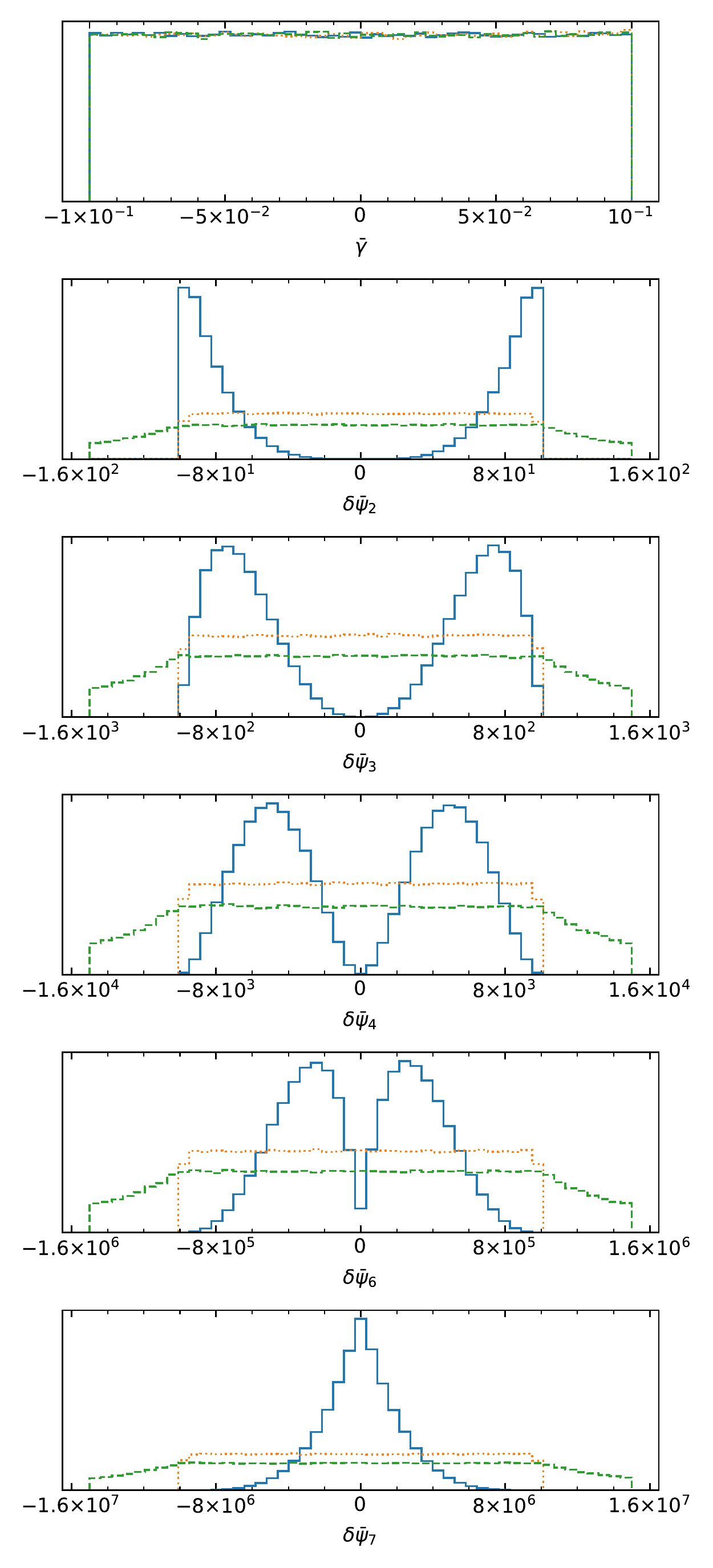}
     \includegraphics[width=.45\linewidth]{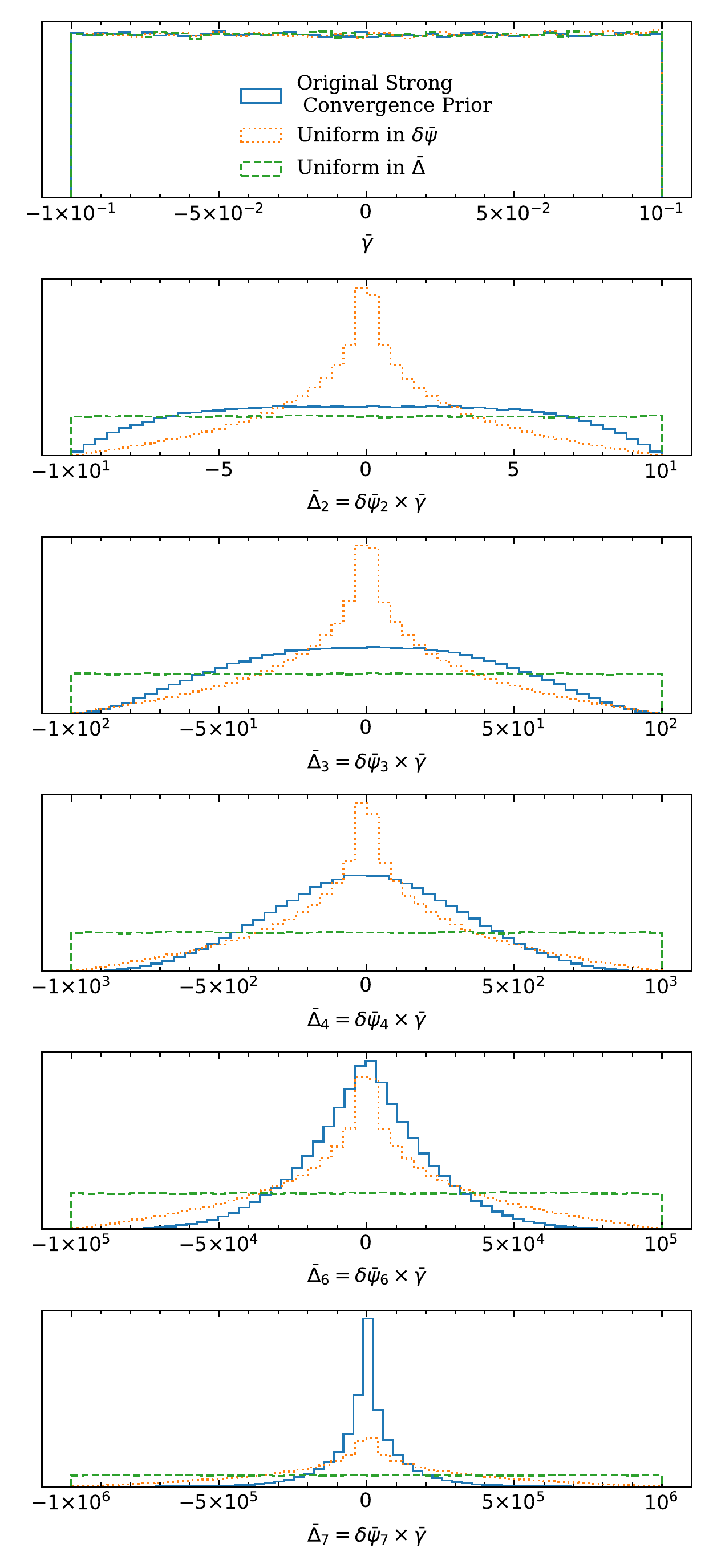}
    \caption{
    The one dimensional, marginalized prior on a set of deformations in the three models considered in Sec.~\ref{sec:alt_params}.
    The leading-order deformation enters at Newtonian order relative to GR, and each of the subsequent deformations are labeled as the $i/2$PN order relative to this leading order deformation.
    Included in this figure is the original model described in Sec.~\ref{sec:improved_methods}, shown as a solid blue line. 
    The first alternative model uses the $\delta\bar{\psi}_i$ parametrization described in Eq.~\eqref{eq:PN_form_2} but with simple, uniform ranges with fixed (increasingly larger with higher PN order) boundaries, shown as a dotted orange line. 
    The second alternative model is described by a series of $\bar{\Delta}_i$ parameters with uniform distributions with fixed (increasingly larger with higher PN order) boundaries, defined in Eq.~\eqref{eq:PN_form_alternate}, shown as a dashed green line.
    Note that the range of the prior on the $\delta \bar{\psi}_i$ parameters in the second alternative parametrization extends beyond the frame of the figure, but the range was restricted for visual purposes.
    }\label{fig:prior_example_alt}
\end{figure*}

The first model uses the same functional form as the majority of our work here, defining the deformations in the phase as a series of coefficients exactly as shown in Eq.~\eqref{eq:PN_form_2}, but with a \textit{different} prior. We modify the prior to have simple, fixed boundaries, as opposed to the more complicated prior used in the main body of this work (previously, we enforced the convergence criteria at every point in parameter space through Eq.~\eqref{eq:prior_boundary} and Eq.~\eqref{eq:prior_boundary_NLO}). In particular, the prior for the new model will have fixed boundaries on $\bar{\gamma}$ and the $\delta \bar{\psi}_i$'s with progressively larger ranges, namely
\begin{align}\nonumber
    |\bar{\gamma} | < 10^{-1} \,, \\ \nonumber
    |\delta \bar{\psi}_2 | < 10^{2} \,, \\ \nonumber
    |\delta \bar{\psi}_3 | < 10^{3} \,, \\ \nonumber
    |\delta \bar{\psi}_4 | < 10^{4} \,, \\ \nonumber
    |\delta \bar{\psi}_6 | < 10^{6} \,, \\ 
    |\delta \bar{\psi}_7 | < 10^{7} \,.
\end{align}
This prior ensures that the convergence criteria is \emph{generally} satisfied but not guaranteed, while removing some of the complicated structure from our original prior (shown in Fig~\ref{fig:prior_example}). By reanalyzing our synthetic data with this new model, we will quantify how strong of an impact our enforcement of a strict convergence criteria had on our results above.

We note in passing that the prior range on $\bar{\gamma}$ never played much of a role in our analysis, as the prior bounds on the higher-order parameters, $\delta \bar{\psi}_i$'s, were independent of $\bar{\gamma}$. 
As we will see below, that will no longer be the case, and we therefore updated our prior bound on $\bar{\gamma}$ to a much more reasonable range of $|\bar{\gamma}| < 10^{-1}$ (informed by our original experiments). 
We have verified that we obtain the same results with this new prior range and with the original parametrization.

The second parametrization we investigate is written in the form of Eq.~\eqref{eq:PN_form_alternate}, where our model is parametrized by $\bar{\gamma}$ and the series of $\bar{\Delta}_i$'s.
In other words, we are moving from parametrizing our deformations as a relative series of terms to working with a model described by the \emph{absolute} deformations.
With this parametrization, we enforce similar priors as our first alternative parametrizations, with fixed boundaries of successively larger sizes, namely
\begin{align}\nonumber
    |\bar{\gamma} | < 10^{-1} \,, \\ \nonumber
    |\bar{\Delta}_2 | < 10^{1} \,, \\ \nonumber
    |\bar{\Delta}_3 | < 10^{2} \,, \\ \nonumber
    |\bar{\Delta}_4 | < 10^{3} \,, \\ \nonumber
    |\bar{\Delta}_6 | < 10^{5} \,, \\ 
    |\bar{\Delta}_7 | < 10^{6} \,. 
\end{align}
With this model, we will investigate the impact that our choice of parameterization had on our conclusions of the previous section.

As the prior distribution is the major reason for testing these alternative parametrizations, we show all three prior distributions on $\bar{\gamma}$, $\delta \bar{\psi}_i$, and $\bar{\Delta}_i$ in Fig.~\ref{fig:prior_example_alt}.
From this figure, we can see that the first alternative parametrization produces the strongest prior on the absolute deformations, $\bar{\Delta}_i$, for low orders, but the original prior still places the most prior weight at small values of $\bar{\Delta}_i$ for large PN orders.
Our first alternative parametrization does alleviate the ``pressure'' on the $\delta \bar{\psi}_i$ parameters seen in the original parametrization, which was pushing low PN deformations away from zero (and away from GR) and pushing high PN deformations towards zero.
The second alternative parametrization has a uniform distribution in the absolute deformations, or the $\bar{\Delta}_i$ parameters, giving a variation of an uninformative prior. 
From this figure, we would expect comparable results from the original parametrization and the first alternative model, as the prior distributions are relatively similar.
The second alternative parametrization, uniform in $\bar{\Delta}_i$, will most likely result in the most degradation between the six-deformation and single-deformation model, as the prior is the least informative and most similar to the current, state-of-the art methodology.

\begin{figure}[h!]
    \centering
    \includegraphics[width=\linewidth]{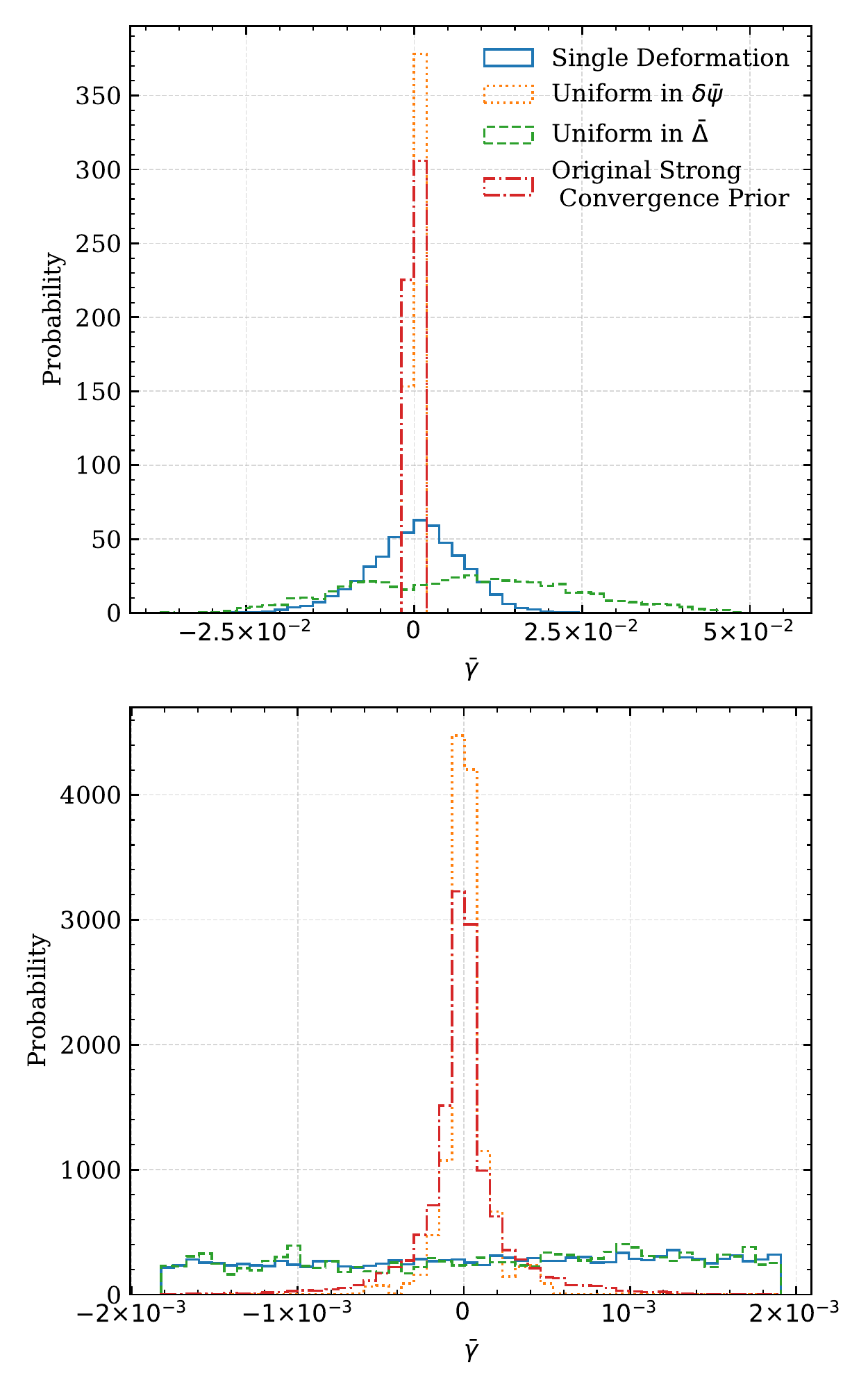}
    \caption{
    The final, marginalized posterior distributions on $\bar{\gamma}$ coming from the three models discussed in Sec.~\ref{sec:alt_params} using the synthetic data from the ``light'' source as observed by the 2g detector network.
    The top panel shows a larger range of $\bar{\gamma}$ and the lower panel zooms-in the range of $\bar{\gamma}$ to present details in the two narrower distributions.
    The probabilities on the y-axis are normalized to the shown range of $\bar{\gamma}$ in both panels.
    The distribution coming from the model using a single deformation parameter at Newtonian order is shown as a solid blue line.
    The distribution coming from the model described by the first alternative parametrization with six deformations (with a prior uniform in $\delta \bar{\psi}_i$) is shown as the dotted orange line.
    The distribution coming from the model described by the second alternative parametrization with six deformations(with a prior uniform in $ \bar{\Delta}_i$) is shown as the dashed green line.
    The distribution coming from the model described by the original parametrization with six deformations (with a prior that strictly enforces our notion of convergence) is shown as the dotted-dashed red line.
    Note that the first alternative parametrization and the original parametrization are almost totally overlapping in the top panel.
    }
    \label{fig:param_comp}
\end{figure}

We now use these two additional parametrizations to analyze the ``light'' source, as seen by the 2g detector network, and compare the constraints on $\bar{\gamma}$ to the posterior coming from a model that has a single deformation at Newtonian order and to the constraint coming from the original parametrization with six deformations.
As all three models (the original model and the two alternative parametrizations) reduce to the same form as the number of deformations is taken to one, this will provide a consistent metric for comparison.
The results for those analyses are shown in Fig.~\ref{fig:param_comp}, and as you can observe, they are consistent with the expectations presented above.
The two models described by a relative series of deformations at higher orders (the original model and the first alternative model) produce comparable results.
The second alternative model (uniform in $\bar{\Delta}_i$) produces a constraint that is mildly weaker than the original, single deformation model. We therefore conclude that the conclusions presented in Sec.~\ref{sec:improved_methods} are robust to our parameterization and choice of priors. 

\section{A Concrete Example: ssGB theory}\label{sec:ssGB}

The methodology we have proposed here seems to accomplish the purpose it was designed for: allowing for realistic uncertainty in our limited modeling while illustrating the current robustness of the bounds placed on modified theories of gravity.
A critical question, however, is the accuracy to which it actually relates to known, interesting theories of gravity actively being studied.
One particular theory currently of interest is scalar Gauss Bonnet, or sGB, which is inspired by low-energy limits of higher energy theories~\cite{Kanti:1995vq,Boulware:1985wk,1987NuPhB.291...41G}.
sGB contains a scalar field, $\phi$, that  couples to a curvature invariant called the Gauss Bonnet invariant, $\mathcal{G}$.
The Gauss Bonnet invariant is defined by $\mathcal{G} = R^2 -4 R_{\mu \nu} R^{\mu \nu} + R_{\mu \nu \rho \sigma } R^{\mu \nu \rho \sigma }$, where $R_{\mu \nu \rho \sigma }$, $R_{\mu \nu}$, and $R$ are the Riemann tensor, Ricci tensor, and Ricci scalar, respectively.
The full action can be written as~\cite{Shiralilou:2021mfl}
\begin{equation}\label{eq:sgb_action}
    S = \frac{1}{16 \pi} \int d^4 x \sqrt{-g}\left[ R - 2 \partial^\mu \phi \partial_\mu \phi + \alpha f(\phi) \mathcal{G} \right] \,,
\end{equation}
where $g$ is the determinant of the metric, $\alpha$ is the (dimensionful) coupling constant of the theory, and $f(\phi)$ is a coupling function. 
The theory reduces to GR minimally coupled to the scalar field in the $\alpha \to 0$.
Therefore, constraints on this theory typically relate to the magnitude of $\alpha$.
Certain choices of the coupling function $f(\phi)$ result in different flavors of sGB.
In particular, an interesting choice motivated by string theory is Einstein-dilaton-Gauss-Bonnet, where $f(\phi) = e^{2 \phi}/4$~\cite{Kanti:1995vq}.
Another option, which will be the focus of this paper, is shift-symmetric sGB or ssGB, obtained by making the choice of $f(\phi) = 2 \phi$~\cite{Yunes:2011we,Yagi:2015oca,Sotiriou:2014pfa}, which results from a small-field expansion of Einstein-dilaton-Gauss-Bonnet gravity~\cite{Yagi:2015oca}. 

The GW phase in ssGB was recently calculated beyond the leading PN order modification (at -1PN order, relative to Newtonian order in GR) in~\cite{Shiralilou:2021mfl}, giving us an example theory to explore. 
In their work, they classified the inspiral into two distinct regimes: that which is dominated by the emission of dipolar radiation by the scalar field (which is absent in GR), and that which is dominated by quadrupolar radiation.
The boundary between dipolar and quadrupolar radiation is typically well before the binary reaches a separation of $100 m$, so the regime of interest to ground-based GW detectors is the quadrupolar driven one.
To evaluate the convergence properties of the series, we take the expression for the GW phase derived in Eq.~\pseudoeqref{93}--\pseudoeqref{96} of~\cite{Shiralilou:2021mfl} and isolate terms involving deviations from GR, i.e.~terms that involve $\alpha$, the coupling constant in ssGB.
This gives the following form for the phase modification
\begin{equation}
    \Psi_{\rm GW} - \Psi_{\rm{GW},\rm{ GR}} = \delta \psi_{\rm LO, \rm ssGB} v^{-7} + \delta \psi_{\rm NLO, \rm ssGB} v^{-5} + \delta \psi_{\rm NNLO, \rm ssGB} v^{-3}  \,,
\end{equation}
where the coefficients take the form 
\begin{align}\label{eq:ssGB_coeff1}
    \delta \psi_{\rm LO, \rm ssGB} &= -\frac{5}{7168 }\zeta  \frac{\left(4 \eta - 1 \right)}{\eta^5}\,,\\\label{eq:ssGB_coeff2}
    \delta \psi_{\rm NLO, \rm ssGB} &= - \frac{5}{688128} \zeta\frac{\left( 685 - 3916 \eta + 2016 \eta^2 \right)}{\eta^5}\,,\\\label{eq:ssGB_coeff3}
    \delta \psi_{\rm NNLO, \rm ssGB} &= \frac{5}{387072} \zeta \frac{\left( 1-2 \eta \right)^2 \left(  995 + 952 \eta \right)}{\eta^5} \,,
\end{align}
where $\zeta \equiv {\alpha^2}/{m^4}$ is the dimensionless coupling constant of the theory\footnote{The expressions presented here are not valid in the $\eta \ll 1$ regime, because then the curvature of the small black hole becomes very large, and the effective field theory treatment used to derive these expressions breaks down.}. We have verified that this expression reduce exactly to the leading PN order results first obtained in~\cite{Yagi:2011xp,Yunes:2016jcc} in the non-spinning limit\footnote{Note that the results of~\cite{Shiralilou:2021mfl} are formally complete only to 1PN order, implying that the $v^{-3}$ coefficient has a 2PN correction coming from the dipolar term that has not yet been computed.}. 
Observe that the phase deformation has the exact structure that we anticipated in Eq.~\eqref{eq:PN_form_2}.
Namely, the above phase deformation can be rewritten as 
\begin{align}\nonumber
    \Psi_{\rm GW} - \Psi_{\rm GW,\rm GR} &= \zeta  \, \delta \psi_{\rm LO,ssGB}(\eta) \, v^{-7} \\
    &\times \left[ 1 + \delta \psi'_{\rm NLO,\rm ssGB}(\eta) \, v^2 + \delta \psi'_{\rm NNLO,\rm ssGB}(\eta) \, v^4\right]\,,
\end{align}
with the following re-definitions of the coefficients
\begin{align}
    \delta \psi'_{\rm NLO,\rm ssGB} &= \frac{\delta\psi_{\rm NLO,\rm ssGB}}{\delta \psi_{\rm LO,ssGB}} =  \frac{685 - 3916 \eta + 2016 \eta^2 }{96 \left( 4 \eta - 1\right)}\,
    \\
    \delta \psi'_{\rm NNLO,\rm ssGB} &=
    \frac{\delta \psi_{\rm NNLO,\rm ssGB}}{\delta \psi_{\rm LO,ssGB}} =  - \frac{\left(1-2\eta \right)^2 (995 + 952 \eta)}{54 \left( 4\eta - 1\right)}\,.
\end{align}
The phase modification has therefore been written as a series in $v$, all proportional to the coupling constant $\alpha$ appearing in the coefficient of the overall controlling factor of the series $\delta \psi_{\rm LO,ssGB}$.

\begin{figure*}
    \centering
    \includegraphics[width=0.45\linewidth]{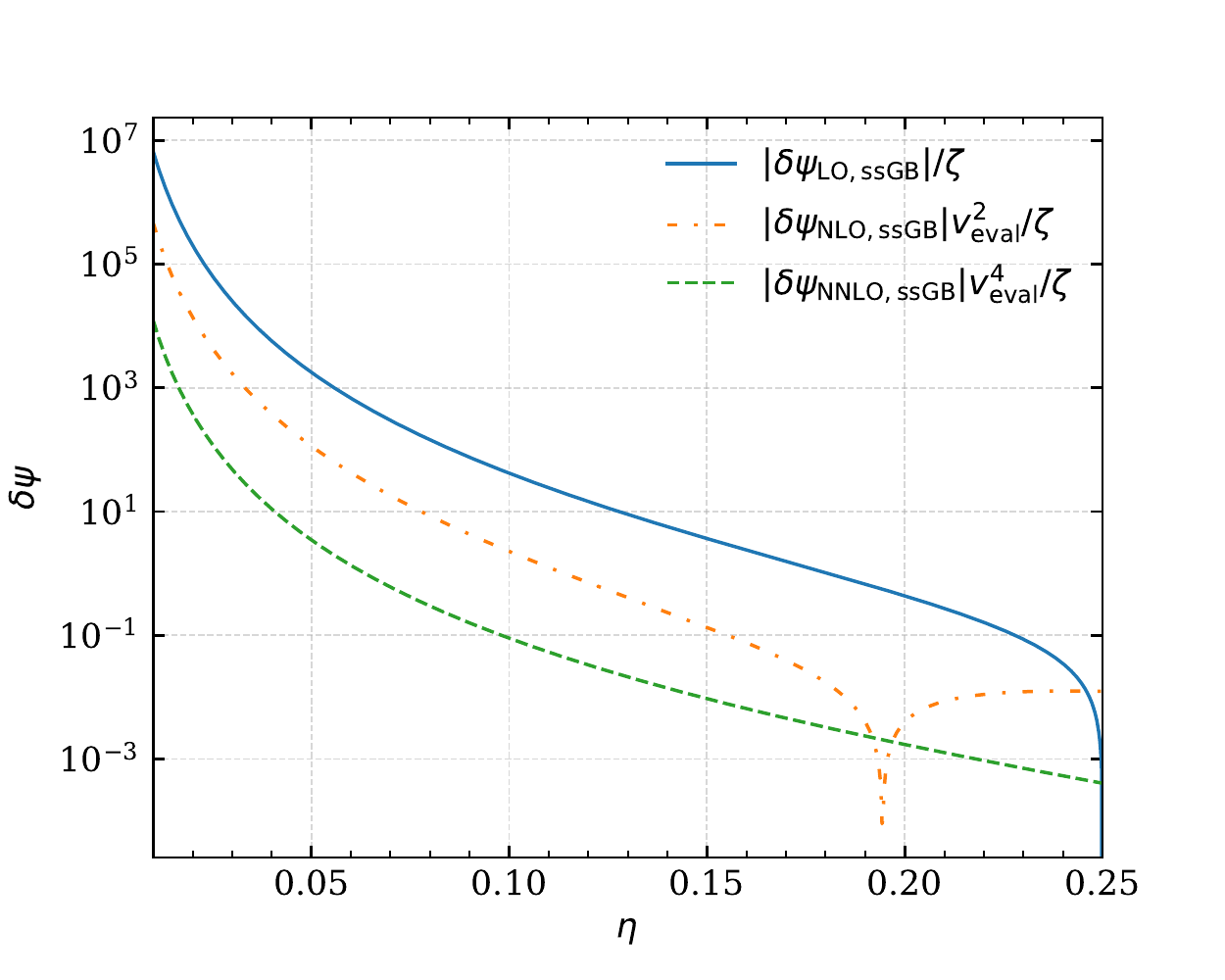}
    \includegraphics[width=0.45\linewidth]{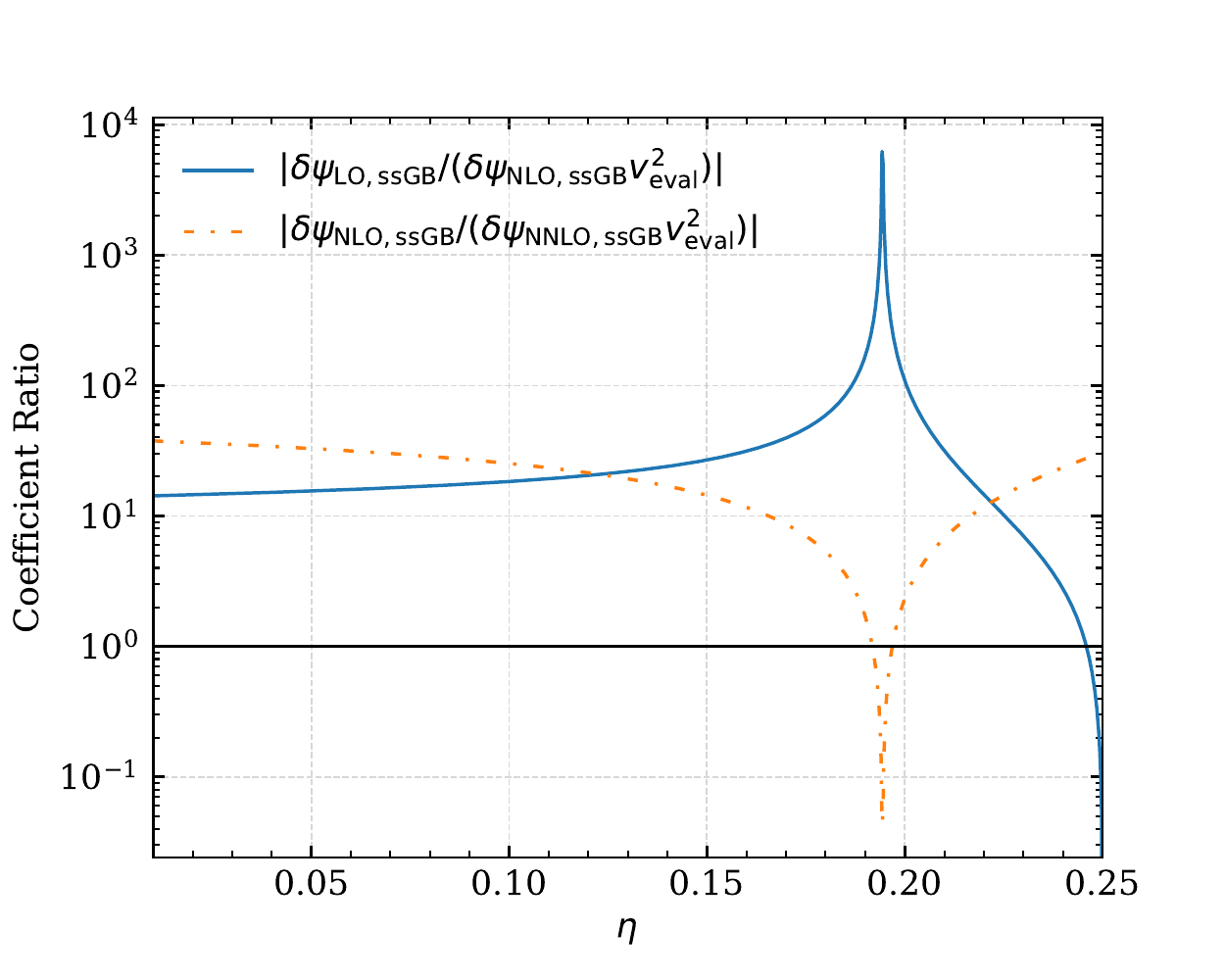}
    \caption{Absolute value of the terms in the phase deformation at different PN order (left) and ratio of terms (right) in ssGB as a function of symmetric mass ratio, evaluated at $v_{\rm eval} = 0.1$.
    Observe that, for this choice of $v_{\rm eval}$, the leading PN order term is larger than the next-to-leading order one, which in turn is larger than the next-to-next-to-leading order, except at $\eta \sim 0.19$, where the 1PN term vanishes identically. Observe also that the ratios are all larger than unity, except again for a specific value of $\eta$ at which the 1PN term vanishes.
    }
    \label{fig:ssGB_coeff}
\end{figure*}

With this in hand, we can now compare the above results to the general framework we developed in the previous section when studying the convergence properties of the PN series in modified theories.
To do so, we need to look at the ratios of these coefficients, which we show below.
For the LO and NLO terms, we arrive at the following expression for their ratio
\begin{equation}\label{eq:ssGB_ratio1}
\frac{\delta \psi_{\rm LO, \rm ssGB}}{\delta \psi_{\rm NLO,\rm ssGB} v_{\rm eval}^2} =  \frac{96 (4 \eta -1)}{\left[685 + 4 \eta \left(-979 + 504\eta\right)\right]v_{\rm eval}^2}\,.
\end{equation}
For the NLO and NNLO terms, we evaluate their ratio to 
\begin{equation}\label{eq:ssGB_ratio2}
\frac{\delta \psi_{\rm NLO, \rm ssGB}}{\delta \psi_{\rm NNLO,\rm ssGB} v_{\rm eval}^2} = \frac{9\left[685 + 4 \eta \left(-979 + 504\eta\right)\right]}{  16 \left( 1-2 \eta\right)^2 \left( 995 + 952 \eta\right)v_{\rm eval}^2}\,.
\end{equation}
As we are only considering non-spinning binaries, these ratios are only a function $\eta$, so we can easily plot them to determine their convergence properties.
The absolute deviations (divided by $\zeta$), as shown in Eqs.~\eqref{eq:ssGB_coeff1},~\eqref{eq:ssGB_coeff2}, and~\eqref{eq:ssGB_coeff3} and including appropriate values of $v_{\rm eval}$, are shown in the left panel of Fig.~\ref{fig:ssGB_coeff}, while the ratios of these coefficients, as defined in Eqs.~\eqref{eq:ssGB_ratio1} and~\eqref{eq:ssGB_ratio2}, are shown in the right panel.

Figure~\ref{fig:ssGB_coeff} illustrates that the PN expansion of ssGB theory conforms, on average, to the criteria we have outlined in Sec.~\ref{sec:improved_methods}. This is most easily evidenced by the right panel of this figure, which shows that the ratio of coefficients is always above unity, except for a specific value of $\eta$ at which the 1PN term $\delta \psi_{\rm NLO, \rm ssGB}$ vanishes. Since this happens at a single point, averaging over all values of $\eta$, it is clear that the convergence criteria is satisfied. 

\begin{figure}
\includegraphics[width=\linewidth]{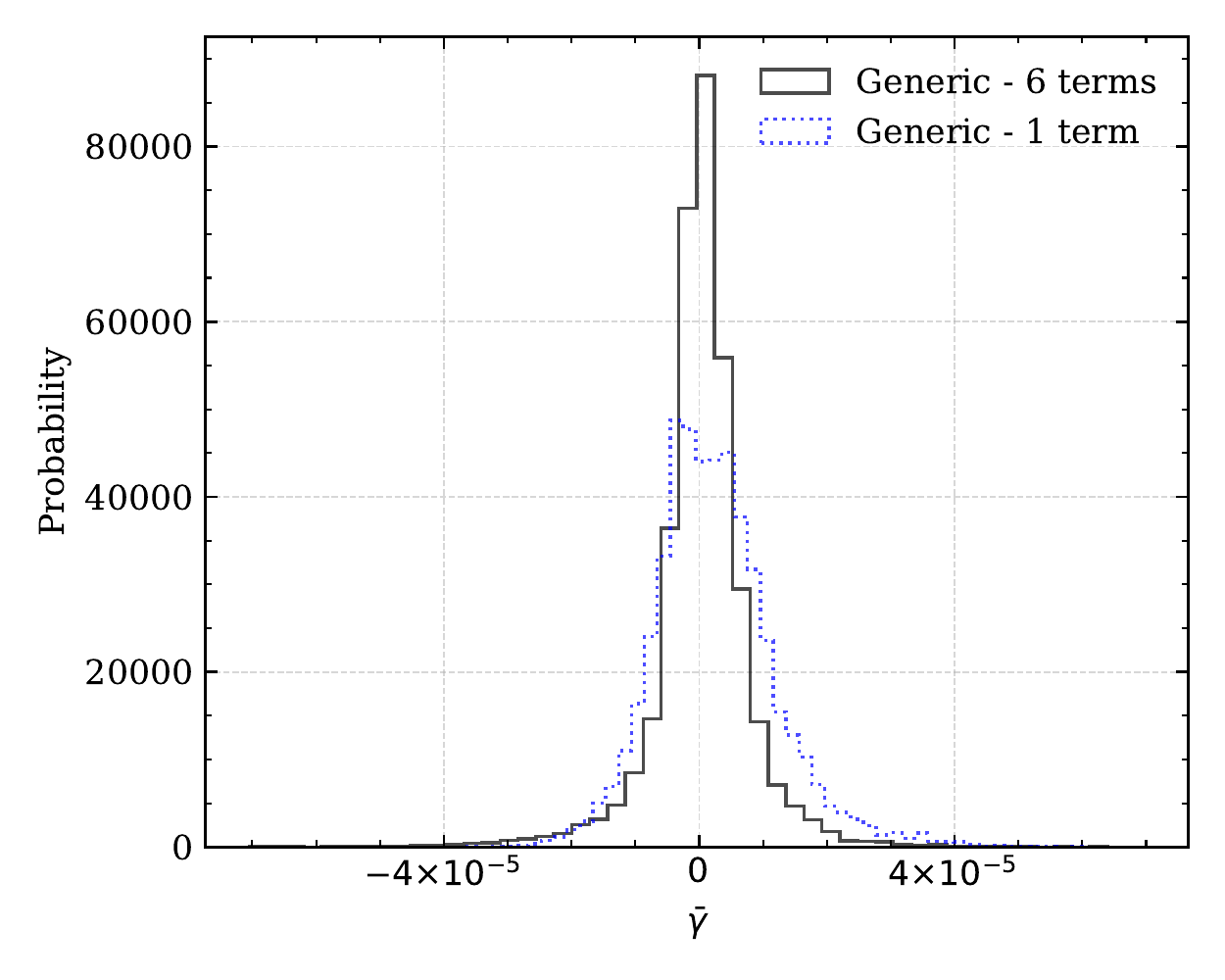}
\caption{
    For reference, we show the two distributions on $\bar{\gamma}$ coming from the two generic deformation models (models 3 and 4).
    This figure shows that the distributions on the generic parameter $\bar{\gamma}$ for these two models are comparable, with a slight enhancement when using the 6 deformation variation.
    This same information is shown graphically in Fig.~\ref{fig:heavy_constraints_2g} and tabulated in Table~\ref{tab:heavy_2g}.
    When compared with Fig.~\ref{fig:sgb_comparison}, we see that this relation persists. 
    Even once the posteriors are transformed to theory-specific constants, constraints are not meaningfully changed when using these two models.
    Furthermore, note that the posterior on $\bar{\gamma}$ is fully consistent with GR (the model that truly describes the injected data), which is further evidence that the deviation away from GR in Fig~\ref{fig:sgb_comparison} for the generic models is a artifact of the transformation. 
}\label{fig:EdGB_generic}
\end{figure}

But are our conclusions from Sec.~\ref{sec:exper_results} still valid in this specific theory?
To answer this question, we completed another series of Bayesian studies, but now within the context of ssGB theory.
We used the ``light'' injection from Table~\ref{tab:inj_param}, and recovered with a IMRPhenomD model with one of the following phase deformations appended to its inspiral phase:
\begin{itemize}
\setlength\parskip{0.0cm}
\setlength\itemsep{0.1cm}
    \item Model 1: only the leading PN order deformation in ssGB theory (defined in Eq.~\eqref{eq:ssGB_coeff1}),
    \item Model 2: all the phase deformation terms in ssGB theory (defined in Eq.~\eqref{eq:ssGB_coeff2} and Eq.~\eqref{eq:ssGB_coeff3})
    \item Model 3: a single ppE deformation term (as in Eq.~\eqref{eq:PN_form_2} but with $\delta \bar{\psi}_{i>0}  =0$), 
    \item Model 4: 6-parameter ppE deformations (as in Eq.~\eqref{eq:PN_form_2} but with $\delta \bar{\psi}_{i>6}  =0$).
\end{itemize}
For models 3 and 4, after carrying out a Bayesian parameter estimation study, we mapped the constraints on $\bar{\gamma}$ to constraints on $\alpha$ through
\begin{equation}
    \bar{\gamma} = \delta \psi_{\rm LO,ssGB}, \qquad b=-7\,,
\end{equation}
to enable comparisons with the results using models 1 and 2. 
For reference, the posteriors for $\bar{\gamma}$ for models 3 and 4 are shown in Fig.~\ref{fig:EdGB_generic}. 
This figure simply reiterates in more detail the conclusions from Sec.~\ref{sec:exper_results} (Fig.~\ref{fig:heavy_constraints_2g} and Table~\ref{tab:heavy_2g}). 
Namely, the difference between using a single deformation and six deformations within the model established in Sec.~\ref{sec:improved_methods} is a mild enhancement of constraints.

\begin{figure}
\includegraphics[width=\linewidth]{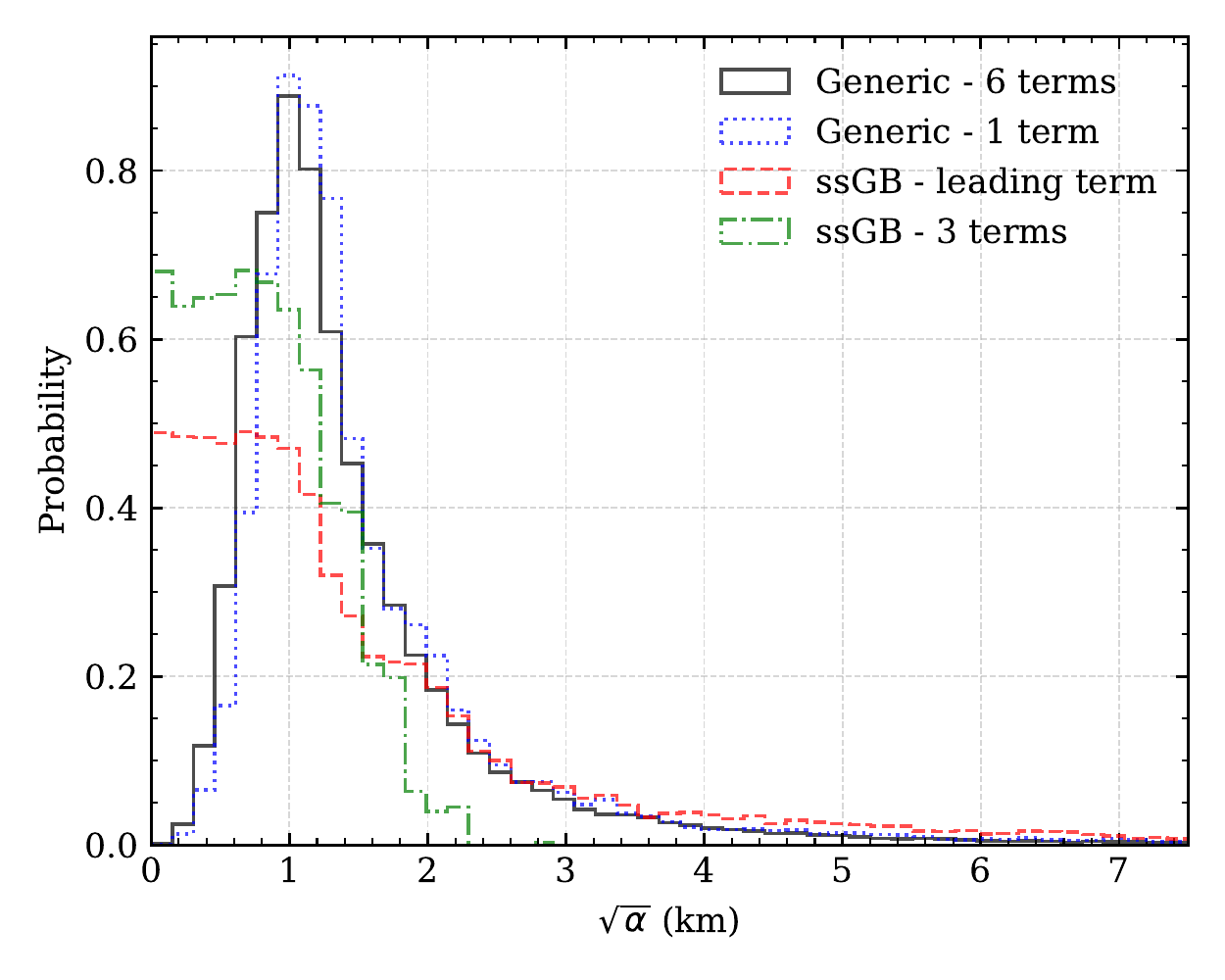}
\caption{
    Marginalized posterior distributions on $\sqrt{\alpha}$ for a GR injection extracted with the four models described in the text. We here considered the ``light'' source, defined by Table~\ref{tab:inj_param}, and a 2g detector network. When carrying out the Bayesian studies with the multi-parameter ppE model, we employed the PN prior with $r_{12,\rm eval} = 100$, as done in the rest of this paper. The constraints on $\bar{\gamma}$ obtained with models 3 and 4 were mapped to constraints on $\sqrt{\alpha}$ to enable comparisons with the results obtained with models 1 and 2. Observe that clearly the posterior distributions are all consistent with each other (modulo singularities in the transformation at $\alpha = 0$, discussed in detail in~\cite{Nair:2019iur,Perkins:2021mhb}). This shows clearly that in ssGB theory, leading PN order constraints are conservative and sufficient to place bounds on the theory. 
}\label{fig:sgb_comparison}
\end{figure}

Marginalized posterior distributions of $\sqrt{\alpha}$ obtained using these four models are shown in Fig.~\ref{fig:sgb_comparison}.
While the transformation leaves certain artifacts in the posterior distributions because of singularities in the transformation (as discussed in great detail in~\cite{Perkins:2021mhb} and~\cite{Nair:2019iur}), the upper limits on $\sqrt{\alpha}$ obtained with the four models are consistent with each other.
The $90\%$ confidence upper limit on $\sqrt{\alpha}$ with a single deformation to the phase (the red curve in Fig.~\ref{fig:sgb_comparison}) is $5.0$~km, while that obtained using the full, three term modification to the GW phase (the green curve in Fig.~\ref{fig:sgb_comparison}) is $1.5$~km, leading to a strengthening factor of about 3. 
The additional information incorporated into the waveform through the higher order deformation only serve to improve the constraint on $\sqrt{\alpha}$.
This comparison provides further evidence that constraints derived from leading PN order deformations are robust to future work on deriving higher PN order corrections.

The bias in the distribution on $\sqrt{\alpha}$ coming from the generic parametrizations is a known issue~\cite{Nair:2019iur,Perkins:2021mhb}, and is not of much concern in the present context.
The issue fundamentally lies with the Jacobian of the transformation between the two parametrizations, and causes the derived prior distribution on $\sqrt{\alpha}$ to go to zero in the GR limit when mapping the generic parametrization to ssGB.
When transforming distributions between two different basis, one must account for the Jacobian of the transformation as
\begin{equation}\label{eq:jacobian_transform}
p(\paramvec_1) = p(\paramvec_2) \frac{d \paramvec_2}{d \paramvec_1}\,,
\end{equation}
In the present case, we are using one parametrization $\paramvec_1 \equiv [\Xi \cup \sqrt{\alpha}]$ in ssGB and one parametrization $\paramvec_2 \equiv [\Xi \cup \bar \gamma \cup \sum_i \delta\bar \psi_i]$ in the generic framework.
From Eq.~\eqref{eq:jacobian_transform}, we know we need the Jacobian, $d \paramvec_1 / d \paramvec_2$, to transform from parametrization 2 to parametrization 1.
Here lies the issue, as the component of the Jacobian related to $\sqrt{\alpha}$ and $\bar \gamma$ is given as 
\begin{equation}\label{eq:jac}
    \frac{d\bar \gamma }{d \sqrt{\alpha}} \propto (\alpha)^{3/2}\,,
\end{equation}
using the relation in Eq.~\eqref{eq:ssGB_coeff1}.
As this expression goes to 0 in the $\sqrt{\alpha} \rightarrow 0$ limit, the prior will have zero weight for the GR limit.
Now the discrepancy is clear: the lack of agreement with the posterior on $\sqrt{\alpha}$ coming from the generic parametrization is not an indication that GR is lacking, but instead an inherent flaw of the parametrization. 
While this is an indication of a failing of the parametrization, the fact that the prior seems to be widening the constraint on $\sqrt{\alpha}$ instead of artificially shrinking it indicates that any constraints derived from this method related to \emph{upper limits on $\sqrt{\alpha}$} are actually \emph{conservative}.

\begin{figure}
\includegraphics[width=\linewidth]{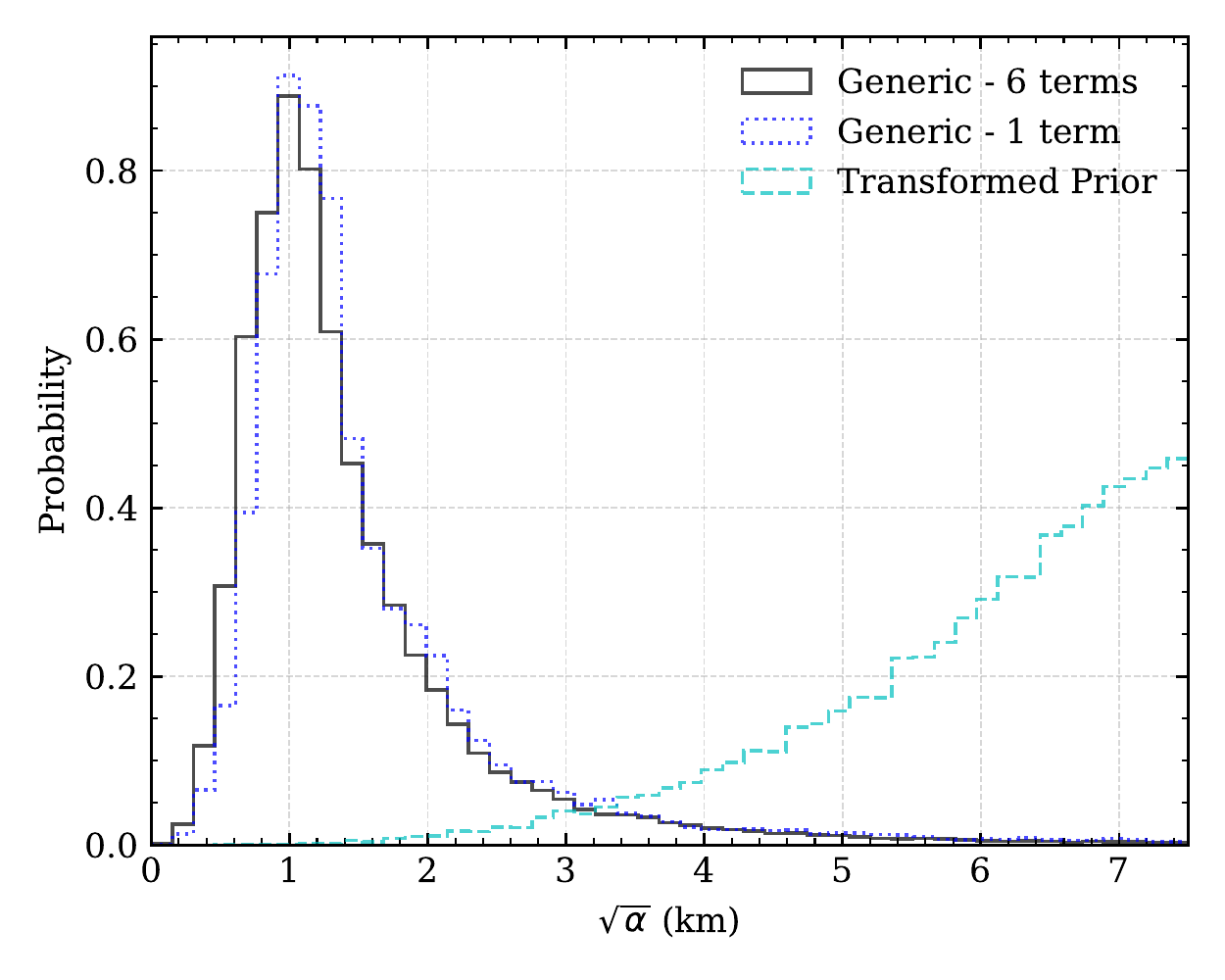}
\caption{
    Marginalized posterior distributions on $\sqrt{\alpha}$ for a GR injection extracted with the two generic models described in the text. One of the models uses a single deformation while the second model uses a series of six deformations. We here considered the ``light'' source, defined by Table~\ref{tab:inj_param}, and a 2g detector network. These two data sets are models 3 and 4, described in the text and shown in Fig.~\ref{fig:sgb_comparison}. For comparison, we also show the derived prior on $\sqrt{\alpha}$, calculated by taking a uniform prior on $\bar \gamma$ and the usual priors on source parameters, then mapping it to $\sqrt{\alpha}$ with the same prescription as was used for transforming the two generic model constraints. This figure illustrates the interesting behavior around $\sqrt{\alpha} \rightarrow 0$ (the GR limit).
    The fact that the derived prior on $\sqrt{\alpha}$, as transformed from a uniform prior on $\bar \gamma$, disallows $\sqrt{\alpha}=0$ explains the bias in the posteriors for models 3 and 4.
    The issue is related to the Jacobian of the transformation, discussed in the text, and is not of serious concern in analysis such as these.
}\label{fig:sgb_comparison_prior}
\end{figure}

Graphically, this is illustrated in Fig.~\ref{fig:sgb_comparison_prior} where the posteriors on $\sqrt{\alpha}$ coming from mapping constraints from the generic models are plotted along side samples from the full, derived prior on $\sqrt{\alpha}$.
The samples from the prior were drawn from a uniform distribution for $\bar \gamma$ and the usual priors for the source parameters, then mapped to $\sqrt{\alpha}$ in the same way as the analysis using models 3 and 4.
The original posterior and prior are completely consistent with GR, as shown in this figure, so the lack of support in the GR limit for $\sqrt{\alpha}$ must come from the transformation itself.

\section{Conclusions}\label{sec:conclusions}

We have here studied whether the inclusion of higher PN order terms in the modified gravity deformations to the GW phase of inspiraling binaries affects the constraints one can place on these theories. First, we focus on a particular implementation of multi-parameter phase deformation, in which an overall controlling factor contains all of the coupling-constant information of the modified theory. Then, we develop a novel, PN-based prior to impose on parametric deviations, which ensures that the terms added obey certain convergence criteria, as they must if they derive from a PN expansion in the inspiral phase (even in a modified theory). 

Our analysis shows that the constraints placed on modified theories with single-parameter ppE waveforms are robust and reliable. More specifically, the inclusion of higher PN order terms in the inspiral phase do not weaken the constraints we can place with single-parameter ppE models. 
In fact, the inclusion of these terms actually improves the constraints on coupling constants of modified theories, and typically the strengthening of the bound is mild. We verified that these conclusions with an array of Bayesian parameter estimation studies, in which we injected synthetic GR signals and extracted with a variety of single- and multi-parameter ppE models. We further considered a specific theory, ssGB gravity, to exemplify our findings with a concrete set of deformations to the phase. As expected from our generic analysis, the inclusion of higher PN order terms in the ssGB inspiral phase does not weaken bounds obtained with leading order waveforms. Instead, the higher PN order terms improve the bounds on the ssGB coupling parameter, but only by about a factor of 3. These results are consistent with a very recent analysis of ssGB theory with the same higher PN order model we use here on real aLIGO/Virgo data~\cite{Lyu:2022gdr}.   

One can compare our methodology to other techniques to measure multiple phase deformations simultaneously. One such methodology is through the use of a principal component analysis (PCA) on tests of GR~\cite{Saleem:2021nsb,Shoom:2021mdj,Pai:2012mv}. In this method, the phase deformations are combined through certain linear transformations, so that covariances are minimized. To do this, samples are first drawn from the posterior distribution with a waveform model that includes multiple ppE phase deformations, and then this posterior distribution is decomposed into its eigenvectors. Constraints are placed on linear combinations of deformations that produce the tightest constraints (the eigenvectors with the largest eigenvalues).
Effectively, this just equates to finding a parametrization that optimizes the constraint you can place on these deformations, but it comes with serious drawbacks related to tests of fundamental physics.
All information is lost about the physical meaning of these deformations, as this basis has no connection to calculations performed in modified gravity.
This can still be an effective consistency test of GR (in the same category of ``residual tests'' discussed in Sec.~\ref{sec:intro}), but it does not provide information about  fundamental physics beyond this specific type of test. 
Our approach, instead, attempts to stay close to theoretical physics, using physical insight to concentrate our attention on the ``relevant'' deformation, the controlling factor $\bar{\gamma}$ that encodes the coupling constants of the theory, at the expense of losing less relevant information about the higher order corrections (which are just functions of the system parameters).
With this, we still retain the mapping between theories of modified gravity and $\bar{\gamma}$, allowing constraints in the latter to lead to constraints on the coupling constants of modified theories. 

These results encourage the use of leading PN order deformations in the inspiral phase to constrain theories of gravity beyond GR.
The results will certainly improve with the inclusion of higher PN order deformations, but the bounds will not degrade through their inclusion. 
This does, however, come with the caveat that we have only considered theories of gravity that allow for well-behaved, series solutions to the field equations, both in the PN expansion and the expansion in the coupling.
There are theories that do not conform to these criteria, such as theories of gravity exhibiting spontaneous scalarization~\cite{Sampson:2013jpa,Khalil:2019wyy,Silva:2020omi,Doneva:2017bvd,Silva:2017uqg} or other abrupt, discontinuous transformations.
Future work could focus on those theories to attempt to develop a generic framework that can also be applied to them.

\acknowledgements

The authors would like to thank Walter del Pozzo and Emanuele Berti for useful comments on the first draft of this manuscript. We would also like to especially thank Jonathan Gair for detailed comments on the first draft, which led us to add Sec. VI to the final version. This work was partially supported by the Center for AstroPhysical Surveys (CAPS) at the National Center for Supercomputing Applications (NCSA), University of Illinois at Urbana-Champaign. S. P. acknowledges support from the Illinois Center for Advanced Studies of the Universe (ICASU)/CAPS Graduate Fellowship. N. Y. acknowledges support from the Simons Foundation through Grant No. 896696. This work made use of the Illinois Campus Cluster, a computing resource that is operated by the Illinois Campus Cluster Program (ICCP) in conjunction with the National Center for Supercomputing Applications (NCSA) and which is supported by funds from the University of Illinois at Urbana- Champaign.

\bibliography{refs}

\begin{thebibliography}{80}%
\makeatletter
\providecommand \@ifxundefined [1]{%
 \@ifx{#1\undefined}
}%
\providecommand \@ifnum [1]{%
 \ifnum #1\expandafter \@firstoftwo
 \else \expandafter \@secondoftwo
 \fi
}%
\providecommand \@ifx [1]{%
 \ifx #1\expandafter \@firstoftwo
 \else \expandafter \@secondoftwo
 \fi
}%
\providecommand \natexlab [1]{#1}%
\providecommand \enquote  [1]{``#1''}%
\providecommand \bibnamefont  [1]{#1}%
\providecommand \bibfnamefont [1]{#1}%
\providecommand \citenamefont [1]{#1}%
\providecommand \href@noop [0]{\@secondoftwo}%
\providecommand \href [0]{\begingroup \@sanitize@url \@href}%
\providecommand \@href[1]{\@@startlink{#1}\@@href}%
\providecommand \@@href[1]{\endgroup#1\@@endlink}%
\providecommand \@sanitize@url [0]{\catcode `\\12\catcode `\$12\catcode
  `\&12\catcode `\#12\catcode `\^12\catcode `\_12\catcode `\%12\relax}%
\providecommand \@@startlink[1]{}%
\providecommand \@@endlink[0]{}%
\providecommand \url  [0]{\begingroup\@sanitize@url \@url }%
\providecommand \@url [1]{\endgroup\@href {#1}{\urlprefix }}%
\providecommand \urlprefix  [0]{URL }%
\providecommand \Eprint [0]{\href }%
\providecommand \doibase [0]{https://doi.org/}%
\providecommand \selectlanguage [0]{\@gobble}%
\providecommand \bibinfo  [0]{\@secondoftwo}%
\providecommand \bibfield  [0]{\@secondoftwo}%
\providecommand \translation [1]{[#1]}%
\providecommand \BibitemOpen [0]{}%
\providecommand \bibitemStop [0]{}%
\providecommand \bibitemNoStop [0]{.\EOS\space}%
\providecommand \EOS [0]{\spacefactor3000\relax}%
\providecommand \BibitemShut  [1]{\csname bibitem#1\endcsname}%
\let\auto@bib@innerbib\@empty
\bibitem [{\citenamefont {Will}(2014)}]{Will:2014kxa}%
  \BibitemOpen
  \bibfield  {author} {\bibinfo {author} {\bibfnamefont {C.~M.}\ \bibnamefont
  {Will}},\ }\bibfield  {title} {\bibinfo {title} {{The Confrontation between
  General Relativity and Experiment}},\ }\href
  {https://doi.org/10.12942/lrr-2014-4} {\bibfield  {journal} {\bibinfo
  {journal} {Living Rev. Rel.}\ }\textbf {\bibinfo {volume} {17}},\ \bibinfo
  {pages} {4} (\bibinfo {year} {2014})},\ \Eprint
  {https://arxiv.org/abs/1403.7377} {arXiv:1403.7377 [gr-qc]} \BibitemShut
  {NoStop}%
\bibitem [{\citenamefont {Berti}\ \emph {et~al.}(2015)\citenamefont {Berti}
  \emph {et~al.}}]{Berti:2015itd}%
  \BibitemOpen
  \bibfield  {author} {\bibinfo {author} {\bibfnamefont {E.}~\bibnamefont
  {Berti}} \emph {et~al.},\ }\bibfield  {title} {\bibinfo {title} {{Testing
  General Relativity with Present and Future Astrophysical Observations}},\
  }\href {https://doi.org/10.1088/0264-9381/32/24/243001} {\bibfield  {journal}
  {\bibinfo  {journal} {Class. Quant. Grav.}\ }\textbf {\bibinfo {volume}
  {32}},\ \bibinfo {pages} {243001} (\bibinfo {year} {2015})},\ \Eprint
  {https://arxiv.org/abs/1501.07274} {arXiv:1501.07274 [gr-qc]} \BibitemShut
  {NoStop}%
\bibitem [{\citenamefont {Shomer}(2007)}]{Shomer:2007vq}%
  \BibitemOpen
  \bibfield  {author} {\bibinfo {author} {\bibfnamefont {A.}~\bibnamefont
  {Shomer}},\ }\bibfield  {title} {\bibinfo {title} {{A Pedagogical explanation
  for the non-renormalizability of gravity}},\ }\Eprint
  {https://arxiv.org/abs/0709.3555} {arXiv:0709.3555 [hep-th]}  (\bibinfo
  {year} {2007})\BibitemShut {NoStop}%
\bibitem [{\citenamefont {Senovilla}\ and\ \citenamefont
  {Garfinkle}(2015)}]{Senovilla:2014gza}%
  \BibitemOpen
  \bibfield  {author} {\bibinfo {author} {\bibfnamefont {J.~M.~M.}\
  \bibnamefont {Senovilla}}\ and\ \bibinfo {author} {\bibfnamefont
  {D.}~\bibnamefont {Garfinkle}},\ }\bibfield  {title} {\bibinfo {title} {{The
  1965 Penrose singularity theorem}},\ }\href
  {https://doi.org/10.1088/0264-9381/32/12/124008} {\bibfield  {journal}
  {\bibinfo  {journal} {Class. Quant. Grav.}\ }\textbf {\bibinfo {volume}
  {32}},\ \bibinfo {pages} {124008} (\bibinfo {year} {2015})},\ \Eprint
  {https://arxiv.org/abs/1410.5226} {arXiv:1410.5226 [gr-qc]} \BibitemShut
  {NoStop}%
\bibitem [{\citenamefont {Penrose}(1965)}]{Penrose:1964wq}%
  \BibitemOpen
  \bibfield  {author} {\bibinfo {author} {\bibfnamefont {R.}~\bibnamefont
  {Penrose}},\ }\bibfield  {title} {\bibinfo {title} {{Gravitational collapse
  and space-time singularities}},\ }\href
  {https://doi.org/10.1103/PhysRevLett.14.57} {\bibfield  {journal} {\bibinfo
  {journal} {Phys. Rev. Lett.}\ }\textbf {\bibinfo {volume} {14}},\ \bibinfo
  {pages} {57} (\bibinfo {year} {1965})}\BibitemShut {NoStop}%
\bibitem [{\citenamefont {Riess}\ \emph {et~al.}(1998)\citenamefont {Riess}
  \emph {et~al.}}]{SupernovaSearchTeam:1998fmf}%
  \BibitemOpen
  \bibfield  {author} {\bibinfo {author} {\bibfnamefont {A.~G.}\ \bibnamefont
  {Riess}} \emph {et~al.} (\bibinfo {collaboration} {Supernova Search Team}),\
  }\bibfield  {title} {\bibinfo {title} {{Observational evidence from
  supernovae for an accelerating universe and a cosmological constant}},\
  }\href {https://doi.org/10.1086/300499} {\bibfield  {journal} {\bibinfo
  {journal} {Astron. J.}\ }\textbf {\bibinfo {volume} {116}},\ \bibinfo {pages}
  {1009} (\bibinfo {year} {1998})},\ \Eprint
  {https://arxiv.org/abs/astro-ph/9805201} {arXiv:astro-ph/9805201}
  \BibitemShut {NoStop}%
\bibitem [{\citenamefont {Perlmutter}\ \emph {et~al.}(1999)\citenamefont
  {Perlmutter} \emph {et~al.}}]{SupernovaCosmologyProject:1998vns}%
  \BibitemOpen
  \bibfield  {author} {\bibinfo {author} {\bibfnamefont {S.}~\bibnamefont
  {Perlmutter}} \emph {et~al.} (\bibinfo {collaboration} {Supernova Cosmology
  Project}),\ }\bibfield  {title} {\bibinfo {title} {{Measurements of $\Omega$
  and $\Lambda$ from 42 high redshift supernovae}},\ }\href
  {https://doi.org/10.1086/307221} {\bibfield  {journal} {\bibinfo  {journal}
  {Astrophys. J.}\ }\textbf {\bibinfo {volume} {517}},\ \bibinfo {pages} {565}
  (\bibinfo {year} {1999})},\ \Eprint {https://arxiv.org/abs/astro-ph/9812133}
  {arXiv:astro-ph/9812133} \BibitemShut {NoStop}%
\bibitem [{\citenamefont {Sofue}\ and\ \citenamefont
  {Rubin}(2001)}]{Sofue:2000jx}%
  \BibitemOpen
  \bibfield  {author} {\bibinfo {author} {\bibfnamefont {Y.}~\bibnamefont
  {Sofue}}\ and\ \bibinfo {author} {\bibfnamefont {V.}~\bibnamefont {Rubin}},\
  }\bibfield  {title} {\bibinfo {title} {{Rotation curves of spiral
  galaxies}},\ }\href {https://doi.org/10.1146/annurev.astro.39.1.137}
  {\bibfield  {journal} {\bibinfo  {journal} {Ann. Rev. Astron. Astrophys.}\
  }\textbf {\bibinfo {volume} {39}},\ \bibinfo {pages} {137} (\bibinfo {year}
  {2001})},\ \Eprint {https://arxiv.org/abs/astro-ph/0010594}
  {arXiv:astro-ph/0010594} \BibitemShut {NoStop}%
\bibitem [{\citenamefont {Bertone}\ and\ \citenamefont
  {Hooper}(2018)}]{Bertone:2016nfn}%
  \BibitemOpen
  \bibfield  {author} {\bibinfo {author} {\bibfnamefont {G.}~\bibnamefont
  {Bertone}}\ and\ \bibinfo {author} {\bibfnamefont {D.}~\bibnamefont
  {Hooper}},\ }\bibfield  {title} {\bibinfo {title} {{History of dark
  matter}},\ }\href {https://doi.org/10.1103/RevModPhys.90.045002} {\bibfield
  {journal} {\bibinfo  {journal} {Rev. Mod. Phys.}\ }\textbf {\bibinfo {volume}
  {90}},\ \bibinfo {pages} {045002} (\bibinfo {year} {2018})},\ \Eprint
  {https://arxiv.org/abs/1605.04909} {arXiv:1605.04909 [astro-ph.CO]}
  \BibitemShut {NoStop}%
\bibitem [{\citenamefont {Canetti}\ \emph {et~al.}(2012)\citenamefont
  {Canetti}, \citenamefont {Drewes},\ and\ \citenamefont
  {Shaposhnikov}}]{Canetti:2012zc}%
  \BibitemOpen
  \bibfield  {author} {\bibinfo {author} {\bibfnamefont {L.}~\bibnamefont
  {Canetti}}, \bibinfo {author} {\bibfnamefont {M.}~\bibnamefont {Drewes}},\
  and\ \bibinfo {author} {\bibfnamefont {M.}~\bibnamefont {Shaposhnikov}},\
  }\bibfield  {title} {\bibinfo {title} {{Matter and Antimatter in the
  Universe}},\ }\href {https://doi.org/10.1088/1367-2630/14/9/095012}
  {\bibfield  {journal} {\bibinfo  {journal} {New J. Phys.}\ }\textbf {\bibinfo
  {volume} {14}},\ \bibinfo {pages} {095012} (\bibinfo {year} {2012})},\
  \Eprint {https://arxiv.org/abs/1204.4186} {arXiv:1204.4186 [hep-ph]}
  \BibitemShut {NoStop}%
\bibitem [{\citenamefont {Carson}\ and\ \citenamefont
  {Yagi}(2020)}]{Carson:2020rea}%
  \BibitemOpen
  \bibfield  {author} {\bibinfo {author} {\bibfnamefont {Z.}~\bibnamefont
  {Carson}}\ and\ \bibinfo {author} {\bibfnamefont {K.}~\bibnamefont {Yagi}},\
  }\href {https://doi.org/10.1007/978-981-15-4702-7_41-1} {\emph {\bibinfo
  {title} {{Testing General Relativity with Gravitational Waves}}}}\ (\bibinfo
  {year} {2020})\ \Eprint {https://arxiv.org/abs/2011.02938} {arXiv:2011.02938
  [gr-qc]} \BibitemShut {NoStop}%
\bibitem [{\citenamefont {Yunes}\ \emph {et~al.}(2016)\citenamefont {Yunes},
  \citenamefont {Yagi},\ and\ \citenamefont {Pretorius}}]{Yunes:2016jcc}%
  \BibitemOpen
  \bibfield  {author} {\bibinfo {author} {\bibfnamefont {N.}~\bibnamefont
  {Yunes}}, \bibinfo {author} {\bibfnamefont {K.}~\bibnamefont {Yagi}},\ and\
  \bibinfo {author} {\bibfnamefont {F.}~\bibnamefont {Pretorius}},\ }\bibfield
  {title} {\bibinfo {title} {{Theoretical Physics Implications of the Binary
  Black-Hole Mergers GW150914 and GW151226}},\ }\href
  {https://doi.org/10.1103/PhysRevD.94.084002} {\bibfield  {journal} {\bibinfo
  {journal} {Phys. Rev.}\ }\textbf {\bibinfo {volume} {D94}},\ \bibinfo {pages}
  {084002} (\bibinfo {year} {2016})},\ \Eprint
  {https://arxiv.org/abs/1603.08955} {arXiv:1603.08955 [gr-qc]} \BibitemShut
  {NoStop}%
\bibitem [{\citenamefont {Abbott}\ \emph {et~al.}(2016)\citenamefont {Abbott}
  \emph {et~al.}}]{LIGOScientific:2016lio}%
  \BibitemOpen
  \bibfield  {author} {\bibinfo {author} {\bibfnamefont {B.~P.}\ \bibnamefont
  {Abbott}} \emph {et~al.} (\bibinfo {collaboration} {LIGO Scientific,
  Virgo}),\ }\bibfield  {title} {\bibinfo {title} {{Tests of general relativity
  with GW150914}},\ }\href {https://doi.org/10.1103/PhysRevLett.116.221101}
  {\bibfield  {journal} {\bibinfo  {journal} {Phys. Rev. Lett.}\ }\textbf
  {\bibinfo {volume} {116}},\ \bibinfo {pages} {221101} (\bibinfo {year}
  {2016})},\ \bibinfo {note} {[Erratum: Phys.Rev.Lett. 121, 129902 (2018)]},\
  \Eprint {https://arxiv.org/abs/1602.03841} {arXiv:1602.03841 [gr-qc]}
  \BibitemShut {NoStop}%
\bibitem [{\citenamefont {Cornish}\ and\ \citenamefont
  {Littenberg}(2015)}]{Cornish:2014kda}%
  \BibitemOpen
  \bibfield  {author} {\bibinfo {author} {\bibfnamefont {N.~J.}\ \bibnamefont
  {Cornish}}\ and\ \bibinfo {author} {\bibfnamefont {T.~B.}\ \bibnamefont
  {Littenberg}},\ }\bibfield  {title} {\bibinfo {title} {{BayesWave: Bayesian
  Inference for Gravitational Wave Bursts and Instrument Glitches}},\ }\href
  {https://doi.org/10.1088/0264-9381/32/13/135012} {\bibfield  {journal}
  {\bibinfo  {journal} {Class. Quant. Grav.}\ }\textbf {\bibinfo {volume}
  {32}},\ \bibinfo {pages} {135012} (\bibinfo {year} {2015})},\ \Eprint
  {https://arxiv.org/abs/1410.3835} {arXiv:1410.3835 [gr-qc]} \BibitemShut
  {NoStop}%
\bibitem [{\citenamefont {Cornish}\ \emph {et~al.}(2021)\citenamefont
  {Cornish}, \citenamefont {Littenberg}, \citenamefont {B\'ecsy}, \citenamefont
  {Chatziioannou}, \citenamefont {Clark}, \citenamefont {Ghonge},\ and\
  \citenamefont {Millhouse}}]{Cornish:2020dwh}%
  \BibitemOpen
  \bibfield  {author} {\bibinfo {author} {\bibfnamefont {N.~J.}\ \bibnamefont
  {Cornish}}, \bibinfo {author} {\bibfnamefont {T.~B.}\ \bibnamefont
  {Littenberg}}, \bibinfo {author} {\bibfnamefont {B.}~\bibnamefont {B\'ecsy}},
  \bibinfo {author} {\bibfnamefont {K.}~\bibnamefont {Chatziioannou}}, \bibinfo
  {author} {\bibfnamefont {J.~A.}\ \bibnamefont {Clark}}, \bibinfo {author}
  {\bibfnamefont {S.}~\bibnamefont {Ghonge}},\ and\ \bibinfo {author}
  {\bibfnamefont {M.}~\bibnamefont {Millhouse}},\ }\bibfield  {title} {\bibinfo
  {title} {{BayesWave analysis pipeline in the era of gravitational wave
  observations}},\ }\href {https://doi.org/10.1103/PhysRevD.103.044006}
  {\bibfield  {journal} {\bibinfo  {journal} {Phys. Rev. D}\ }\textbf {\bibinfo
  {volume} {103}},\ \bibinfo {pages} {044006} (\bibinfo {year} {2021})},\
  \Eprint {https://arxiv.org/abs/2011.09494} {arXiv:2011.09494 [gr-qc]}
  \BibitemShut {NoStop}%
\bibitem [{\citenamefont {Chatziioannou}\ \emph {et~al.}(2021)\citenamefont
  {Chatziioannou}, \citenamefont {Isi}, \citenamefont {Haster},\ and\
  \citenamefont {Littenberg}}]{Chatziioannou:2021mij}%
  \BibitemOpen
  \bibfield  {author} {\bibinfo {author} {\bibfnamefont {K.}~\bibnamefont
  {Chatziioannou}}, \bibinfo {author} {\bibfnamefont {M.}~\bibnamefont {Isi}},
  \bibinfo {author} {\bibfnamefont {C.-J.}\ \bibnamefont {Haster}},\ and\
  \bibinfo {author} {\bibfnamefont {T.~B.}\ \bibnamefont {Littenberg}},\
  }\bibfield  {title} {\bibinfo {title} {{Morphology-independent test of the
  mixed polarization content of transient gravitational wave signals}},\ }\href
  {https://doi.org/10.1103/PhysRevD.104.044005} {\bibfield  {journal} {\bibinfo
   {journal} {Phys. Rev. D}\ }\textbf {\bibinfo {volume} {104}},\ \bibinfo
  {pages} {044005} (\bibinfo {year} {2021})},\ \Eprint
  {https://arxiv.org/abs/2105.01521} {arXiv:2105.01521 [gr-qc]} \BibitemShut
  {NoStop}%
\bibitem [{\citenamefont {Cornish}\ \emph {et~al.}(2011)\citenamefont
  {Cornish}, \citenamefont {Sampson}, \citenamefont {Yunes},\ and\
  \citenamefont {Pretorius}}]{Cornish:2011ys}%
  \BibitemOpen
  \bibfield  {author} {\bibinfo {author} {\bibfnamefont {N.}~\bibnamefont
  {Cornish}}, \bibinfo {author} {\bibfnamefont {L.}~\bibnamefont {Sampson}},
  \bibinfo {author} {\bibfnamefont {N.}~\bibnamefont {Yunes}},\ and\ \bibinfo
  {author} {\bibfnamefont {F.}~\bibnamefont {Pretorius}},\ }\bibfield  {title}
  {\bibinfo {title} {{Gravitational Wave Tests of General Relativity with the
  Parameterized Post-Einsteinian Framework}},\ }\href
  {https://doi.org/10.1103/PhysRevD.84.062003} {\bibfield  {journal} {\bibinfo
  {journal} {Phys. Rev. D}\ }\textbf {\bibinfo {volume} {84}},\ \bibinfo
  {pages} {062003} (\bibinfo {year} {2011})},\ \Eprint
  {https://arxiv.org/abs/1105.2088} {arXiv:1105.2088 [gr-qc]} \BibitemShut
  {NoStop}%
\bibitem [{\citenamefont {Vallisneri}(2012)}]{Vallisneri:2012qq}%
  \BibitemOpen
  \bibfield  {author} {\bibinfo {author} {\bibfnamefont {M.}~\bibnamefont
  {Vallisneri}},\ }\bibfield  {title} {\bibinfo {title} {{Testing general
  relativity with gravitational waves: a reality check}},\ }\href
  {https://doi.org/10.1103/PhysRevD.86.082001} {\bibfield  {journal} {\bibinfo
  {journal} {Phys. Rev. D}\ }\textbf {\bibinfo {volume} {86}},\ \bibinfo
  {pages} {082001} (\bibinfo {year} {2012})},\ \Eprint
  {https://arxiv.org/abs/1207.4759} {arXiv:1207.4759 [gr-qc]} \BibitemShut
  {NoStop}%
\bibitem [{\citenamefont {Yunes}\ and\ \citenamefont
  {Pretorius}(2009)}]{Yunes:2009ke}%
  \BibitemOpen
  \bibfield  {author} {\bibinfo {author} {\bibfnamefont {N.}~\bibnamefont
  {Yunes}}\ and\ \bibinfo {author} {\bibfnamefont {F.}~\bibnamefont
  {Pretorius}},\ }\bibfield  {title} {\bibinfo {title} {{Fundamental
  Theoretical Bias in Gravitational Wave Astrophysics and the Parameterized
  Post-Einsteinian Framework}},\ }\href
  {https://doi.org/10.1103/PhysRevD.80.122003} {\bibfield  {journal} {\bibinfo
  {journal} {Phys. Rev. D}\ }\textbf {\bibinfo {volume} {80}},\ \bibinfo
  {pages} {122003} (\bibinfo {year} {2009})},\ \Eprint
  {https://arxiv.org/abs/0909.3328} {arXiv:0909.3328 [gr-qc]} \BibitemShut
  {NoStop}%
\bibitem [{\citenamefont {Sampson}\ \emph {et~al.}(2013)\citenamefont
  {Sampson}, \citenamefont {Cornish},\ and\ \citenamefont
  {Yunes}}]{Sampson:2013lpa}%
  \BibitemOpen
  \bibfield  {author} {\bibinfo {author} {\bibfnamefont {L.}~\bibnamefont
  {Sampson}}, \bibinfo {author} {\bibfnamefont {N.}~\bibnamefont {Cornish}},\
  and\ \bibinfo {author} {\bibfnamefont {N.}~\bibnamefont {Yunes}},\ }\bibfield
   {title} {\bibinfo {title} {{Gravitational Wave Tests of Strong Field General
  Relativity with Binary Inspirals: Realistic Injections and Optimal Model
  Selection}},\ }\href {https://doi.org/10.1103/PhysRevD.87.102001} {\bibfield
  {journal} {\bibinfo  {journal} {Phys. Rev. D}\ }\textbf {\bibinfo {volume}
  {87}},\ \bibinfo {pages} {102001} (\bibinfo {year} {2013})},\ \Eprint
  {https://arxiv.org/abs/1303.1185} {arXiv:1303.1185 [gr-qc]} \BibitemShut
  {NoStop}%
\bibitem [{\citenamefont {Chatziioannou}\ \emph {et~al.}(2012)\citenamefont
  {Chatziioannou}, \citenamefont {Yunes},\ and\ \citenamefont
  {Cornish}}]{Chatziioannou:2012rf}%
  \BibitemOpen
  \bibfield  {author} {\bibinfo {author} {\bibfnamefont {K.}~\bibnamefont
  {Chatziioannou}}, \bibinfo {author} {\bibfnamefont {N.}~\bibnamefont
  {Yunes}},\ and\ \bibinfo {author} {\bibfnamefont {N.}~\bibnamefont
  {Cornish}},\ }\bibfield  {title} {\bibinfo {title} {{Model-Independent Test
  of General Relativity: An Extended post-Einsteinian Framework with Complete
  Polarization Content}},\ }\href {https://doi.org/10.1103/PhysRevD.86.022004}
  {\bibfield  {journal} {\bibinfo  {journal} {Phys. Rev. D}\ }\textbf {\bibinfo
  {volume} {86}},\ \bibinfo {pages} {022004} (\bibinfo {year} {2012})},\
  \bibinfo {note} {[Erratum: Phys.Rev.D 95, 129901 (2017)]},\ \Eprint
  {https://arxiv.org/abs/1204.2585} {arXiv:1204.2585 [gr-qc]} \BibitemShut
  {NoStop}%
\bibitem [{\citenamefont {Mishra}\ \emph {et~al.}(2010)\citenamefont {Mishra},
  \citenamefont {Arun}, \citenamefont {Iyer},\ and\ \citenamefont
  {Sathyaprakash}}]{Mishra:2010tp}%
  \BibitemOpen
  \bibfield  {author} {\bibinfo {author} {\bibfnamefont {C.~K.}\ \bibnamefont
  {Mishra}}, \bibinfo {author} {\bibfnamefont {K.~G.}\ \bibnamefont {Arun}},
  \bibinfo {author} {\bibfnamefont {B.~R.}\ \bibnamefont {Iyer}},\ and\
  \bibinfo {author} {\bibfnamefont {B.~S.}\ \bibnamefont {Sathyaprakash}},\
  }\bibfield  {title} {\bibinfo {title} {{Parametrized tests of post-Newtonian
  theory using Advanced LIGO and Einstein Telescope}},\ }\href
  {https://doi.org/10.1103/PhysRevD.82.064010} {\bibfield  {journal} {\bibinfo
  {journal} {Phys. Rev. D}\ }\textbf {\bibinfo {volume} {82}},\ \bibinfo
  {pages} {064010} (\bibinfo {year} {2010})},\ \Eprint
  {https://arxiv.org/abs/1005.0304} {arXiv:1005.0304 [gr-qc]} \BibitemShut
  {NoStop}%
\bibitem [{\citenamefont {Arun}\ \emph
  {et~al.}(2006{\natexlab{a}})\citenamefont {Arun}, \citenamefont {Iyer},
  \citenamefont {Qusailah},\ and\ \citenamefont {Sathyaprakash}}]{Arun:2006yw}%
  \BibitemOpen
  \bibfield  {author} {\bibinfo {author} {\bibfnamefont {K.~G.}\ \bibnamefont
  {Arun}}, \bibinfo {author} {\bibfnamefont {B.~R.}\ \bibnamefont {Iyer}},
  \bibinfo {author} {\bibfnamefont {M.~S.~S.}\ \bibnamefont {Qusailah}},\ and\
  \bibinfo {author} {\bibfnamefont {B.~S.}\ \bibnamefont {Sathyaprakash}},\
  }\bibfield  {title} {\bibinfo {title} {{Testing post-Newtonian theory with
  gravitational wave observations}},\ }\href
  {https://doi.org/10.1088/0264-9381/23/9/L01} {\bibfield  {journal} {\bibinfo
  {journal} {Class. Quant. Grav.}\ }\textbf {\bibinfo {volume} {23}},\ \bibinfo
  {pages} {L37} (\bibinfo {year} {2006}{\natexlab{a}})},\ \Eprint
  {https://arxiv.org/abs/gr-qc/0604018} {arXiv:gr-qc/0604018} \BibitemShut
  {NoStop}%
\bibitem [{\citenamefont {Abbott}\ \emph
  {et~al.}(2019{\natexlab{a}})\citenamefont {Abbott} \emph
  {et~al.}}]{LIGOScientific:2019fpa}%
  \BibitemOpen
  \bibfield  {author} {\bibinfo {author} {\bibfnamefont {B.~P.}\ \bibnamefont
  {Abbott}} \emph {et~al.} (\bibinfo {collaboration} {LIGO Scientific,
  Virgo}),\ }\bibfield  {title} {\bibinfo {title} {{Tests of General Relativity
  with the Binary Black Hole Signals from the LIGO-Virgo Catalog GWTC-1}},\
  }\href {https://doi.org/10.1103/PhysRevD.100.104036} {\bibfield  {journal}
  {\bibinfo  {journal} {Phys. Rev. D}\ }\textbf {\bibinfo {volume} {100}},\
  \bibinfo {pages} {104036} (\bibinfo {year} {2019}{\natexlab{a}})},\ \Eprint
  {https://arxiv.org/abs/1903.04467} {arXiv:1903.04467 [gr-qc]} \BibitemShut
  {NoStop}%
\bibitem [{\citenamefont {Abbott}\ \emph
  {et~al.}(2021{\natexlab{a}})\citenamefont {Abbott} \emph
  {et~al.}}]{LIGOScientific:2020tif}%
  \BibitemOpen
  \bibfield  {author} {\bibinfo {author} {\bibfnamefont {R.}~\bibnamefont
  {Abbott}} \emph {et~al.} (\bibinfo {collaboration} {LIGO Scientific,
  Virgo}),\ }\bibfield  {title} {\bibinfo {title} {{Tests of general relativity
  with binary black holes from the second LIGO-Virgo gravitational-wave
  transient catalog}},\ }\href {https://doi.org/10.1103/PhysRevD.103.122002}
  {\bibfield  {journal} {\bibinfo  {journal} {Phys. Rev. D}\ }\textbf {\bibinfo
  {volume} {103}},\ \bibinfo {pages} {122002} (\bibinfo {year}
  {2021}{\natexlab{a}})},\ \Eprint {https://arxiv.org/abs/2010.14529}
  {arXiv:2010.14529 [gr-qc]} \BibitemShut {NoStop}%
\bibitem [{\citenamefont {Abbott}\ \emph {et~al.}(2020)\citenamefont {Abbott}
  \emph {et~al.}}]{Abbott:2020jks}%
  \BibitemOpen
  \bibfield  {author} {\bibinfo {author} {\bibfnamefont {R.}~\bibnamefont
  {Abbott}} \emph {et~al.} (\bibinfo {collaboration} {LIGO Scientific,
  Virgo}),\ }\bibfield  {title} {\bibinfo {title} {{Tests of General Relativity
  with Binary Black Holes from the second LIGO-Virgo Gravitational-Wave
  Transient Catalog}},\ }\Eprint {https://arxiv.org/abs/2010.14529}
  {arXiv:2010.14529 [gr-qc]}  (\bibinfo {year} {2020})\BibitemShut {NoStop}%
\bibitem [{\citenamefont {Abbott}\ \emph
  {et~al.}(2021{\natexlab{b}})\citenamefont {Abbott} \emph
  {et~al.}}]{LIGOScientific:2021sio}%
  \BibitemOpen
  \bibfield  {author} {\bibinfo {author} {\bibfnamefont {R.}~\bibnamefont
  {Abbott}} \emph {et~al.} (\bibinfo {collaboration} {LIGO Scientific, VIRGO,
  KAGRA}),\ }\bibfield  {title} {\bibinfo {title} {{Tests of General Relativity
  with GWTC-3}},\ }\Eprint {https://arxiv.org/abs/2112.06861} {arXiv:2112.06861
  [gr-qc]}  (\bibinfo {year} {2021}{\natexlab{b}})\BibitemShut {NoStop}%
\bibitem [{\citenamefont {Li}\ \emph {et~al.}(2012)\citenamefont {Li},
  \citenamefont {Del~Pozzo}, \citenamefont {Vitale}, \citenamefont {Van
  Den~Broeck}, \citenamefont {Agathos}, \citenamefont {Veitch}, \citenamefont
  {Grover}, \citenamefont {Sidery}, \citenamefont {Sturani},\ and\
  \citenamefont {Vecchio}}]{Li:2011cg}%
  \BibitemOpen
  \bibfield  {author} {\bibinfo {author} {\bibfnamefont {T.~G.~F.}\
  \bibnamefont {Li}}, \bibinfo {author} {\bibfnamefont {W.}~\bibnamefont
  {Del~Pozzo}}, \bibinfo {author} {\bibfnamefont {S.}~\bibnamefont {Vitale}},
  \bibinfo {author} {\bibfnamefont {C.}~\bibnamefont {Van Den~Broeck}},
  \bibinfo {author} {\bibfnamefont {M.}~\bibnamefont {Agathos}}, \bibinfo
  {author} {\bibfnamefont {J.}~\bibnamefont {Veitch}}, \bibinfo {author}
  {\bibfnamefont {K.}~\bibnamefont {Grover}}, \bibinfo {author} {\bibfnamefont
  {T.}~\bibnamefont {Sidery}}, \bibinfo {author} {\bibfnamefont
  {R.}~\bibnamefont {Sturani}},\ and\ \bibinfo {author} {\bibfnamefont
  {A.}~\bibnamefont {Vecchio}},\ }\bibfield  {title} {\bibinfo {title}
  {{Towards a generic test of the strong field dynamics of general relativity
  using compact binary coalescence}},\ }\href
  {https://doi.org/10.1103/PhysRevD.85.082003} {\bibfield  {journal} {\bibinfo
  {journal} {Phys. Rev. D}\ }\textbf {\bibinfo {volume} {85}},\ \bibinfo
  {pages} {082003} (\bibinfo {year} {2012})},\ \Eprint
  {https://arxiv.org/abs/1110.0530} {arXiv:1110.0530 [gr-qc]} \BibitemShut
  {NoStop}%
\bibitem [{\citenamefont {Agathos}\ \emph {et~al.}(2014)\citenamefont
  {Agathos}, \citenamefont {Del~Pozzo}, \citenamefont {Li}, \citenamefont {Van
  Den~Broeck}, \citenamefont {Veitch},\ and\ \citenamefont
  {Vitale}}]{Agathos:2013upa}%
  \BibitemOpen
  \bibfield  {author} {\bibinfo {author} {\bibfnamefont {M.}~\bibnamefont
  {Agathos}}, \bibinfo {author} {\bibfnamefont {W.}~\bibnamefont {Del~Pozzo}},
  \bibinfo {author} {\bibfnamefont {T.~G.~F.}\ \bibnamefont {Li}}, \bibinfo
  {author} {\bibfnamefont {C.}~\bibnamefont {Van Den~Broeck}}, \bibinfo
  {author} {\bibfnamefont {J.}~\bibnamefont {Veitch}},\ and\ \bibinfo {author}
  {\bibfnamefont {S.}~\bibnamefont {Vitale}},\ }\bibfield  {title} {\bibinfo
  {title} {{TIGER: A data analysis pipeline for testing the strong-field
  dynamics of general relativity with gravitational wave signals from
  coalescing compact binaries}},\ }\href
  {https://doi.org/10.1103/PhysRevD.89.082001} {\bibfield  {journal} {\bibinfo
  {journal} {Phys. Rev. D}\ }\textbf {\bibinfo {volume} {89}},\ \bibinfo
  {pages} {082001} (\bibinfo {year} {2014})},\ \Eprint
  {https://arxiv.org/abs/1311.0420} {arXiv:1311.0420 [gr-qc]} \BibitemShut
  {NoStop}%
\bibitem [{\citenamefont {Meidam}\ \emph {et~al.}(2018)\citenamefont {Meidam}
  \emph {et~al.}}]{Meidam:2017dgf}%
  \BibitemOpen
  \bibfield  {author} {\bibinfo {author} {\bibfnamefont {J.}~\bibnamefont
  {Meidam}} \emph {et~al.},\ }\bibfield  {title} {\bibinfo {title}
  {{Parametrized tests of the strong-field dynamics of general relativity using
  gravitational wave signals from coalescing binary black holes: Fast
  likelihood calculations and sensitivity of the method}},\ }\href
  {https://doi.org/10.1103/PhysRevD.97.044033} {\bibfield  {journal} {\bibinfo
  {journal} {Phys. Rev. D}\ }\textbf {\bibinfo {volume} {97}},\ \bibinfo
  {pages} {044033} (\bibinfo {year} {2018})},\ \Eprint
  {https://arxiv.org/abs/1712.08772} {arXiv:1712.08772 [gr-qc]} \BibitemShut
  {NoStop}%
\bibitem [{\citenamefont {Abbott}\ \emph
  {et~al.}(2019{\natexlab{b}})\citenamefont {Abbott} \emph
  {et~al.}}]{LIGOScientific:2018dkp}%
  \BibitemOpen
  \bibfield  {author} {\bibinfo {author} {\bibfnamefont {B.~P.}\ \bibnamefont
  {Abbott}} \emph {et~al.} (\bibinfo {collaboration} {LIGO Scientific,
  Virgo}),\ }\bibfield  {title} {\bibinfo {title} {{Tests of General Relativity
  with GW170817}},\ }\href {https://doi.org/10.1103/PhysRevLett.123.011102}
  {\bibfield  {journal} {\bibinfo  {journal} {Phys. Rev. Lett.}\ }\textbf
  {\bibinfo {volume} {123}},\ \bibinfo {pages} {011102} (\bibinfo {year}
  {2019}{\natexlab{b}})},\ \Eprint {https://arxiv.org/abs/1811.00364}
  {arXiv:1811.00364 [gr-qc]} \BibitemShut {NoStop}%
\bibitem [{\citenamefont {Tahura}\ and\ \citenamefont
  {Yagi}(2018)}]{Tahura:2018zuq}%
  \BibitemOpen
  \bibfield  {author} {\bibinfo {author} {\bibfnamefont {S.}~\bibnamefont
  {Tahura}}\ and\ \bibinfo {author} {\bibfnamefont {K.}~\bibnamefont {Yagi}},\
  }\bibfield  {title} {\bibinfo {title} {{Parameterized Post-Einsteinian
  Gravitational Waveforms in Various Modified Theories of Gravity}},\ }\href
  {https://doi.org/10.1103/PhysRevD.98.084042} {\bibfield  {journal} {\bibinfo
  {journal} {Phys. Rev. D}\ }\textbf {\bibinfo {volume} {98}},\ \bibinfo
  {pages} {084042} (\bibinfo {year} {2018})},\ \bibinfo {note} {[Erratum:
  Phys.Rev.D 101, 109902 (2020)]},\ \Eprint {https://arxiv.org/abs/1809.00259}
  {arXiv:1809.00259 [gr-qc]} \BibitemShut {NoStop}%
\bibitem [{\citenamefont {Perkins}\ \emph
  {et~al.}(2021{\natexlab{a}})\citenamefont {Perkins}, \citenamefont {Yunes},\
  and\ \citenamefont {Berti}}]{Perkins:2020tra}%
  \BibitemOpen
  \bibfield  {author} {\bibinfo {author} {\bibfnamefont {S.~E.}\ \bibnamefont
  {Perkins}}, \bibinfo {author} {\bibfnamefont {N.}~\bibnamefont {Yunes}},\
  and\ \bibinfo {author} {\bibfnamefont {E.}~\bibnamefont {Berti}},\ }\bibfield
   {title} {\bibinfo {title} {{Probing Fundamental Physics with Gravitational
  Waves: The Next Generation}},\ }\href
  {https://doi.org/10.1103/PhysRevD.103.044024} {\bibfield  {journal} {\bibinfo
   {journal} {Phys. Rev. D}\ }\textbf {\bibinfo {volume} {103}},\ \bibinfo
  {pages} {044024} (\bibinfo {year} {2021}{\natexlab{a}})},\ \Eprint
  {https://arxiv.org/abs/2010.09010} {arXiv:2010.09010 [gr-qc]} \BibitemShut
  {NoStop}%
\bibitem [{\citenamefont {Arun}\ \emph
  {et~al.}(2006{\natexlab{b}})\citenamefont {Arun}, \citenamefont {Iyer},
  \citenamefont {Qusailah},\ and\ \citenamefont {Sathyaprakash}}]{Arun:2006hn}%
  \BibitemOpen
  \bibfield  {author} {\bibinfo {author} {\bibfnamefont {K.~G.}\ \bibnamefont
  {Arun}}, \bibinfo {author} {\bibfnamefont {B.~R.}\ \bibnamefont {Iyer}},
  \bibinfo {author} {\bibfnamefont {M.~S.~S.}\ \bibnamefont {Qusailah}},\ and\
  \bibinfo {author} {\bibfnamefont {B.~S.}\ \bibnamefont {Sathyaprakash}},\
  }\bibfield  {title} {\bibinfo {title} {{Probing the non-linear structure of
  general relativity with black hole binaries}},\ }\href
  {https://doi.org/10.1103/PhysRevD.74.024006} {\bibfield  {journal} {\bibinfo
  {journal} {Phys. Rev. D}\ }\textbf {\bibinfo {volume} {74}},\ \bibinfo
  {pages} {024006} (\bibinfo {year} {2006}{\natexlab{b}})},\ \Eprint
  {https://arxiv.org/abs/gr-qc/0604067} {arXiv:gr-qc/0604067} \BibitemShut
  {NoStop}%
\bibitem [{\citenamefont {Perkins}\ \emph
  {et~al.}(2021{\natexlab{b}})\citenamefont {Perkins}, \citenamefont {Nair},
  \citenamefont {Silva},\ and\ \citenamefont {Yunes}}]{Perkins:2021mhb}%
  \BibitemOpen
  \bibfield  {author} {\bibinfo {author} {\bibfnamefont {S.~E.}\ \bibnamefont
  {Perkins}}, \bibinfo {author} {\bibfnamefont {R.}~\bibnamefont {Nair}},
  \bibinfo {author} {\bibfnamefont {H.~O.}\ \bibnamefont {Silva}},\ and\
  \bibinfo {author} {\bibfnamefont {N.}~\bibnamefont {Yunes}},\ }\bibfield
  {title} {\bibinfo {title} {{Improved gravitational-wave constraints on
  higher-order curvature theories of gravity}},\ }\Eprint
  {https://arxiv.org/abs/2104.11189} {arXiv:2104.11189 [gr-qc]}  (\bibinfo
  {year} {2021}{\natexlab{b}})\BibitemShut {NoStop}%
\bibitem [{\citenamefont {Blanchet}(2014)}]{Blanchet:2013haa}%
  \BibitemOpen
  \bibfield  {author} {\bibinfo {author} {\bibfnamefont {L.}~\bibnamefont
  {Blanchet}},\ }\bibfield  {title} {\bibinfo {title} {{Gravitational Radiation
  from Post-Newtonian Sources and Inspiralling Compact Binaries}},\ }\href
  {https://doi.org/10.12942/lrr-2014-2} {\bibfield  {journal} {\bibinfo
  {journal} {Living Rev. Rel.}\ }\textbf {\bibinfo {volume} {17}},\ \bibinfo
  {pages} {2} (\bibinfo {year} {2014})},\ \Eprint
  {https://arxiv.org/abs/1310.1528} {arXiv:1310.1528 [gr-qc]} \BibitemShut
  {NoStop}%
\bibitem [{\citenamefont {Damour}\ and\ \citenamefont
  {Nagar}(2016)}]{Damour2016}%
  \BibitemOpen
  \bibfield  {author} {\bibinfo {author} {\bibfnamefont {T.}~\bibnamefont
  {Damour}}\ and\ \bibinfo {author} {\bibfnamefont {A.}~\bibnamefont {Nagar}},\
  }\bibinfo {title} {The effective-one-body approach to the general
  relativistic two body problem},\ in\ \href
  {https://doi.org/10.1007/978-3-319-19416-5_7} {\emph {\bibinfo {booktitle}
  {Astrophysical Black Holes}}},\ \bibinfo {editor} {edited by\ \bibinfo
  {editor} {\bibfnamefont {F.}~\bibnamefont {Haardt}}, \bibinfo {editor}
  {\bibfnamefont {V.}~\bibnamefont {Gorini}}, \bibinfo {editor} {\bibfnamefont
  {U.}~\bibnamefont {Moschella}}, \bibinfo {editor} {\bibfnamefont
  {A.}~\bibnamefont {Treves}},\ and\ \bibinfo {editor} {\bibfnamefont
  {M.}~\bibnamefont {Colpi}}}\ (\bibinfo  {publisher} {Springer International
  Publishing},\ \bibinfo {address} {Cham},\ \bibinfo {year} {2016})\ pp.\
  \bibinfo {pages} {273--312}\BibitemShut {NoStop}%
\bibitem [{\citenamefont {Shiralilou}\ \emph {et~al.}(2021)\citenamefont
  {Shiralilou}, \citenamefont {Hinderer}, \citenamefont {Nissanke},
  \citenamefont {Ortiz},\ and\ \citenamefont {Witek}}]{Shiralilou:2021mfl}%
  \BibitemOpen
  \bibfield  {author} {\bibinfo {author} {\bibfnamefont {B.}~\bibnamefont
  {Shiralilou}}, \bibinfo {author} {\bibfnamefont {T.}~\bibnamefont
  {Hinderer}}, \bibinfo {author} {\bibfnamefont {S.}~\bibnamefont {Nissanke}},
  \bibinfo {author} {\bibfnamefont {N.}~\bibnamefont {Ortiz}},\ and\ \bibinfo
  {author} {\bibfnamefont {H.}~\bibnamefont {Witek}},\ }\bibfield  {title}
  {\bibinfo {title} {{Post-Newtonian Gravitational and Scalar Waves in
  Scalar-Gauss-Bonnet Gravity}},\ }\Eprint {https://arxiv.org/abs/2105.13972}
  {arXiv:2105.13972 [gr-qc]}  (\bibinfo {year} {2021})\BibitemShut {NoStop}%
\bibitem [{\citenamefont {de~Rham}(2014)}]{deRham:2014zqa}%
  \BibitemOpen
  \bibfield  {author} {\bibinfo {author} {\bibfnamefont {C.}~\bibnamefont
  {de~Rham}},\ }\bibfield  {title} {\bibinfo {title} {{Massive Gravity}},\
  }\href {https://doi.org/10.12942/lrr-2014-7} {\bibfield  {journal} {\bibinfo
  {journal} {Living Rev. Rel.}\ }\textbf {\bibinfo {volume} {17}},\ \bibinfo
  {pages} {7} (\bibinfo {year} {2014})},\ \Eprint
  {https://arxiv.org/abs/1401.4173} {arXiv:1401.4173 [hep-th]} \BibitemShut
  {NoStop}%
\bibitem [{\citenamefont {Damour}\ and\ \citenamefont
  {Esposito-Farese}(1993)}]{Damour:1993hw}%
  \BibitemOpen
  \bibfield  {author} {\bibinfo {author} {\bibfnamefont {T.}~\bibnamefont
  {Damour}}\ and\ \bibinfo {author} {\bibfnamefont {G.}~\bibnamefont
  {Esposito-Farese}},\ }\bibfield  {title} {\bibinfo {title} {{Nonperturbative
  strong field effects in tensor - scalar theories of gravitation}},\ }\href
  {https://doi.org/10.1103/PhysRevLett.70.2220} {\bibfield  {journal} {\bibinfo
   {journal} {Phys. Rev. Lett.}\ }\textbf {\bibinfo {volume} {70}},\ \bibinfo
  {pages} {2220} (\bibinfo {year} {1993})}\BibitemShut {NoStop}%
\bibitem [{\citenamefont {Damour}\ and\ \citenamefont
  {Esposito-Farese}(1996)}]{Damour:1996ke}%
  \BibitemOpen
  \bibfield  {author} {\bibinfo {author} {\bibfnamefont {T.}~\bibnamefont
  {Damour}}\ and\ \bibinfo {author} {\bibfnamefont {G.}~\bibnamefont
  {Esposito-Farese}},\ }\bibfield  {title} {\bibinfo {title} {{Tensor - scalar
  gravity and binary pulsar experiments}},\ }\href
  {https://doi.org/10.1103/PhysRevD.54.1474} {\bibfield  {journal} {\bibinfo
  {journal} {Phys. Rev. D}\ }\textbf {\bibinfo {volume} {54}},\ \bibinfo
  {pages} {1474} (\bibinfo {year} {1996})},\ \Eprint
  {https://arxiv.org/abs/gr-qc/9602056} {arXiv:gr-qc/9602056} \BibitemShut
  {NoStop}%
\bibitem [{\citenamefont {Doneva}\ and\ \citenamefont
  {Yazadjiev}(2018)}]{Doneva:2017bvd}%
  \BibitemOpen
  \bibfield  {author} {\bibinfo {author} {\bibfnamefont {D.~D.}\ \bibnamefont
  {Doneva}}\ and\ \bibinfo {author} {\bibfnamefont {S.~S.}\ \bibnamefont
  {Yazadjiev}},\ }\bibfield  {title} {\bibinfo {title} {{New Gauss-Bonnet Black
  Holes with Curvature-Induced Scalarization in Extended Scalar-Tensor
  Theories}},\ }\href {https://doi.org/10.1103/PhysRevLett.120.131103}
  {\bibfield  {journal} {\bibinfo  {journal} {Phys. Rev. Lett.}\ }\textbf
  {\bibinfo {volume} {120}},\ \bibinfo {pages} {131103} (\bibinfo {year}
  {2018})},\ \Eprint {https://arxiv.org/abs/1711.01187} {arXiv:1711.01187
  [gr-qc]} \BibitemShut {NoStop}%
\bibitem [{\citenamefont {Silva}\ \emph {et~al.}(2018)\citenamefont {Silva},
  \citenamefont {Sakstein}, \citenamefont {Gualtieri}, \citenamefont
  {Sotiriou},\ and\ \citenamefont {Berti}}]{Silva:2017uqg}%
  \BibitemOpen
  \bibfield  {author} {\bibinfo {author} {\bibfnamefont {H.~O.}\ \bibnamefont
  {Silva}}, \bibinfo {author} {\bibfnamefont {J.}~\bibnamefont {Sakstein}},
  \bibinfo {author} {\bibfnamefont {L.}~\bibnamefont {Gualtieri}}, \bibinfo
  {author} {\bibfnamefont {T.~P.}\ \bibnamefont {Sotiriou}},\ and\ \bibinfo
  {author} {\bibfnamefont {E.}~\bibnamefont {Berti}},\ }\bibfield  {title}
  {\bibinfo {title} {{Spontaneous scalarization of black holes and compact
  stars from a Gauss-Bonnet coupling}},\ }\href
  {https://doi.org/10.1103/PhysRevLett.120.131104} {\bibfield  {journal}
  {\bibinfo  {journal} {Phys. Rev. Lett.}\ }\textbf {\bibinfo {volume} {120}},\
  \bibinfo {pages} {131104} (\bibinfo {year} {2018})},\ \Eprint
  {https://arxiv.org/abs/1711.02080} {arXiv:1711.02080 [gr-qc]} \BibitemShut
  {NoStop}%
\bibitem [{\citenamefont {Ramazano\u{g}lu}(2017)}]{Ramazanoglu:2017xbl}%
  \BibitemOpen
  \bibfield  {author} {\bibinfo {author} {\bibfnamefont {F.~M.}\ \bibnamefont
  {Ramazano\u{g}lu}},\ }\bibfield  {title} {\bibinfo {title} {{Spontaneous
  growth of vector fields in gravity}},\ }\href
  {https://doi.org/10.1103/PhysRevD.96.064009} {\bibfield  {journal} {\bibinfo
  {journal} {Phys. Rev. D}\ }\textbf {\bibinfo {volume} {96}},\ \bibinfo
  {pages} {064009} (\bibinfo {year} {2017})},\ \Eprint
  {https://arxiv.org/abs/1706.01056} {arXiv:1706.01056 [gr-qc]} \BibitemShut
  {NoStop}%
\bibitem [{\citenamefont {Silva}\ \emph
  {et~al.}(2021{\natexlab{a}})\citenamefont {Silva}, \citenamefont {Coates},
  \citenamefont {Ramazano\u{g}lu},\ and\ \citenamefont
  {Sotiriou}}]{Silva:2021jya}%
  \BibitemOpen
  \bibfield  {author} {\bibinfo {author} {\bibfnamefont {H.~O.}\ \bibnamefont
  {Silva}}, \bibinfo {author} {\bibfnamefont {A.}~\bibnamefont {Coates}},
  \bibinfo {author} {\bibfnamefont {F.~M.}\ \bibnamefont {Ramazano\u{g}lu}},\
  and\ \bibinfo {author} {\bibfnamefont {T.~P.}\ \bibnamefont {Sotiriou}},\
  }\bibfield  {title} {\bibinfo {title} {{The ghost of vector fields in compact
  stars}},\ }\Eprint {https://arxiv.org/abs/2110.04594} {arXiv:2110.04594
  [gr-qc]}  (\bibinfo {year} {2021}{\natexlab{a}})\BibitemShut {NoStop}%
\bibitem [{\citenamefont {Sampson}\ \emph {et~al.}(2014)\citenamefont
  {Sampson}, \citenamefont {Cornish},\ and\ \citenamefont
  {Yunes}}]{Sampson:2013jpa}%
  \BibitemOpen
  \bibfield  {author} {\bibinfo {author} {\bibfnamefont {L.}~\bibnamefont
  {Sampson}}, \bibinfo {author} {\bibfnamefont {N.}~\bibnamefont {Cornish}},\
  and\ \bibinfo {author} {\bibfnamefont {N.}~\bibnamefont {Yunes}},\ }\bibfield
   {title} {\bibinfo {title} {{Mismodeling in gravitational-wave astronomy: The
  trouble with templates}},\ }\href
  {https://doi.org/10.1103/PhysRevD.89.064037} {\bibfield  {journal} {\bibinfo
  {journal} {Phys. Rev. D}\ }\textbf {\bibinfo {volume} {89}},\ \bibinfo
  {pages} {064037} (\bibinfo {year} {2014})},\ \Eprint
  {https://arxiv.org/abs/1311.4898} {arXiv:1311.4898 [gr-qc]} \BibitemShut
  {NoStop}%
\bibitem [{\citenamefont {Swendsen}\ and\ \citenamefont
  {Wang}(1986)}]{PhysRevLett.57.2607}%
  \BibitemOpen
  \bibfield  {author} {\bibinfo {author} {\bibfnamefont {R.~H.}\ \bibnamefont
  {Swendsen}}\ and\ \bibinfo {author} {\bibfnamefont {J.-S.}\ \bibnamefont
  {Wang}},\ }\bibfield  {title} {\bibinfo {title} {Replica monte carlo
  simulation of spin-glasses},\ }\href
  {https://doi.org/10.1103/PhysRevLett.57.2607} {\bibfield  {journal} {\bibinfo
   {journal} {Phys. Rev. Lett.}\ }\textbf {\bibinfo {volume} {57}},\ \bibinfo
  {pages} {2607} (\bibinfo {year} {1986})}\BibitemShut {NoStop}%
\bibitem [{\citenamefont {Earl}\ and\ \citenamefont
  {Deem}(2005)}]{Earl2005ParallelTT}%
  \BibitemOpen
  \bibfield  {author} {\bibinfo {author} {\bibfnamefont {D.~J.}\ \bibnamefont
  {Earl}}\ and\ \bibinfo {author} {\bibfnamefont {M.~W.}\ \bibnamefont
  {Deem}},\ }\bibfield  {title} {\bibinfo {title} {Parallel tempering: theory,
  applications, and new perspectives.},\ }\href@noop {} {\bibfield  {journal}
  {\bibinfo  {journal} {Physical chemistry chemical physics : PCCP}\ }\textbf
  {\bibinfo {volume} {7 23}},\ \bibinfo {pages} {3910} (\bibinfo {year}
  {2005})}\BibitemShut {NoStop}%
\bibitem [{\citenamefont {{Vousden}}\ \emph {et~al.}(2016)\citenamefont
  {{Vousden}}, \citenamefont {{Farr}},\ and\ \citenamefont
  {{Mandel}}}]{2016MNRAS.455.1919V}%
  \BibitemOpen
  \bibfield  {author} {\bibinfo {author} {\bibfnamefont {W.~D.}\ \bibnamefont
  {{Vousden}}}, \bibinfo {author} {\bibfnamefont {W.~M.}\ \bibnamefont
  {{Farr}}},\ and\ \bibinfo {author} {\bibfnamefont {I.}~\bibnamefont
  {{Mandel}}},\ }\bibfield  {title} {\bibinfo {title} {{Dynamic temperature
  selection for parallel tempering in Markov chain Monte Carlo simulations}},\
  }\href {https://doi.org/10.1093/mnras/stv2422} {\bibfield  {journal}
  {\bibinfo  {journal} {\mnras}\ }\textbf {\bibinfo {volume} {455}},\ \bibinfo
  {pages} {1919} (\bibinfo {year} {2016})},\ \Eprint
  {https://arxiv.org/abs/1501.05823} {arXiv:1501.05823 [astro-ph.IM]}
  \BibitemShut {NoStop}%
\bibitem [{\citenamefont {Harris}\ \emph {et~al.}(2020)\citenamefont {Harris},
  \citenamefont {Millman}, \citenamefont {van~der Walt}, \citenamefont
  {Gommers}, \citenamefont {Virtanen}, \citenamefont {Cournapeau},
  \citenamefont {Wieser}, \citenamefont {Taylor}, \citenamefont {Berg},
  \citenamefont {Smith}, \citenamefont {Kern}, \citenamefont {Picus},
  \citenamefont {Hoyer}, \citenamefont {van Kerkwijk}, \citenamefont {Brett},
  \citenamefont {Haldane}, \citenamefont {del R{\'{i}}o}, \citenamefont
  {Wiebe}, \citenamefont {Peterson}, \citenamefont {G{\'{e}}rard-Marchant},
  \citenamefont {Sheppard}, \citenamefont {Reddy}, \citenamefont {Weckesser},
  \citenamefont {Abbasi}, \citenamefont {Gohlke},\ and\ \citenamefont
  {Oliphant}}]{harris2020array}%
  \BibitemOpen
  \bibfield  {author} {\bibinfo {author} {\bibfnamefont {C.~R.}\ \bibnamefont
  {Harris}}, \bibinfo {author} {\bibfnamefont {K.~J.}\ \bibnamefont {Millman}},
  \bibinfo {author} {\bibfnamefont {S.~J.}\ \bibnamefont {van~der Walt}},
  \bibinfo {author} {\bibfnamefont {R.}~\bibnamefont {Gommers}}, \bibinfo
  {author} {\bibfnamefont {P.}~\bibnamefont {Virtanen}}, \bibinfo {author}
  {\bibfnamefont {D.}~\bibnamefont {Cournapeau}}, \bibinfo {author}
  {\bibfnamefont {E.}~\bibnamefont {Wieser}}, \bibinfo {author} {\bibfnamefont
  {J.}~\bibnamefont {Taylor}}, \bibinfo {author} {\bibfnamefont
  {S.}~\bibnamefont {Berg}}, \bibinfo {author} {\bibfnamefont {N.~J.}\
  \bibnamefont {Smith}}, \bibinfo {author} {\bibfnamefont {R.}~\bibnamefont
  {Kern}}, \bibinfo {author} {\bibfnamefont {M.}~\bibnamefont {Picus}},
  \bibinfo {author} {\bibfnamefont {S.}~\bibnamefont {Hoyer}}, \bibinfo
  {author} {\bibfnamefont {M.~H.}\ \bibnamefont {van Kerkwijk}}, \bibinfo
  {author} {\bibfnamefont {M.}~\bibnamefont {Brett}}, \bibinfo {author}
  {\bibfnamefont {A.}~\bibnamefont {Haldane}}, \bibinfo {author} {\bibfnamefont
  {J.~F.}\ \bibnamefont {del R{\'{i}}o}}, \bibinfo {author} {\bibfnamefont
  {M.}~\bibnamefont {Wiebe}}, \bibinfo {author} {\bibfnamefont
  {P.}~\bibnamefont {Peterson}}, \bibinfo {author} {\bibfnamefont
  {P.}~\bibnamefont {G{\'{e}}rard-Marchant}}, \bibinfo {author} {\bibfnamefont
  {K.}~\bibnamefont {Sheppard}}, \bibinfo {author} {\bibfnamefont
  {T.}~\bibnamefont {Reddy}}, \bibinfo {author} {\bibfnamefont
  {W.}~\bibnamefont {Weckesser}}, \bibinfo {author} {\bibfnamefont
  {H.}~\bibnamefont {Abbasi}}, \bibinfo {author} {\bibfnamefont
  {C.}~\bibnamefont {Gohlke}},\ and\ \bibinfo {author} {\bibfnamefont {T.~E.}\
  \bibnamefont {Oliphant}},\ }\bibfield  {title} {\bibinfo {title} {Array
  programming with {NumPy}},\ }\href
  {https://doi.org/10.1038/s41586-020-2649-2} {\bibfield  {journal} {\bibinfo
  {journal} {Nature}\ }\textbf {\bibinfo {volume} {585}},\ \bibinfo {pages}
  {357} (\bibinfo {year} {2020})}\BibitemShut {NoStop}%
\bibitem [{\citenamefont {Husa}\ \emph {et~al.}(2016)\citenamefont {Husa},
  \citenamefont {Khan}, \citenamefont {Hannam}, \citenamefont {P\"urrer},
  \citenamefont {Ohme}, \citenamefont {Jim\'enez~Forteza},\ and\ \citenamefont
  {Boh\'e}}]{Husa:2015iqa}%
  \BibitemOpen
  \bibfield  {author} {\bibinfo {author} {\bibfnamefont {S.}~\bibnamefont
  {Husa}}, \bibinfo {author} {\bibfnamefont {S.}~\bibnamefont {Khan}}, \bibinfo
  {author} {\bibfnamefont {M.}~\bibnamefont {Hannam}}, \bibinfo {author}
  {\bibfnamefont {M.}~\bibnamefont {P\"urrer}}, \bibinfo {author}
  {\bibfnamefont {F.}~\bibnamefont {Ohme}}, \bibinfo {author} {\bibfnamefont
  {X.}~\bibnamefont {Jim\'enez~Forteza}},\ and\ \bibinfo {author}
  {\bibfnamefont {A.}~\bibnamefont {Boh\'e}},\ }\bibfield  {title} {\bibinfo
  {title} {{Frequency-domain gravitational waves from nonprecessing black-hole
  binaries. I. New numerical waveforms and anatomy of the signal}},\ }\href
  {https://doi.org/10.1103/PhysRevD.93.044006} {\bibfield  {journal} {\bibinfo
  {journal} {Phys. Rev. D}\ }\textbf {\bibinfo {volume} {93}},\ \bibinfo
  {pages} {044006} (\bibinfo {year} {2016})},\ \Eprint
  {https://arxiv.org/abs/1508.07250} {arXiv:1508.07250 [gr-qc]} \BibitemShut
  {NoStop}%
\bibitem [{\citenamefont {Khan}\ \emph {et~al.}(2016)\citenamefont {Khan},
  \citenamefont {Husa}, \citenamefont {Hannam}, \citenamefont {Ohme},
  \citenamefont {P\"urrer}, \citenamefont {Jim\'enez~Forteza},\ and\
  \citenamefont {Boh\'e}}]{Khan:2015jqa}%
  \BibitemOpen
  \bibfield  {author} {\bibinfo {author} {\bibfnamefont {S.}~\bibnamefont
  {Khan}}, \bibinfo {author} {\bibfnamefont {S.}~\bibnamefont {Husa}}, \bibinfo
  {author} {\bibfnamefont {M.}~\bibnamefont {Hannam}}, \bibinfo {author}
  {\bibfnamefont {F.}~\bibnamefont {Ohme}}, \bibinfo {author} {\bibfnamefont
  {M.}~\bibnamefont {P\"urrer}}, \bibinfo {author} {\bibfnamefont
  {X.}~\bibnamefont {Jim\'enez~Forteza}},\ and\ \bibinfo {author}
  {\bibfnamefont {A.}~\bibnamefont {Boh\'e}},\ }\bibfield  {title} {\bibinfo
  {title} {{Frequency-domain gravitational waves from nonprecessing black-hole
  binaries. II. A phenomenological model for the advanced detector era}},\
  }\href {https://doi.org/10.1103/PhysRevD.93.044007} {\bibfield  {journal}
  {\bibinfo  {journal} {Phys. Rev. D}\ }\textbf {\bibinfo {volume} {93}},\
  \bibinfo {pages} {044007} (\bibinfo {year} {2016})},\ \Eprint
  {https://arxiv.org/abs/1508.07253} {arXiv:1508.07253 [gr-qc]} \BibitemShut
  {NoStop}%
\bibitem [{\citenamefont {Chatziioannou}\ \emph {et~al.}(2017)\citenamefont
  {Chatziioannou}, \citenamefont {Klein}, \citenamefont {Yunes},\ and\
  \citenamefont {Cornish}}]{Chatziioannou:2017tdw}%
  \BibitemOpen
  \bibfield  {author} {\bibinfo {author} {\bibfnamefont {K.}~\bibnamefont
  {Chatziioannou}}, \bibinfo {author} {\bibfnamefont {A.}~\bibnamefont
  {Klein}}, \bibinfo {author} {\bibfnamefont {N.}~\bibnamefont {Yunes}},\ and\
  \bibinfo {author} {\bibfnamefont {N.}~\bibnamefont {Cornish}},\ }\bibfield
  {title} {\bibinfo {title} {{Constructing Gravitational Waves from Generic
  Spin-Precessing Compact Binary Inspirals}},\ }\href
  {https://doi.org/10.1103/PhysRevD.95.104004} {\bibfield  {journal} {\bibinfo
  {journal} {Phys. Rev. D}\ }\textbf {\bibinfo {volume} {95}},\ \bibinfo
  {pages} {104004} (\bibinfo {year} {2017})},\ \Eprint
  {https://arxiv.org/abs/1703.03967} {arXiv:1703.03967 [gr-qc]} \BibitemShut
  {NoStop}%
\bibitem [{\citenamefont {Gupta}\ \emph {et~al.}(2020)\citenamefont {Gupta},
  \citenamefont {Datta}, \citenamefont {Kastha}, \citenamefont {Borhanian},
  \citenamefont {Arun},\ and\ \citenamefont {Sathyaprakash}}]{Gupta:2020lxa}%
  \BibitemOpen
  \bibfield  {author} {\bibinfo {author} {\bibfnamefont {A.}~\bibnamefont
  {Gupta}}, \bibinfo {author} {\bibfnamefont {S.}~\bibnamefont {Datta}},
  \bibinfo {author} {\bibfnamefont {S.}~\bibnamefont {Kastha}}, \bibinfo
  {author} {\bibfnamefont {S.}~\bibnamefont {Borhanian}}, \bibinfo {author}
  {\bibfnamefont {K.~G.}\ \bibnamefont {Arun}},\ and\ \bibinfo {author}
  {\bibfnamefont {B.~S.}\ \bibnamefont {Sathyaprakash}},\ }\bibfield  {title}
  {\bibinfo {title} {{Multiparameter tests of general relativity using
  multiband gravitational-wave observations}},\ }\href
  {https://doi.org/10.1103/PhysRevLett.125.201101} {\bibfield  {journal}
  {\bibinfo  {journal} {Phys. Rev. Lett.}\ }\textbf {\bibinfo {volume} {125}},\
  \bibinfo {pages} {201101} (\bibinfo {year} {2020})},\ \Eprint
  {https://arxiv.org/abs/2005.09607} {arXiv:2005.09607 [gr-qc]} \BibitemShut
  {NoStop}%
\bibitem [{\citenamefont {Blanchet}(2002)}]{Blanchet:2002xy}%
  \BibitemOpen
  \bibfield  {author} {\bibinfo {author} {\bibfnamefont {L.}~\bibnamefont
  {Blanchet}},\ }\bibfield  {title} {\bibinfo {title} {{On the accuracy of the
  postNewtonian approximation}},\ }in\ \href
  {https://doi.org/10.1142/9789812791368_0022} {\emph {\bibinfo {booktitle}
  {{25th Johns Hopkins Workshop on Current Problems in Particle Theory: 2001: A
  Relativistic Spacetime Odyssey. Experiments and Theoretical Viewpoints on
  General Relativity and Quantum Gravity}}}}\ (\bibinfo {year} {2002})\ pp.\
  \bibinfo {pages} {411--430},\ \Eprint {https://arxiv.org/abs/gr-qc/0207037}
  {arXiv:gr-qc/0207037} \BibitemShut {NoStop}%
\bibitem [{\citenamefont {Yunes}\ and\ \citenamefont
  {Berti}(2008)}]{Yunes:2008tw}%
  \BibitemOpen
  \bibfield  {author} {\bibinfo {author} {\bibfnamefont {N.}~\bibnamefont
  {Yunes}}\ and\ \bibinfo {author} {\bibfnamefont {E.}~\bibnamefont {Berti}},\
  }\bibfield  {title} {\bibinfo {title} {{Accuracy of the post-Newtonian
  approximation: Optimal asymptotic expansion for quasicircular, extreme-mass
  ratio inspirals}},\ }\href {https://doi.org/10.1103/PhysRevD.77.124006}
  {\bibfield  {journal} {\bibinfo  {journal} {Phys. Rev. D}\ }\textbf {\bibinfo
  {volume} {77}},\ \bibinfo {pages} {124006} (\bibinfo {year} {2008})},\
  \bibinfo {note} {[Erratum: Phys.Rev.D 83, 109901 (2011)]},\ \Eprint
  {https://arxiv.org/abs/0803.1853} {arXiv:0803.1853 [gr-qc]} \BibitemShut
  {NoStop}%
\bibitem [{\citenamefont {Damour}\ \emph {et~al.}(2001)\citenamefont {Damour},
  \citenamefont {Iyer},\ and\ \citenamefont {Sathyaprakash}}]{Damour:2000zb}%
  \BibitemOpen
  \bibfield  {author} {\bibinfo {author} {\bibfnamefont {T.}~\bibnamefont
  {Damour}}, \bibinfo {author} {\bibfnamefont {B.~R.}\ \bibnamefont {Iyer}},\
  and\ \bibinfo {author} {\bibfnamefont {B.~S.}\ \bibnamefont
  {Sathyaprakash}},\ }\bibfield  {title} {\bibinfo {title} {{A Comparison of
  search templates for gravitational waves from binary inspiral}},\ }\href
  {https://doi.org/10.1103/PhysRevD.63.044023} {\bibfield  {journal} {\bibinfo
  {journal} {Phys. Rev. D}\ }\textbf {\bibinfo {volume} {63}},\ \bibinfo
  {pages} {044023} (\bibinfo {year} {2001})},\ \bibinfo {note} {[Erratum:
  Phys.Rev.D 72, 029902 (2005)]},\ \Eprint
  {https://arxiv.org/abs/gr-qc/0010009} {arXiv:gr-qc/0010009} \BibitemShut
  {NoStop}%
\bibitem [{\citenamefont {Damour}\ \emph {et~al.}(2002)\citenamefont {Damour},
  \citenamefont {Iyer},\ and\ \citenamefont {Sathyaprakash}}]{Damour:2002kr}%
  \BibitemOpen
  \bibfield  {author} {\bibinfo {author} {\bibfnamefont {T.}~\bibnamefont
  {Damour}}, \bibinfo {author} {\bibfnamefont {B.~R.}\ \bibnamefont {Iyer}},\
  and\ \bibinfo {author} {\bibfnamefont {B.~S.}\ \bibnamefont
  {Sathyaprakash}},\ }\bibfield  {title} {\bibinfo {title} {{A Comparison of
  search templates for gravitational waves from binary inspiral - 3.5PN
  update}},\ }\href {https://doi.org/10.1103/PhysRevD.66.027502} {\bibfield
  {journal} {\bibinfo  {journal} {Phys. Rev. D}\ }\textbf {\bibinfo {volume}
  {66}},\ \bibinfo {pages} {027502} (\bibinfo {year} {2002})},\ \Eprint
  {https://arxiv.org/abs/gr-qc/0207021} {arXiv:gr-qc/0207021} \BibitemShut
  {NoStop}%
\bibitem [{\citenamefont {Arun}\ \emph {et~al.}(2005)\citenamefont {Arun},
  \citenamefont {Iyer}, \citenamefont {Sathyaprakash},\ and\ \citenamefont
  {Sundararajan}}]{Arun:2004hn}%
  \BibitemOpen
  \bibfield  {author} {\bibinfo {author} {\bibfnamefont {K.~G.}\ \bibnamefont
  {Arun}}, \bibinfo {author} {\bibfnamefont {B.~R.}\ \bibnamefont {Iyer}},
  \bibinfo {author} {\bibfnamefont {B.~S.}\ \bibnamefont {Sathyaprakash}},\
  and\ \bibinfo {author} {\bibfnamefont {P.~A.}\ \bibnamefont {Sundararajan}},\
  }\bibfield  {title} {\bibinfo {title} {{Parameter estimation of inspiralling
  compact binaries using 3.5 post-Newtonian gravitational wave phasing: The
  Non-spinning case}},\ }\href {https://doi.org/10.1103/PhysRevD.71.084008}
  {\bibfield  {journal} {\bibinfo  {journal} {Phys. Rev. D}\ }\textbf {\bibinfo
  {volume} {71}},\ \bibinfo {pages} {084008} (\bibinfo {year} {2005})},\
  \bibinfo {note} {[Erratum: Phys.Rev.D 72, 069903 (2005)]},\ \Eprint
  {https://arxiv.org/abs/gr-qc/0411146} {arXiv:gr-qc/0411146} \BibitemShut
  {NoStop}%
\bibitem [{\citenamefont {Aasi}\ \emph {et~al.}(2015)\citenamefont {Aasi} \emph
  {et~al.}}]{TheLIGOScientific:2014jea}%
  \BibitemOpen
  \bibfield  {author} {\bibinfo {author} {\bibfnamefont {J.}~\bibnamefont
  {Aasi}} \emph {et~al.} (\bibinfo {collaboration} {LIGO Scientific}),\
  }\bibfield  {title} {\bibinfo {title} {{Advanced LIGO}},\ }\href
  {https://doi.org/10.1088/0264-9381/32/7/074001} {\bibfield  {journal}
  {\bibinfo  {journal} {Class. Quant. Grav.}\ }\textbf {\bibinfo {volume}
  {32}},\ \bibinfo {pages} {074001} (\bibinfo {year} {2015})},\ \Eprint
  {https://arxiv.org/abs/1411.4547} {arXiv:1411.4547 [gr-qc]} \BibitemShut
  {NoStop}%
\bibitem [{\citenamefont {Acernese}\ \emph {et~al.}(2015)\citenamefont
  {Acernese} \emph {et~al.}}]{VIRGO:2014yos}%
  \BibitemOpen
  \bibfield  {author} {\bibinfo {author} {\bibfnamefont {F.}~\bibnamefont
  {Acernese}} \emph {et~al.} (\bibinfo {collaboration} {VIRGO}),\ }\bibfield
  {title} {\bibinfo {title} {{Advanced Virgo: a second-generation
  interferometric gravitational wave detector}},\ }\href
  {https://doi.org/10.1088/0264-9381/32/2/024001} {\bibfield  {journal}
  {\bibinfo  {journal} {Class. Quant. Grav.}\ }\textbf {\bibinfo {volume}
  {32}},\ \bibinfo {pages} {024001} (\bibinfo {year} {2015})},\ \Eprint
  {https://arxiv.org/abs/1408.3978} {arXiv:1408.3978 [gr-qc]} \BibitemShut
  {NoStop}%
\bibitem [{\citenamefont {O'Reilly}\ \emph {et~al.}(2020)\citenamefont
  {O'Reilly}, \citenamefont {Branchesi}, \citenamefont {Haino},\ and\
  \citenamefont {Gemme}}]{ligo_SN_forecast}%
  \BibitemOpen
  \bibfield  {author} {\bibinfo {author} {\bibfnamefont {B.}~\bibnamefont
  {O'Reilly}}, \bibinfo {author} {\bibfnamefont {M.}~\bibnamefont {Branchesi}},
  \bibinfo {author} {\bibfnamefont {S.}~\bibnamefont {Haino}},\ and\ \bibinfo
  {author} {\bibfnamefont {G.}~\bibnamefont {Gemme}},\ }\href
  {https://dcc.ligo.org/LIGO-T2000012/public} {\emph {\bibinfo {title} {{LIGO
  Document T2000012-v1}}}},\ \bibinfo {type} {Tech. Rep.}\ (\bibinfo {year}
  {2020})\BibitemShut {NoStop}%
\bibitem [{\citenamefont {{Dwyer}}\ \emph {et~al.}(2015)\citenamefont
  {{Dwyer}}, \citenamefont {{Sigg}}, \citenamefont {{Ballmer}}, \citenamefont
  {{Barsotti}}, \citenamefont {{Mavalvala}},\ and\ \citenamefont
  {{Evans}}}]{2015PhRvD..91h2001D}%
  \BibitemOpen
  \bibfield  {author} {\bibinfo {author} {\bibfnamefont {S.}~\bibnamefont
  {{Dwyer}}}, \bibinfo {author} {\bibfnamefont {D.}~\bibnamefont {{Sigg}}},
  \bibinfo {author} {\bibfnamefont {S.~W.}\ \bibnamefont {{Ballmer}}}, \bibinfo
  {author} {\bibfnamefont {L.}~\bibnamefont {{Barsotti}}}, \bibinfo {author}
  {\bibfnamefont {N.}~\bibnamefont {{Mavalvala}}},\ and\ \bibinfo {author}
  {\bibfnamefont {M.}~\bibnamefont {{Evans}}},\ }\bibfield  {title} {\bibinfo
  {title} {{Gravitational wave detector with cosmological reach}},\ }\href
  {https://doi.org/10.1103/PhysRevD.91.082001} {\bibfield  {journal} {\bibinfo
  {journal} {\prd}\ }\textbf {\bibinfo {volume} {91}},\ \bibinfo {eid} {082001}
  (\bibinfo {year} {2015})},\ \Eprint {https://arxiv.org/abs/1410.0612}
  {arXiv:1410.0612 [astro-ph.IM]} \BibitemShut {NoStop}%
\bibitem [{\citenamefont {Punturo}\ \emph {et~al.}(2010)\citenamefont {Punturo}
  \emph {et~al.}}]{Punturo:2010zza}%
  \BibitemOpen
  \bibfield  {author} {\bibinfo {author} {\bibfnamefont {M.}~\bibnamefont
  {Punturo}} \emph {et~al.},\ }\bibfield  {title} {\bibinfo {title} {{The third
  generation of gravitational wave observatories and their science reach}},\
  }\href {https://doi.org/10.1088/0264-9381/27/8/084007} {\bibfield  {journal}
  {\bibinfo  {journal} {Class. Quant. Grav.}\ }\textbf {\bibinfo {volume}
  {27}},\ \bibinfo {pages} {084007} (\bibinfo {year} {2010})}\BibitemShut
  {NoStop}%
\bibitem [{\citenamefont {{Cosmic Explorer}}(2020)}]{CE_psd}%
  \BibitemOpen
  \bibfield  {author} {\bibinfo {author} {\bibnamefont {{Cosmic Explorer}}},\
  }\href {https://cosmicexplorer.org/researchers.html} {\bibinfo {title}
  {{https://cosmicexplorer.org/researchers.html}}} (\bibinfo {year}
  {2020})\BibitemShut {NoStop}%
\bibitem [{\citenamefont {Hild}\ \emph {et~al.}(2011)\citenamefont {Hild} \emph
  {et~al.}}]{Hild:2010id}%
  \BibitemOpen
  \bibfield  {author} {\bibinfo {author} {\bibfnamefont {S.}~\bibnamefont
  {Hild}} \emph {et~al.},\ }\bibfield  {title} {\bibinfo {title} {{Sensitivity
  Studies for Third-Generation Gravitational Wave Observatories}},\ }\href
  {https://doi.org/10.1088/0264-9381/28/9/094013} {\bibfield  {journal}
  {\bibinfo  {journal} {Class. Quant. Grav.}\ }\textbf {\bibinfo {volume}
  {28}},\ \bibinfo {pages} {094013} (\bibinfo {year} {2011})},\ \Eprint
  {https://arxiv.org/abs/1012.0908} {arXiv:1012.0908 [gr-qc]} \BibitemShut
  {NoStop}%
\bibitem [{\citenamefont {Kanti}\ \emph {et~al.}(1996)\citenamefont {Kanti},
  \citenamefont {Mavromatos}, \citenamefont {Rizos}, \citenamefont {Tamvakis},\
  and\ \citenamefont {Winstanley}}]{Kanti:1995vq}%
  \BibitemOpen
  \bibfield  {author} {\bibinfo {author} {\bibfnamefont {P.}~\bibnamefont
  {Kanti}}, \bibinfo {author} {\bibfnamefont {N.~E.}\ \bibnamefont
  {Mavromatos}}, \bibinfo {author} {\bibfnamefont {J.}~\bibnamefont {Rizos}},
  \bibinfo {author} {\bibfnamefont {K.}~\bibnamefont {Tamvakis}},\ and\
  \bibinfo {author} {\bibfnamefont {E.}~\bibnamefont {Winstanley}},\ }\bibfield
   {title} {\bibinfo {title} {{Dilatonic black holes in higher curvature string
  gravity}},\ }\href {https://doi.org/10.1103/PhysRevD.54.5049} {\bibfield
  {journal} {\bibinfo  {journal} {Phys. Rev. D}\ }\textbf {\bibinfo {volume}
  {54}},\ \bibinfo {pages} {5049} (\bibinfo {year} {1996})},\ \Eprint
  {https://arxiv.org/abs/hep-th/9511071} {arXiv:hep-th/9511071} \BibitemShut
  {NoStop}%
\bibitem [{\citenamefont {Boulware}\ and\ \citenamefont
  {Deser}(1985)}]{Boulware:1985wk}%
  \BibitemOpen
  \bibfield  {author} {\bibinfo {author} {\bibfnamefont {D.~G.}\ \bibnamefont
  {Boulware}}\ and\ \bibinfo {author} {\bibfnamefont {S.}~\bibnamefont
  {Deser}},\ }\bibfield  {title} {\bibinfo {title} {{String Generated Gravity
  Models}},\ }\href {https://doi.org/10.1103/PhysRevLett.55.2656} {\bibfield
  {journal} {\bibinfo  {journal} {Phys. Rev. Lett.}\ }\textbf {\bibinfo
  {volume} {55}},\ \bibinfo {pages} {2656} (\bibinfo {year}
  {1985})}\BibitemShut {NoStop}%
\bibitem [{\citenamefont {{Gross}}\ and\ \citenamefont
  {{Sloan}}(1987)}]{1987NuPhB.291...41G}%
  \BibitemOpen
  \bibfield  {author} {\bibinfo {author} {\bibfnamefont {D.~J.}\ \bibnamefont
  {{Gross}}}\ and\ \bibinfo {author} {\bibfnamefont {J.~H.}\ \bibnamefont
  {{Sloan}}},\ }\bibfield  {title} {\bibinfo {title} {{The quartic effective
  action for the heterotic string}},\ }\href
  {https://doi.org/10.1016/0550-3213(87)90465-2} {\bibfield  {journal}
  {\bibinfo  {journal} {Nuclear Physics B}\ }\textbf {\bibinfo {volume}
  {291}},\ \bibinfo {pages} {41} (\bibinfo {year} {1987})}\BibitemShut
  {NoStop}%
\bibitem [{\citenamefont {Yunes}\ and\ \citenamefont
  {Stein}(2011)}]{Yunes:2011we}%
  \BibitemOpen
  \bibfield  {author} {\bibinfo {author} {\bibfnamefont {N.}~\bibnamefont
  {Yunes}}\ and\ \bibinfo {author} {\bibfnamefont {L.~C.}\ \bibnamefont
  {Stein}},\ }\bibfield  {title} {\bibinfo {title} {{Non-Spinning Black Holes
  in Alternative Theories of Gravity}},\ }\href
  {https://doi.org/10.1103/PhysRevD.83.104002} {\bibfield  {journal} {\bibinfo
  {journal} {Phys. Rev. D}\ }\textbf {\bibinfo {volume} {83}},\ \bibinfo
  {pages} {104002} (\bibinfo {year} {2011})},\ \Eprint
  {https://arxiv.org/abs/1101.2921} {arXiv:1101.2921 [gr-qc]} \BibitemShut
  {NoStop}%
\bibitem [{\citenamefont {Yagi}\ \emph {et~al.}(2016)\citenamefont {Yagi},
  \citenamefont {Stein},\ and\ \citenamefont {Yunes}}]{Yagi:2015oca}%
  \BibitemOpen
  \bibfield  {author} {\bibinfo {author} {\bibfnamefont {K.}~\bibnamefont
  {Yagi}}, \bibinfo {author} {\bibfnamefont {L.~C.}\ \bibnamefont {Stein}},\
  and\ \bibinfo {author} {\bibfnamefont {N.}~\bibnamefont {Yunes}},\ }\bibfield
   {title} {\bibinfo {title} {{Challenging the Presence of Scalar Charge and
  Dipolar Radiation in Binary Pulsars}},\ }\href
  {https://doi.org/10.1103/PhysRevD.93.024010} {\bibfield  {journal} {\bibinfo
  {journal} {Phys. Rev. D}\ }\textbf {\bibinfo {volume} {93}},\ \bibinfo
  {pages} {024010} (\bibinfo {year} {2016})},\ \Eprint
  {https://arxiv.org/abs/1510.02152} {arXiv:1510.02152 [gr-qc]} \BibitemShut
  {NoStop}%
\bibitem [{\citenamefont {Sotiriou}\ and\ \citenamefont
  {Zhou}(2014)}]{Sotiriou:2014pfa}%
  \BibitemOpen
  \bibfield  {author} {\bibinfo {author} {\bibfnamefont {T.~P.}\ \bibnamefont
  {Sotiriou}}\ and\ \bibinfo {author} {\bibfnamefont {S.-Y.}\ \bibnamefont
  {Zhou}},\ }\bibfield  {title} {\bibinfo {title} {{Black hole hair in
  generalized scalar-tensor gravity: An explicit example}},\ }\href
  {https://doi.org/10.1103/PhysRevD.90.124063} {\bibfield  {journal} {\bibinfo
  {journal} {Phys. Rev. D}\ }\textbf {\bibinfo {volume} {90}},\ \bibinfo
  {pages} {124063} (\bibinfo {year} {2014})},\ \Eprint
  {https://arxiv.org/abs/1408.1698} {arXiv:1408.1698 [gr-qc]} \BibitemShut
  {NoStop}%
\bibitem [{\citenamefont {Yagi}\ \emph {et~al.}(2012)\citenamefont {Yagi},
  \citenamefont {Stein}, \citenamefont {Yunes},\ and\ \citenamefont
  {Tanaka}}]{Yagi:2011xp}%
  \BibitemOpen
  \bibfield  {author} {\bibinfo {author} {\bibfnamefont {K.}~\bibnamefont
  {Yagi}}, \bibinfo {author} {\bibfnamefont {L.~C.}\ \bibnamefont {Stein}},
  \bibinfo {author} {\bibfnamefont {N.}~\bibnamefont {Yunes}},\ and\ \bibinfo
  {author} {\bibfnamefont {T.}~\bibnamefont {Tanaka}},\ }\bibfield  {title}
  {\bibinfo {title} {{Post-Newtonian, Quasi-Circular Binary Inspirals in
  Quadratic Modified Gravity}},\ }\href
  {https://doi.org/10.1103/PhysRevD.85.064022} {\bibfield  {journal} {\bibinfo
  {journal} {Phys. Rev. D}\ }\textbf {\bibinfo {volume} {85}},\ \bibinfo
  {pages} {064022} (\bibinfo {year} {2012})},\ \bibinfo {note} {[Erratum:
  Phys.Rev.D 93, 029902 (2016)]},\ \Eprint {https://arxiv.org/abs/1110.5950}
  {arXiv:1110.5950 [gr-qc]} \BibitemShut {NoStop}%
\bibitem [{\citenamefont {Nair}\ \emph {et~al.}(2019)\citenamefont {Nair},
  \citenamefont {Perkins}, \citenamefont {Silva},\ and\ \citenamefont
  {Yunes}}]{Nair:2019iur}%
  \BibitemOpen
  \bibfield  {author} {\bibinfo {author} {\bibfnamefont {R.}~\bibnamefont
  {Nair}}, \bibinfo {author} {\bibfnamefont {S.}~\bibnamefont {Perkins}},
  \bibinfo {author} {\bibfnamefont {H.~O.}\ \bibnamefont {Silva}},\ and\
  \bibinfo {author} {\bibfnamefont {N.}~\bibnamefont {Yunes}},\ }\bibfield
  {title} {\bibinfo {title} {{Fundamental Physics Implications for
  Higher-Curvature Theories from Binary Black Hole Signals in the LIGO-Virgo
  Catalog GWTC-1}},\ }\href {https://doi.org/10.1103/PhysRevLett.123.191101}
  {\bibfield  {journal} {\bibinfo  {journal} {Phys. Rev. Lett.}\ }\textbf
  {\bibinfo {volume} {123}},\ \bibinfo {pages} {191101} (\bibinfo {year}
  {2019})},\ \Eprint {https://arxiv.org/abs/1905.00870} {arXiv:1905.00870
  [gr-qc]} \BibitemShut {NoStop}%
\bibitem [{\citenamefont {Lyu}\ \emph {et~al.}(2022)\citenamefont {Lyu},
  \citenamefont {Jiang},\ and\ \citenamefont {Yagi}}]{Lyu:2022gdr}%
  \BibitemOpen
  \bibfield  {author} {\bibinfo {author} {\bibfnamefont {Z.}~\bibnamefont
  {Lyu}}, \bibinfo {author} {\bibfnamefont {N.}~\bibnamefont {Jiang}},\ and\
  \bibinfo {author} {\bibfnamefont {K.}~\bibnamefont {Yagi}},\ }\bibfield
  {title} {\bibinfo {title} {{Constraints on Einstein-dilation-Gauss-Bonnet
  gravity from black hole-neutron star gravitational wave events}},\ }\href
  {https://doi.org/10.1103/PhysRevD.105.064001} {\bibfield  {journal} {\bibinfo
   {journal} {Phys. Rev. D}\ }\textbf {\bibinfo {volume} {105}},\ \bibinfo
  {pages} {064001} (\bibinfo {year} {2022})},\ \Eprint
  {https://arxiv.org/abs/2201.02543} {arXiv:2201.02543 [gr-qc]} \BibitemShut
  {NoStop}%
\bibitem [{\citenamefont {Saleem}\ \emph {et~al.}(2021)\citenamefont {Saleem},
  \citenamefont {Datta}, \citenamefont {Arun},\ and\ \citenamefont
  {Sathyaprakash}}]{Saleem:2021nsb}%
  \BibitemOpen
  \bibfield  {author} {\bibinfo {author} {\bibfnamefont {M.}~\bibnamefont
  {Saleem}}, \bibinfo {author} {\bibfnamefont {S.}~\bibnamefont {Datta}},
  \bibinfo {author} {\bibfnamefont {K.~G.}\ \bibnamefont {Arun}},\ and\
  \bibinfo {author} {\bibfnamefont {B.~S.}\ \bibnamefont {Sathyaprakash}},\
  }\bibfield  {title} {\bibinfo {title} {{Parametrized tests of post-Newtonian
  theory using principal component analysis}},\ }\Eprint
  {https://arxiv.org/abs/2110.10147} {arXiv:2110.10147 [gr-qc]}  (\bibinfo
  {year} {2021})\BibitemShut {NoStop}%
\bibitem [{\citenamefont {Shoom}\ \emph {et~al.}(2021)\citenamefont {Shoom},
  \citenamefont {Gupta}, \citenamefont {Krishnan}, \citenamefont {Nielsen},\
  and\ \citenamefont {Capano}}]{Shoom:2021mdj}%
  \BibitemOpen
  \bibfield  {author} {\bibinfo {author} {\bibfnamefont {A.~A.}\ \bibnamefont
  {Shoom}}, \bibinfo {author} {\bibfnamefont {P.~K.}\ \bibnamefont {Gupta}},
  \bibinfo {author} {\bibfnamefont {B.}~\bibnamefont {Krishnan}}, \bibinfo
  {author} {\bibfnamefont {A.~B.}\ \bibnamefont {Nielsen}},\ and\ \bibinfo
  {author} {\bibfnamefont {C.~D.}\ \bibnamefont {Capano}},\ }\bibfield  {title}
  {\bibinfo {title} {{Testing GR with the Gravitational Wave Inspiral Signal
  GW170817}},\ }\Eprint {https://arxiv.org/abs/2105.02191} {arXiv:2105.02191
  [gr-qc]}  (\bibinfo {year} {2021})\BibitemShut {NoStop}%
\bibitem [{\citenamefont {Pai}\ and\ \citenamefont {Arun}(2013)}]{Pai:2012mv}%
  \BibitemOpen
  \bibfield  {author} {\bibinfo {author} {\bibfnamefont {A.}~\bibnamefont
  {Pai}}\ and\ \bibinfo {author} {\bibfnamefont {K.~G.}\ \bibnamefont {Arun}},\
  }\bibfield  {title} {\bibinfo {title} {{Singular value decomposition in
  parametrised tests of post-Newtonian theory}},\ }\href
  {https://doi.org/10.1088/0264-9381/30/2/025011} {\bibfield  {journal}
  {\bibinfo  {journal} {Class. Quant. Grav.}\ }\textbf {\bibinfo {volume}
  {30}},\ \bibinfo {pages} {025011} (\bibinfo {year} {2013})},\ \Eprint
  {https://arxiv.org/abs/1207.1943} {arXiv:1207.1943 [gr-qc]} \BibitemShut
  {NoStop}%
\bibitem [{\citenamefont {Khalil}\ \emph {et~al.}(2019)\citenamefont {Khalil},
  \citenamefont {Sennett}, \citenamefont {Steinhoff},\ and\ \citenamefont
  {Buonanno}}]{Khalil:2019wyy}%
  \BibitemOpen
  \bibfield  {author} {\bibinfo {author} {\bibfnamefont {M.}~\bibnamefont
  {Khalil}}, \bibinfo {author} {\bibfnamefont {N.}~\bibnamefont {Sennett}},
  \bibinfo {author} {\bibfnamefont {J.}~\bibnamefont {Steinhoff}},\ and\
  \bibinfo {author} {\bibfnamefont {A.}~\bibnamefont {Buonanno}},\ }\bibfield
  {title} {\bibinfo {title} {{Theory-agnostic framework for dynamical
  scalarization of compact binaries}},\ }\href
  {https://doi.org/10.1103/PhysRevD.100.124013} {\bibfield  {journal} {\bibinfo
   {journal} {Phys. Rev. D}\ }\textbf {\bibinfo {volume} {100}},\ \bibinfo
  {pages} {124013} (\bibinfo {year} {2019})},\ \Eprint
  {https://arxiv.org/abs/1906.08161} {arXiv:1906.08161 [gr-qc]} \BibitemShut
  {NoStop}%
\bibitem [{\citenamefont {Silva}\ \emph
  {et~al.}(2021{\natexlab{b}})\citenamefont {Silva}, \citenamefont {Witek},
  \citenamefont {Elley},\ and\ \citenamefont {Yunes}}]{Silva:2020omi}%
  \BibitemOpen
  \bibfield  {author} {\bibinfo {author} {\bibfnamefont {H.~O.}\ \bibnamefont
  {Silva}}, \bibinfo {author} {\bibfnamefont {H.}~\bibnamefont {Witek}},
  \bibinfo {author} {\bibfnamefont {M.}~\bibnamefont {Elley}},\ and\ \bibinfo
  {author} {\bibfnamefont {N.}~\bibnamefont {Yunes}},\ }\bibfield  {title}
  {\bibinfo {title} {{Dynamical Descalarization in Binary Black Hole
  Mergers}},\ }\href {https://doi.org/10.1103/PhysRevLett.127.031101}
  {\bibfield  {journal} {\bibinfo  {journal} {Phys. Rev. Lett.}\ }\textbf
  {\bibinfo {volume} {127}},\ \bibinfo {pages} {031101} (\bibinfo {year}
  {2021}{\natexlab{b}})},\ \Eprint {https://arxiv.org/abs/2012.10436}
  {arXiv:2012.10436 [gr-qc]} \BibitemShut {NoStop}%
\end{thebibliography}%

\end{document}